\documentclass[12pt,preprint]{aastex}
\usepackage{epsfig,amssymb}
\usepackage{natbib}
\usepackage{subfigure}
\def\beq{\begin{equation}}
\def\eeq{\end{equation}}
\long\def\symbolfootnote[#1]#2{\begingroup%
\def\thefootnote{\fnsymbol{footnote}}\footnote[#1]{#2}\endgroup}
\begin{document}
\title{Mergers of Black Hole -- Neutron Star binaries. I. Methods and First Results}

\author{Emmanouela Rantsiou\altaffilmark{1}, Shiho Kobayashi\altaffilmark{2}, Pablo Laguna\altaffilmark{3} and Frederic A. Rasio\altaffilmark{1}}
\altaffiltext{1}{Department of Physics and Astronomy, Northwestern University, Evanston, IL 60208, USA.\\
\textit{email:} \texttt{emmanouela@northwestern.edu}, \texttt{rasio@northwestern.edu}}
\altaffiltext{2}{Astrophysics Research Institute, Liverpool John Moores University, Twelve Quays House, Egerton Wharf, Birkenhead CH41 1LD. \\ \textit{email:} \texttt{sk@astro.livjm.ac.uk}}
\altaffiltext{3}{Department of Astronomy and Astrophysics and Department of Physics, Pennsylvania State University, University Park, PA 16802, USA. \\ \textit{email:} \texttt{pablo@astro.psu.edu}}

\begin{abstract}
We use a 3-D relativistic SPH (Smoothed Particle Hydrodynamics)
code to study mergers of black hole -- neutron star (BH--NS) binary systems with low mass
ratios, adopting $q\equiv M_{NS}/M_{BH} \simeq 0.1$ as a representative case. The outcome of such mergers
depends sensitively on both  the magnitude of the BH spin and its obliquity
(i.e., the inclination of the binary orbit with respect to the equatorial plane of the BH). In
particular, only systems with sufficiently high BH spin parameter
$a$ and sufficiently low orbital inclinations allow any NS matter
to escape or to form a long-lived disk outside the BH horizon
after disruption. Mergers of binaries with orbital inclinations
above $\sim60^o$ lead to complete prompt accretion of the entire
NS by the BH, even for the case of an extreme Kerr BH. We find
that the formation of a significant disk or torus of NS material
around the BH always requires a near-maximal BH spin and a low
initial inclination of the NS orbit just prior to merger.

\end{abstract}

\maketitle

\section{Introduction and Motivation}

\subsection{Double Compact Objects as Gravitational Wave Sources}

Over the past two decades, the modelling of double compact objects (DCOs) 
has attracted special interest among theorists,
mainly because such systems are expected to be
strong sources of gravitational waves (GWs). Their inspiral and
merger GW signals cover a wide frequency band, from
$\sim 10^{-4}-10^{-1}\,$Hz for supermassive BH 
binaries of $\sim 10^4-10^7\,M_{\odot}$ \citep{Arun} all the way up to $\sim 1000\,$Hz
for mergers of NS--NS binaries, providing potential sources both
for ground-based interferometers (LIGO, VIRGO, etc.) and space-based
detectors (LISA). The inspiral signals can provide
information on the spins and masses of the compact objects
\citep[e.g.,][]{poisson will}. Moreover the
merger signals from BH--NS and NS--NS binaries can carry 
information about the NS internal structure and the equation of state (EOS) of
matter at nuclear densities \citep{faber_rasioI, faber_rasioII, faber_rasioIII}. 
Note that for BH--NS mergers with fairly massive BHs, with $M_{BH}\gtrsim100\,M_\odot$, 
where the NS is expected to plunge into the BH as a whole, not much information on the 
NS EOS will be carried by the GW signal. In contrast, the merger of a NS with a 
stellar-mass BH ($M_{BH}\sim10\,M_\odot$) makes it possible for the NS to be disrupted 
outside the BH's 
innermost stable circular orbit (ISCO), and this will potentially enrich the GW signal with 
a lot of information on the detailed behavior of the NS matter.
Even though the GW signals from the inspiral of
NS--NS binaries are accessible to ground-based interferometers, covering the frequency range
$\sim 40-1000\,$Hz, the signals from their final mergers
will probably be lost in the high-frequency noise level \citep{vallisneriI,faber02}. 
On the other hand, the GW merger signals for typical BH--NS binaries with stellar-mass
BHs are expected to lie well within the
 sensitivity band of LIGO, at frequencies $\sim 100-500\,$Hz.
 
Although double NS binaries have been observed
\citep{thorsett,burgay}, BH--NS and BH--BH binaries remain
undetected. Moreover, the few observed NS--NS binaries (binary pulsars with NS companions
and one double pulsar) are subject to considerable selection effects.
Therefore it is not currently possible to infer much empirically about
the general properties of DCOs based on the 
observed sample. 
Theorists rely instead on binary evolution and population synthesis models to
make predictions about the formation, evolution and
properties of such binaries \citep[e.g.,][]{belcz,nelemans}. These models give estimates of  merger rates for DCOs and the corresponding detection rates for
the various GW interferometers. NS--NS binaries are expected to
merge with rates $\sim1-145\,$Myr$^{-1}$ per MWEG (Milky Way Equivalent
Galaxy) and the equivalent rate for BH--NS binaries is
$\sim 0.07-5\,$Myr$^{-1}$ per MWEG \citep{kim, belcz}. For NS--NS binaries the detection rate
estimates are $\sim (0.4-60) \times 10^{-3}\,$yr$^{-1}$ and
$\sim 2-330  \,$yr$^{-1}$ for LIGO  and Advanced LIGO,  respectively \citep{kim}, while for BH--NS binaries the equivalent rates are $\sim 3 \times
10^{-3}-2 \times10^{-2}$\,yr$^{-1}$ and $0.7-40 \,$
yr$^{-1}$ \citep{belcz}. Although the detection rates are quite low for the current LIGO stage, these
predictions are very promising for Advanced LIGO.
One should remember that, although the merger rates for NS--NS binaries have been 
empirically constrained \citep{kim}, no such constraints have been set for BH--BH and 
BH--NS mergers. The lack of complete understanding of binary star evolution can lead 
to merger rate estimates for these systems which vary significantly, depending on the 
exact physical assumptions adopted in the various binary evolution codes. Current effort 
is focusing on decreasing these uncertainties and setting more solid constraints on the 
merger rates of BH--NS and BH--BH binaries \citep{oshau05, oshau06}.

Along with the binary evolution studies that provide merger rates of compact binaries, general relativistic calculations of binary mergers try to guide the search for GW signals by predicting the exact shape of the signals and generating  GW search templates \citep[e.g.,][]{buonanno, baker,abbott06, apostolatos}.  For BH--NS and BH--BH binaries it is expected that both 
the BH spins and their possible misalignment angle with respect to the orbital angular 
momentum will affect significantly the shape of the GW signals and their detectability
 \citep{apostolatos,grandclement}.

\subsection{Connection to Gamma Ray Bursts}

Another interesting aspect of BH--NS and NS--NS binaries is their
possible connection to the observed short gamma-ray bursts (GRBs).
This hypothesis has gained widespread support over the last few years, both
because of the rapid progress in theoretical modelling and from the 
recent {\em Swift\/} observations of short GRBs  \citep[see][for a recent review]{nakar}.

GRBs are classified into two duration classes,
separated at $\sim 2\,$s (Kouveliotou et al.\ 1993). Long bursts are found
to be predominantly in active star-forming regions. It is now believed
that long bursts are produced when a massive star reaches the end of its
life, its core collapsing to form a BH and, in the process,
ejecting an ultra-relativistic outflow (e.g., Woosley \& Bloom 2006). The
standard collapsar model predicts that a broad-lined and luminous Type
I-c core collapse supernova (SN) accompanies long bursts (MacFadyen \&
Woosley 1999). This association has been confirmed in observations of
several nearby GRBs (e.g., Galama et al.\ 1998; Hjorth et al.\ 2003; Pian et al.\ 2006; Gehrels et al.\ 2006).

Until recently, afterglow of short bursts have been extremely elusive.
This situation changed dramatically in 2005. {\em Swift\/} and {\em HETE-2\/} detected
X-ray afterglows from short bursts (Gehrels et al.\ 2005; Villasenor et
al. 2005). This has led to the identification of host galaxies and to redshift
measurements. More than 10 short burst afterglows have been detected so far,
and distinctive features have emerged. While long bursts occur only in star
forming spiral galaxies, short bursts appear also in elliptical galaxies,
which are dominated by an old stellar population. The low level of star
formation makes it unlikely that the bursts originated in a SN explosion.
Even though a short burst, GRB 050709, was seen in a galaxy with current
star formation, optical observations ruled out a SN association (Fox et
al.\ 2005). The isotropic energy for short bursts is 2-3 orders of
magnitude lower than that for long bursts $E_{iso} \sim 10^{52-54}\,$erg
(Barthelmy et al.\ 2005). These results suggest that compact
stellar mergers are the progenitors of short bursts.

The similarity of X-ray afterglow light curves of long and short bursts
indicates that afterglows of both classes can be described by the same
paradigm, despite the differences in the progenitors. This view is
supported by the fact that the decay rate of short burst afterglows
is the value expected from the standard fireball model (e.g., Piran 2004;
M\'esz\'aros 2006), and that at least in two short bursts (GRB 050709 and
GRB 01221A) there is evidence for a jet break  (Fox et al.\ 2005;
Soderberg et al.\ 2006; Burrows et al.\ 2006). In the standard afterglow
model, these breaks are interpreted as a signature of collimation of a
fireball into a jet with an opening angle $\theta \simeq 6-12$ degrees and
imply a  beaming-corrected energy of $E\sim (0.5-3)\times 10^{49}\,$erg,
much less than that of long bursts, which have $E \sim 10^{51}\,$erg (Frail et
al.\ 2001). The lower energy implies that the mass of the debris torus
formed during the merger could be smaller than that of the torus formed
in the collapse of the core of massive stars.

Combined with the lack of a jet break in GRB 050724, which gives lower
limits of $\theta > 25$ degrees and $E > 4\times 10^{49}\,$erg (Grupe et
al.\ 2006), the current small sample indicates that the outflow of short
bursts is less strongly collimated than most previously reported long GRBs
with the median value $\theta \simeq 5$ degrees (Frail et al.\ 2001;
however see Monfardini et al.\ 2006). The wider jet angle is
consistent with a merger progenitor scenario (e.g., M\'esz\'aros, Rees \&
Wijers 1999), since there is no extended massive stellar envelope (as in
long GRBs) that serves to naturally collimate the outflow. Many more
bright short bursts will be needed to improve the jet break statistics
substantially.

One of the unexpected results from {\em Swift\/} is that early X-ray afterglows of long
bursts show a canonical behavior, where light curves include three
components: (1) a steep decay component, (2) a shallow decay component
and (3) a ``normal'' decay component. On top of this canonical behavior,
many events have superimposed X-ray flares (e.g., Zhang et al.\ 2006). The
X-ray afterglow of a short burst, GRB 050724, associated with an
elliptical host galaxy, also resembles the canonical light curve, and
it suggests a long-lasting engine. A flare at $\sim 100\,$s in the X-ray
light curve decays too sharply to be interpreted as the afterglow
emission from a forward shock, but is consistent with the high latitude
emission from a fireball (Barthelmy et al.\ 2005; Kumar \& Panaitescu
2000). This is appropriate for the late internal shock scenario as
invoked to interpret X-ray flares in long GRBs. This interpretation
requires that the central engine remains active up to at least $\sim
100\,$s, and challenges simple merger models, because the predicted typical
time scale for energy release is much shorter. 

Another interesting scenario based on BH--NS and NS--NS mergers
and connected to GRBs is the production of r-process
elements (heavy nuclei with $A>90-100$) through the nuclear physics
of decompressed NS matter in the ejecta produced by the merger.
It is still not clear today what is the astrophysical site that can
provide the appropriate conditions for r-process nucleosynthesis
to take place, although the conditions themselves can be estimated
\citep{jaikumar}. The possible ejection of extremely neutron-rich
($Y_e\sim 0.1$) material from NS disruptions in compact binary mergers 
is believed to be a promising source for
r-process elements \citep{lattimer}. As such mergers are expected
to happen in the outskirts of galaxies \citep{perna_a}, it is 
possible that the high-velocity ejecta will also enrich the
intergalactic medium with high mass r-process elements
\citep{rosswog_b}.

\subsection{Results from Previous Studies}

In recent years, various groups have performed 3-D hydrodynamic simulations 
of BH-NS mergers, providing  very interesting---if somewhat preliminary---results. 
In parallel with this work there has been an 
ongoing effort to improve both the computational techniques and the overall accuracy of the merger simulations.

Newtonian studies of BH--NS mergers, where the BH is simply represented by a 
point mass, were first used by different groups  \citep[e.g.,][]{lee99b,rosswog} 
to cover a fairly wide range of mass ratios while attempting to determine the merger's dependence on the NS EOS. The general picture that comes out of these Newtonian calculations is a clear connection between mass ratio and NS EOS, and the ability to form a massive enough accretion disk around the BH at the end of the merger.
As an overall trend, for higher mass ratios (with $q \sim 0.7-0.28$)
the survival of NS material and formation of a disk around a
Schwarzschild BH seems possible for soft EOS. For these high mass
ratio mergers the tidal disruption radius lies outside the ISCO,
which allows for more NS material to settle into a disk after the
tidal disruption. However, those studies also suggested that, for a stiffer EOS, the core
of the NS always survives the merger, inhibiting the formation of a massive 
disk around the BH and leading instead to a period of quasi-stable
mass transfer from the NS to the BH \citep{lee99a, kluzniak98}.
A stiffer EOS could, in that case, prevent the
total disruption of the NS, leading to the eventual formation of a ``mini-NS''
along with the disk or even lead to multiple disruptions of the
surviving NS core. \citet{rosswog} suggested that the survival of
a mini-NS for the case of a stiff polytropic EOS ($\Gamma=3$) 
 is connected to the difficulty of forming a massive disk in
their calculations, arguing that the survival of an orbiting NS core acts as
 a storage mechanism that prevents further inflow of material
towards the BH.

 In their full-GR treatment of BH--NS mergers for a non-spinning BH of arbitrary mass,
\citet{shibataa, shibatab} concluded that the disruption of a NS
by a low-mass BH ($M=3.2-4\,M_\odot$) can lead to the formation of a low-mass disk
(of mass $\sim 0.1\,M_\odot$) around the BH, which could potentially power a
short GRB. No formation of more massive disks was ever observed in their simulations, 
indicating that systems with $\sim M_\odot$ disks around a BH cannot be formed through BH--NS mergers with non-spinning BHs. Survival of a NS core, or quasi-stable mass transfer,
as predicted by Newtonian calculations, are never seen in these full-GR calculations.

Also limited to Schwarzschild BHs, the fully relativistic calculations of
\citet{faber06b} focused on BH--NS mergers with a much more extreme mass ratio 
$q=0.1$. For such a mass ratio and a non-spinning BH, the tidal disruption limit of a NS 
with 
canonical mass and radius (i.e., compactness ratio $\mathcal{C}=M_{NS}/R_{NS}\simeq 
0.2$) is reached inside the ISCO.
For this reason, \citet{faber06b} considered artificially undercompact models for the NS 
(with compactness as small as  $\mathcal{C}\simeq 0.04$), in order to study cases where 
the disruption of the NS takes place outside the ISCO. They suggested that their study for these undecompact NSs can serve as an analogue for binaries with lower-mass BHs and more 
compact NSs (which their numerical code could not handle directly), where the tidal radius 
is located well outside the ISCO.
They found that only a small fraction ($\sim5-7\%$)
of the NS mass becomes unbound and escapes from the system
and that, while most of the infalling mass is accreted
promptly by the BH, part of it ($\sim25\%$) remains
bound outside the horizon, forming a disk. 

Motivated by all these recent observational and theoretical developments, 
we have embarked on a new numerical investigation of the merger process for BH--NS 
binaries using a 3-D relativistic SPH code developed specifically for the study of
stellar disruptions by BHs. In this paper, the first of a series,
we present our methods and numerical code, as well as the results from a
first set of preliminary calculations aimed at exploring broadly the parameter space of
these mergers. 
Our paper is organized as follows: In \S 2 we describe the SPH
code used for our simulations; we develop some analytic
considerations regarding the metric used by the code; and
we also discuss the various test calculations that we have performed. In \S 3 we 
present results from our simulations of
equatorial mergers (\S 3.1) and inclined mergers (\S 3.2), including 
a discussion of how to set up initial conditions for the
inclined case. The GW signals extracted from these simulations are 
presented and discussed in \S 4.  Finally, a summary and conclusions
are given in \S 5.

\section{Methods and Tests}

\subsection{Critical Radii}
\begin{figure}[tbp]
\epsfxsize=14cm \centerline{\epsfbox{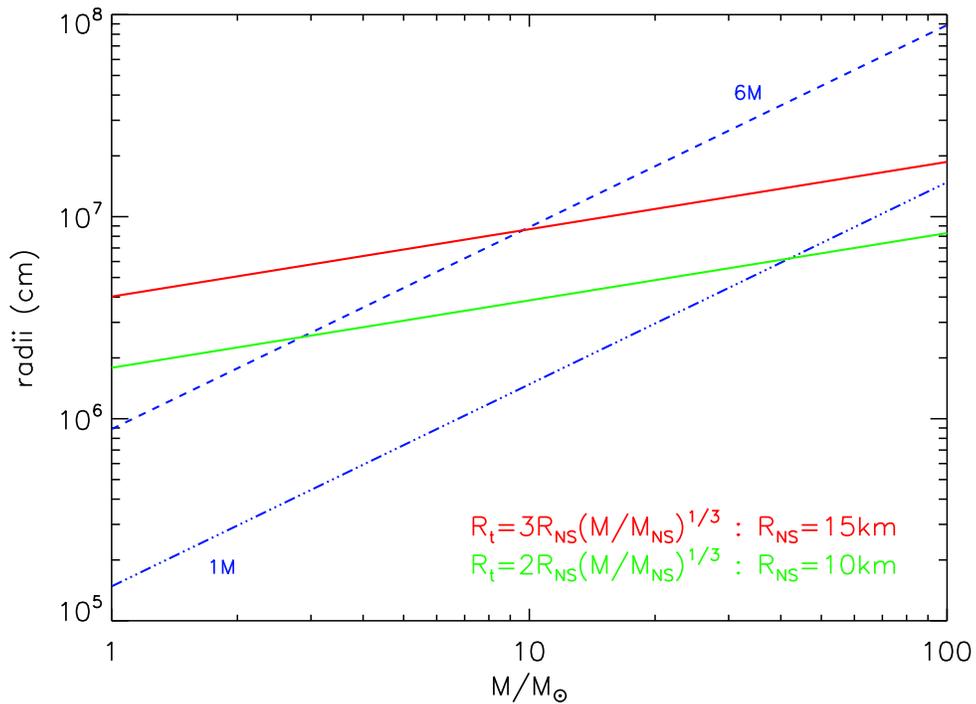}} \caption{Tidal
disruption limits for a $1.4\,M_\odot$ NS in circular orbit around
a BH of mass $M$. 
The radius of the ISCO is also shown for the Schwarzschild (dashed line) and
maximally co-rotating Kerr cases (dash-dotted line). The two solid
lines bracket the tidal limit for NS with different spins and
radii \citep{wiggins, lai}. Orbits are assumed to be in the equatorial plane and prograde.} \label{RL}
\end{figure}

Fig.~\ref{RL} shows
why the final merger of a BH--NS binary (for a typical
stellar-mass BH) is interesting but also particularly
difficult to compute: the tidal (Roche) limit\footnote{see, e.g., \citet{lai} and \citet{wiggins}
for analytical calculations and discussion of tidal radii.} is typically right
around the ISCO and the BH horizon. On the one hand, this implies
that careful, fully relativistic calculations are needed. On the
other hand, it also means that the fluid behavior and the GW
signals could depend sensitively (and carry rich information) on
both the masses and spins of the compact objects, and on the NS
EOS. How much information is carried about the fluid depends on
where exactly the tidal disruption of the NS occurs: for a
sufficiently massive BH, the horizon will always be encountered
well outside the tidal limit, in which case the NS behavior
remains point-like throughout the merger and disruption will never
be observed. In the opposite case where
the tidal limit resides outside the horizon, the GW signal corresponding to
the NS disruption could be detected by ground-based interferometers.

For those cases where disruption occurs outside the BH's horizon,
the final outcome of the merger depends strongly on the relative
positions of the tidal radius $R_t$ (i.e., the point where
disruption takes place) to the ISCO of the BH. Fig.~\ref{RL}
gives an overview of these critical radii and a first idea of what one
could expect the outcome to be for mergers with various mass ratios.
An interesting fact that we notice first is that for a given BH
mass the relative position of $R_t$ to the ISCO changes with the
BH's angular momentum: as the BH spin increases, it drags the
ISCO closer to the BH.
For high enough mass ratios (low BH mass of the order of a
few $M_\odot$) the tidal radius is always encountered outside the
ISCO, even for non-rotating BHs. For a $10\,M_\odot$ BH, the ISCO and
$R_t$ coincide for a Schwarzschild BH, but the ISCO moves inside
$R_t$ for a Kerr (spinning) BH. Finally, for even higher
mass BHs of $\sim15\,M_\odot$ the situation becomes more interesting
as the ISCO lies well outside the tidal radius for a non-spinning
BH and for Kerr BHs with low spin, but it starts migrating inside
$R_t$ as the BH's angular momentum increases. For fast spinning
BHs the tidal radius is encountered well outside the ISCO. Of
course for very massive BHs, $R_t$ is not only inside the ISCO but
inside the horizon as well, even for the case of an extremal Kerr
BH (this happens for $M\gtrsim 100\,M_\odot$).

These simple results are important because disruption
outside the ISCO could lead to the formation of a disk of NS
debris outside the BH's horizon. If the disruption is to happen
inside the ISCO, no such feature is expected. Knowing the
relative position of ISCO and $R_t$ for various binary mass ratios therefore 
gives us a first insight on what to expect as the qualitative
outcome of a merger, and in which ranges of mass ratios we should look for
certain outcomes.

\subsection{Analytic Considerations}

The SPH code makes use of the Kerr-Schild (K-S) form of the Kerr
metric. In this section we summarize the reasons for choosing to
use the K-S metric (i.e., its advantages over the Kerr metric in
Boyer-Lindquist [B-L] coordinates) and how quantities such as the
BH horizons and the angular velocity of equatorial circular orbits
around the BH translate from one coordinate system to the other.
The reader can find extensive discussions of the B-L and K-S
coordinate systems in, e.g., \citet{Chandra, Poisson, Kerr63}.
Here we present a brief overview with emphasis on some useful
aspects of the K-S metric that relate directly to our calculations
and to the presentation of our numerical results.

The Kerr solution in B-L coordinates is given by the familiar
expressions \citep{BL67}

\[ ds^2=-(1-\frac{2Mr}{\Sigma})dt^2+(\frac{\Sigma}{\Delta})dr^2 +\Sigma
d\theta ^2\] \beq
+(r^2+a^2+\frac{2Mra^2}{\Sigma}\sin^2\theta)\sin^2\theta d\phi
^2-\frac{4Mra\sin^2\theta}{\Sigma}d\phi \, dt \label{kerrBL}\eeq

where \beq \Delta=r^2-2Mr+a^2 \eeq
      \beq \Sigma=r^2+a^2\cos^2\theta .\eeq

This form has only one off-diagonal term and is therefore far more
convenient to use than the K-S form of the metric.\footnote{Note
that this form of the Kerr metric reduces to the well known
Schwarzschild solution in the limit $a=0$:
$ds^2=-(1-\frac{2M}{r})dt^2+(1-\frac{2M}{r})^{-1}dr^2+r^2d
\theta^2+r^2\sin^2\theta d\phi ^2$} Yet, it carries some extra
coordinate singularities which correspond to the roots of $\Delta$
: $r_{\pm}=M \pm \sqrt{M^2-a^2}$   ($r_{+}$ and $r_{-}$ are the
future and past horizons respectively, with $r_{+}$ only being an
event horizon). It is useful to observe that: (a) for $a/M=1$ the two coordinate singularities (horizons) coincide, 
(b) for $a/M=0$ there is only one horizon at $r_h/M=2$ and the
curvature singularity at $r/M=0$.
The obvious advantage of casting the Kerr metric into its K-S form is that
one avoids the coordinate singularities at the horizon present in B-L coordinates.

The coordinate singularity at the
horizon present in the B-L form of the metric has the following effect \citep{Poisson}: although it
takes a finite proper time for a particle to cross the event
horizon, it takes infinite coordinate time $t$ to do so. Moreover, since the
angular velocity $d\phi /dt$ tends to a finite limit at the
horizon, $\phi$ has also to increase an infinite amount : $\phi
\rightarrow \infty $ ($\phi$, like $t$, is not a ''good" coordinate
at the horizon). What this practically means for our code is that
it prevents us from extending our calculations all the way to the
horizon of the BH. In that case, one needs instead to place an
absorbing boundary outside the BH's horizon. Fortunately
there is a way to overcome this problem. In order to extend the
Kerr metric beyond the horizon, another coordinate system needs to
be adopted. Keeping in mind that the horizons are null surfaces,
it makes intuitively sense to construct the new coordinates in
terms of null geodesics.

The null geodesics in the Kerr spacetime are given by the tangent
vectors \beq \frac{dt}{d\tau}=\frac{r^2+a^2}{\Delta}E \eeq \beq
\frac{dr}{d\tau}=\pm E \eeq \beq \frac{d\theta}{d\tau}=0
\label{Kerr_nulls}\eeq \beq \frac{d\phi}{d\tau}=\frac{a}{\Delta}E
 \eeq
where $E$ is the specific energy.

The real null vector $\vec{l}$ reads:

\beq l^{\alpha}=\frac{1}{\Delta}(r^2+a^2,\pm \Delta,0,a).\eeq
Setting $E=1$ (and using $\lambda$ for the affine parameter)

\beq \frac{dt}{d\lambda}=\frac{r^2+a^2}{\Delta},
  \frac{dr}{d\lambda}=\pm 1, \eeq

\beq \frac{d\theta}{d\lambda}=0,
     \frac{d\phi}{d\lambda}=\frac{a}{\Delta} .\eeq

By choosing the positive sign for $dr/d\lambda$ we obtain an outgoing congruence
with the tangent vector field defined by \beq l^{\alpha}
\vartheta_{\alpha}=\frac{r^2+a^2}{\Delta}\vartheta_{t}+\vartheta_{r}+\frac{a}
{\Delta}\vartheta_{\phi} .\eeq  The new variables $u$ and
$\tilde{\phi}$ can be introduced in the place of $t$ and $\phi$:
\beq du=dt-\frac{r^2+a^2}{\Delta}dr\eeq \beq d\tilde{\phi}=d\phi
-\frac{a}{\Delta}dr \label{out_cong} . \eeq

 The null geodesic now becomes
\beq l^{\alpha}=(0,1,0,0)\eeq and the metric takes the form

\[  ds^2=-(1-\frac{2M\tilde{r}}{\Sigma})du^2 +\Sigma
d\theta^2-2du \,
d\tilde{r}-\frac{4aM\tilde{r}\sin^2\theta}{\Sigma}du \,
d\tilde{\phi}+2a\sin^2\theta d\tilde{r} \, d\tilde{\phi}\]
\beq+(\tilde{r}^2+a^2+\frac{2M\tilde{r}a^2\sin^2\theta}{\Sigma})\sin^2\theta
d\tilde{\phi}^2 \label{outKS}\eeq

where $\tilde{r}$ is defined by\footnote{a more geometrically
insightful representation of Eq.~(\ref{KSr}) would be
\[\frac{x^2+y^2}{\tilde{r}^2+a^2}+\frac{z^2}{\tilde{r}^2}=1.\]}
\beq \tilde{r}^4-(\rho^2-a^2)\tilde{r}^2-a^2z^2=0 \label{KSr}\eeq
with\footnote{Note that as $a \rightarrow 0$  ,
$\tilde{r}\rightarrow \rho $.} \beq \rho^2=x^2+y^2+z^2 .\eeq

By instead choosing the negative sign for $dr/d\lambda$ we can obtain
an ingoing congruence with the tangent vector field given by \beq
l'^{\alpha}
\vartheta_{\alpha}=\frac{r^2+a^2}{\Delta}\vartheta_{t}-\vartheta_{r}+\frac{a}
{\Delta}\vartheta_{\phi}.\eeq The new variables to be introduced
here are \beq dv=dt+\frac{r^2+a^2}{\Delta}dr\eeq \beq
d\tilde{\phi}'=d\phi +\frac{a}{\Delta}dr \label{in_cong}. \eeq In
this coordinate system the null geodesic simplifies to \beq
l'^{\alpha}=(0,-1,0,0)\eeq

and the metric becomes

\[ ds^2=-(1-\frac{2M\tilde{r}}{\Sigma})dv^2 +\Sigma
d\theta^2+2dvd\tilde{r}-\frac{4aM\tilde{r}\sin^2\theta}{\Sigma}dvd\tilde{\phi}'
-2a\sin^2\theta d\tilde{r}d\tilde{\phi}'\]
\beq+(\tilde{r}^2+a^2+\frac{2M\tilde{r}a^2\sin^2\theta}{\Sigma})\sin^2\theta
d\tilde{\phi}'^2 \label{inKS}.\eeq

\begin{figure}[tbp]
\centering \epsfxsize=12.0cm \epsfbox{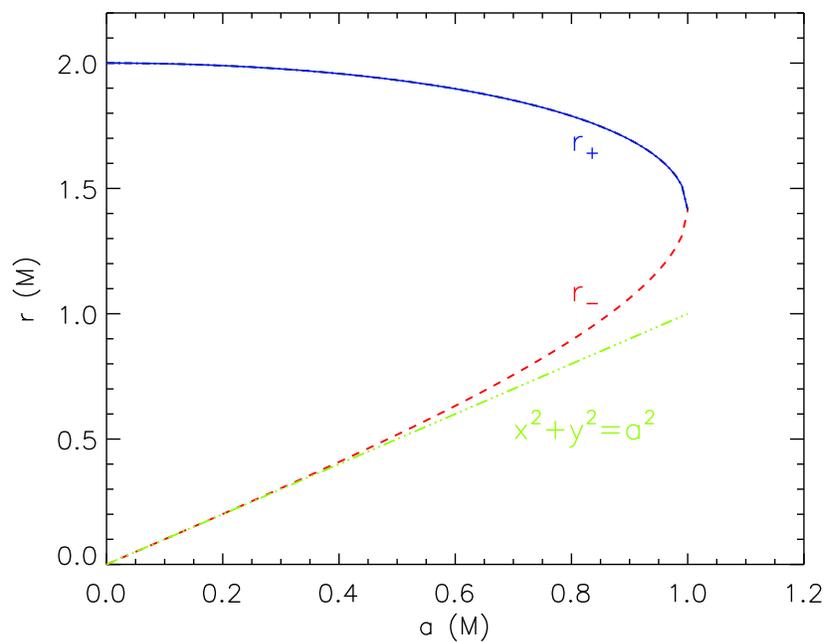}
\caption{The future (solid blue line) and past (dashed red line) horizons of a BH for various values of the Kerr parameter $a$. The straight (dash-dotted green) line represents the ring (curvature) singularity of the K-S metric, which exists only in the equatorial plane.}
\label{horizons}
\end{figure}

What is the use of those two different coordinate sets?  The
coordinates ($v$,$r$,$\theta$,$\tilde{\phi}'$) are well behaved on
the future horizon, yet they are singular on the past horizon,
where ($u$,$r$,$\theta$,$\tilde{\phi}$) are now well behaved.
Therefore the first set of coordinates is used in order to
regularize the past horizon, whereas the second one is used to
regularize the future horizon. If one wants to avoid both horizons
of a Kerr BH, both patches need to be used in order to cover the
entire spacetime around the BH. Then  there is  just the curvature
 singularity of the Kerr metric left, which occurs at
\beq \Sigma =\tilde{r}^2+a^2\cos^2 \theta =0, \eeq implying \beq
x^2+y^2=a^2. \eeq The curvature singularity is not a point but
rather a ring of radius $a$ and exists only in the equatorial
plane. It is always found inside the past horizon, although as
$a/M$ decreases the curvature singularity and the past horizon
approach each other (Fig.~\ref{horizons}).

\subsection{SPH Code}

We employ a 3-D implementation of the Lagrangian SPH technique. Our GRSPH (General Relativistic Smooth Particle Hydrodynamics) code is based on the work by Laguna et al.\ (1993a) in which the
general relativistic hydrodynamic equations were rewritten in a Lagrangian form
similar to their counterparts for non-relativistic
fluids with Newtonian self-gravity.   
The GRSPH code is restricted to {\em fixed curved spacetimes\/}; in particular,
the spacetime of a rotating BH is assumed here. 
The use of a fixed background in our simulations is justified for
low mass ratios ($q\simeq 0.1$ in this paper), i.e., when the mass of
the BH is substantially larger than the NS mass.

Originally the BH metric 
used in the code was in terms of Boyer-Lindquist coordinates. 
For the present work, we use Kerr-Schild coordinates. The main advantage of these coordinates is 
their regularity across the BH horizon, thus allowing SPH particles to freely cross it. 
An important aspect of SPH in curved spacetimes is 
the handling of the local volume-averaging required by the 
smoothing of the equations. 
The smoothing volumes involved are in general not small
enough to ignore curvature effects, especially in the neighborhood of the BH.
Our implementation correctly accounts for these effects.
The most important computational aspect of the GRSPH code is the 
particle neighbor finding algorithm.
The GRSPH code uses an {\it oct-tree\/} data structure as basis for finding 
neighbors \citep{warren-salmon}.
The oct-tree neighbor-searching part of GRSPH has also been successfully 
used for $N$-body large-scale structure 
simulations \citep{heitmann} and is used as the foundation of a different 
SPH code with radiation transport used for supernova core
collapse calculations \citep{fryer06}.
The GRSPH code scales as $O(N \log N)$ with $N$ the total number of SPH particles.
The code was calibrated with three one-dimensional
benchmarks (Laguna et al.\ 1993): (1) relativistic shock tubes,
(2) dust infall onto a BH, and (3) Bondi collapse (see
Laguna et al.\ 1993a for details). The code has been successfully applied
in several previous studies of the tidal disruption of ordinary (main-sequence)
stars by a supermassive BH (Laguna et al.\ 1993b; Kobayashi et al.\
2004; Bogdanovic et al.\ 2004).

 Artificial viscosity is implemented in our code by following the description
presented in \citet{laguna93} and \citet{laguna94}. Artificial
viscosity is a mechanism to account for the possible presence
of shocks and is introduced as a viscous pressure term added to
the SPH equations \citep{laguna93}. As we do not expect any shocks
to appear in our simulations except perhaps at the point where the stream of NS
material accreting onto the BH self-intersects, we keep artificial
viscosity suppressed by lowering the parameters of its two terms 
(the bulk viscosity term and the von Neumann-Richtmyer (1950) viscosity term) to
 the value $0.2$ instead of the more common values $\sim 1$
\citep[see][for details and tests]{laguna93}.

Radiation reaction is implemented in the GRSPH code by following the simple, approximate prescription described in \citet{lee99b}, \S 2.
We use the quadrupole formula for the rate  
of energy radiated  \citep[Eq.~4 in][]{lee99b} to derive a damping force  \citep[Eq.~6 in][]{lee99b} for each  SPH particle. The formula has been slightly modified  as  in our  code we deal with $d\mathbf{u}/dt$ where $\mathbf{u}$ 
is the 4-velocity. We  
have also added a parameter to increase the radiation reaction force  
in order to accelerate the initial inspiral since we do not want to spend too much 
computational time on this initial phase. We have checked that, once the merger starts,  
the precise value of this parameter does not affect our results significantly.
Throughout our calculations and for all the results presented here we use
geometrized units with $G=c=1$.

\subsection{Test Calculations}

\subsubsection{Testing the K-S Metric}
\begin{figure}[ht]
\centering
\includegraphics [width=15cm]{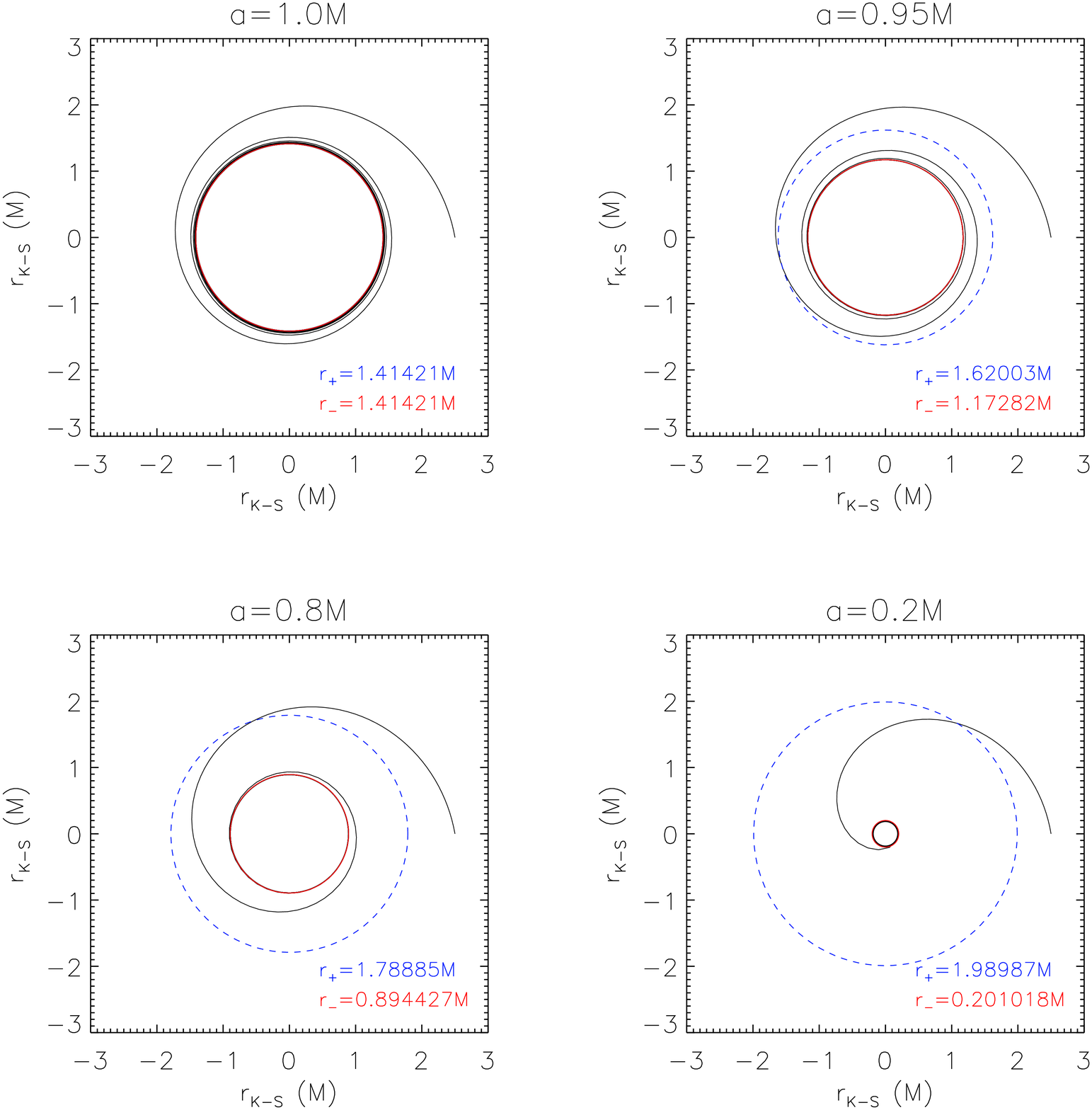}
\caption{A 'horizon-trapped' orbit for different values of $a$.
The initial conditions for the particle are  fixed and only $a$ is
varied in the four orbits. The red circles represent the two
horizons. It is clear that the particle successfully crosses the
outer horizon and is eventually trapped at the inner horizon, except
for the $a=M$ case where the two horizons coincide.} \label{trap4}
\end{figure}

The Kerr-Schild coordinate system used in the SPH code is only
avoiding the outer (future) horizon, since that serves the purpose
of the SPH particles getting as close to the BH's horizon as
possible, without the code crashing or becoming problematic. Yet,
in the extremal Kerr case ($a/M=1$), the two horizons coincide
(Fig.~\ref{horizons})  at $\tilde{r}=\sqrt{2}M$, and as a result,
we see the orbits of the SPH particles being trapped at
$\tilde{r}=\sqrt{2}M$. To check that it is the inner horizon where
our transformation is singular, we set up the following simple
test: we construct a test particle geodesic integrator that makes
use of the K-S metric and we start with an equatorial orbit for
$a/M=1$ which, starting from a finite distance from the BH 
(having $v_z=v_y=y=z=0$), ends being trapped at
$\tilde{r}=\sqrt{2}M$. We then keep following the same orbit as we
reduce the value of $a$ (Fig.~\ref{trap4}). The result is that the
particle ends being trapped at the inner horizon (whichever that
is for the specific value of $a$). The trapping is illusionary: it
reveals the singularity of our coordinate transformation. We thus
avoid using $a/M=1$ in the simulations presented in this paper and
instead we set $a/M=0.99$ for the case of an extremal Kerr B.H. By
doing so, the future horizon moves outwards at
$\tilde{r}_+=1.51067M$ and the past horizon moves inwards at
$\tilde{r}_-=1.31067M$ and are therefore distinguishable.

Another quantity that is going to be affected by the coordinate
transformation (i.e., from B-L to K-S) is the angular velocity
$\Omega_{\phi}$ for a circular equatorial orbit around the BH. In
B-L coordinates $\Omega_{\phi}$ is given by

\beq \Omega_{\phi}= \pm \frac{M^{1/2}}{r^{3/2}\pm aM^{1/2}}
\label{Omega}\eeq where $r^2=x^2+y^2+z^2$ and the upper (lower)
sign corresponds to co-rotating (counter-rotating) orbits. In K-S
coordinates Eq.~(\ref{Omega}) holds when $r$ is replaced by
$\tilde{r}$, as defined by Eq.~(\ref{KSr}) (see Appendix C for an analytic calculation of
 $\Omega_\phi$ in K-S coordinates). In order to check this result numerically, we set up the following test: we use again the geodesic integrator and we place
a test particle at some distance $r_0$ from the BH and give it the
angular velocity that corresponds to an equatorial circular orbit
at the specific $r_0$ as required by Eqs.~(\ref{Omega}) and
(\ref{KSr}). As expected, the test particle remains in this fixed
circular orbit (with $d\tilde{r}/d\tau=0$) as we integrate for a
large number of orbital periods\footnote{If we instead use
Eq.~(\ref{Omega}) with $r^2=x^2+y^2+z^2$, the particle follows an
oscillating orbit ($dr/d\tau \ne 0$) around its initial position,
indicating that the formula used for $\Omega_{\phi}$ needs
modification.}.

\subsubsection{Stable Binary Orbits}

\begin{figure}[ht]
\centering
\includegraphics [width=10cm]{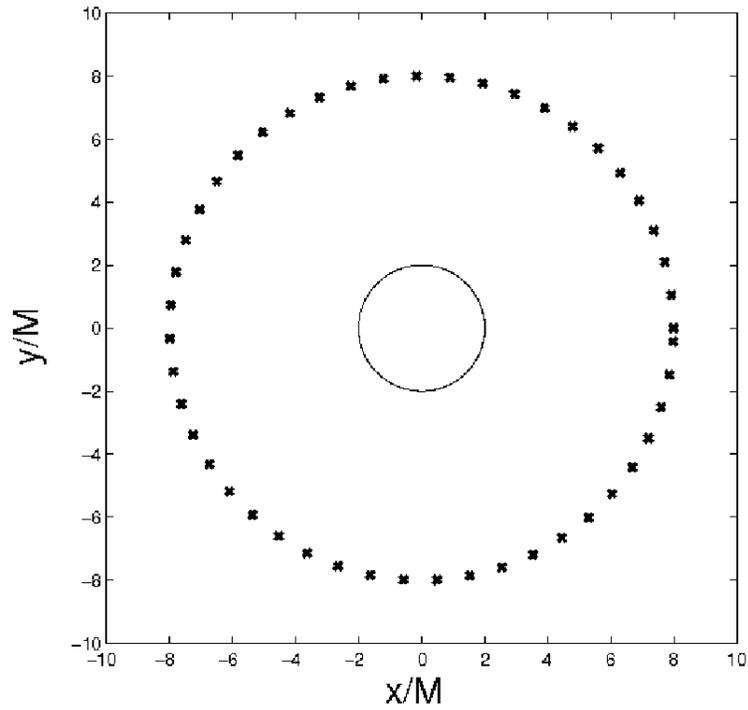}
\caption{Test calculation for a WD orbiting a much more massive BH
at $r=8\,$M. The mass ratio in this case is $q\simeq
4\times10^5$. The constant radius of this circular orbit is
maintained by our code to within better than $10^{-3}$ over one
full period. Here each cross indicates the position of the WD
(center of mass of all SPH particles) at a different time along
the orbit (counter-clockwise).} \label{WD}
\end{figure}

In a first test, we considered a white dwarf (WD) with $M_\ast=0.6\,M_\odot$ and
$R_\ast=1.3\times10^{-2} \,R_\odot \sim 9 \times 10^{8}\,$cm orbiting
around a Schwarzschild BH with $M=2\times10^5\,M_\odot$. Here $5000$
particles are used to represent the WD with a $\Gamma=5/3$ polytropic
EOS. Since the mass ratio is extreme, our approximation (moving the
fluid on a fixed background metric) should be extremely accurate. With
these parameters, the tidal radius $R_t \sim 6.3\times 10^{10}$cm
and the horizon scale $r_h=2M\sim 5.9\times 10^{10}\,$cm are
comparable, and much larger than the WD radius $R_\ast$. If $R_t
\gg r_h$, the point particle approximation  should break down for
orbits near the ISCO because the WD will get disrupted at radii
well outside the ISCO. On the other hand, when $R_t \ll r_h$, the
sound crossing time of the WD is much shorter than the orbital
period, and it is numerically expensive. Thus, we have chosen the
parameters satisfying $R_t \sim r_h$ to test the code for
relativistic orbits. With these parameters we found that we can maintain a circular orbit
at $r=8\,M$ to within $|\Delta r|/r < 10^{-3}$ over one full
orbital period (Fig.~\ref{WD}).

In a second, similar test we considered a NS with $M_{NS}=1.4\,M_\odot$ and $R_{NS}=13.4\,$km
$=1.93\times10^{-5} \, R_\odot$ (represented by $10000$ SPH particles and with a $\Gamma=2$
polytropic EOS) orbiting around a Schwarzschild BH with $M=10\,
M_\odot$.  We found that we can maintain a NS orbiting at $r=20\,M$ stably and
without any noticeable numerical dissipation for more than 20 orbital periods.

\subsection{Initial Conditions for BH--NS Binaries}

We set up initial conditions for BH--NS binaries near the Roche
limit using the SPH code and a relaxation technique similar to
those used for previous SPH studies of close binaries
\citep[e.g.,][]{RS}. First we construct hydrostatic equilibrium NS
models for a simple gamma-law EOS by solving the Lane-Emden
equation.  When the NS with this hydrostatic profile is placed in
orbit near a BH, spurious motions could result as the fluid
responds dynamically to the sudden appearance of a strong tidal
force. Instead, the initial conditions for our dynamical
calculations are obtained by relaxing the NS in the presence of a
BH in the co-rotating frame of the binary. For {\it
synchronized\/} configurations (assumed here), the relaxation is
done by adding an artificial friction term to the Euler equation
of motion in the co-rotating frame. This forces the system to
relax to a minimum-energy state. We numerically determine the
angular velocity $\Omega$ corresponding to a circular orbit at a
given $r$ as part of the relaxation process. The advantage of
using SPH itself for setting up equilibrium solution is that the
dynamical stability of these solutions can then be tested
immediately by using them as initial conditions for dynamical
calculations.

\section{First Results}

\subsection{Equatorial BH--NS Mergers for Spinning and Non-spinning BH}

For all the results presented in this section we ran simulations
using $N=10^4$ SPH particles to represent a NS with a $\Gamma=2$
polytropic EOS, except for one case, for which we have 
performed a crude convergence test by repeating the run with $N=10^5$ particles 
(Run E1 of Table~\ref{table1}).  The number of neighboring particles used in the SPH code is $N_N=80$ and $N_N=140$ for the low and high resolution runs, respectively. The NS mass is $M_{NS}=1.4M_{\odot}$ and its radius is $R_{NS}=15\,$km. The BH has a mass $M=15M_{\odot}$. All SPH particles are 
of equal mass ($m_p=M_{NS}/N$). The NS is initially placed at a distance $r_0=8M$ and for 
 cases with a spinning BH ($a/M \ne 0$) the NS is co-rotating in the equatorial plane. The
angular momentum of the BH is left as a free parameter.

In this first study we would like to get an idea of the possible
outcomes of a BH--NS merger: what percentage---if any---of the NS
mass survives the merger; what are the morphological features of
the merger (e.g., creation of an accretion disk, spread of the
accreting material), and how the BH angular momentum affects these features. 
For the two extreme cases $a/M=0$ and $a/M=0.99$ (we avoid
using $a/M=1$ for the reasons mentioned in \S 2.2), we observe
completely different behaviors.

\begin{figure}
\begin{center}
\epsscale{0.8}
\plottwo{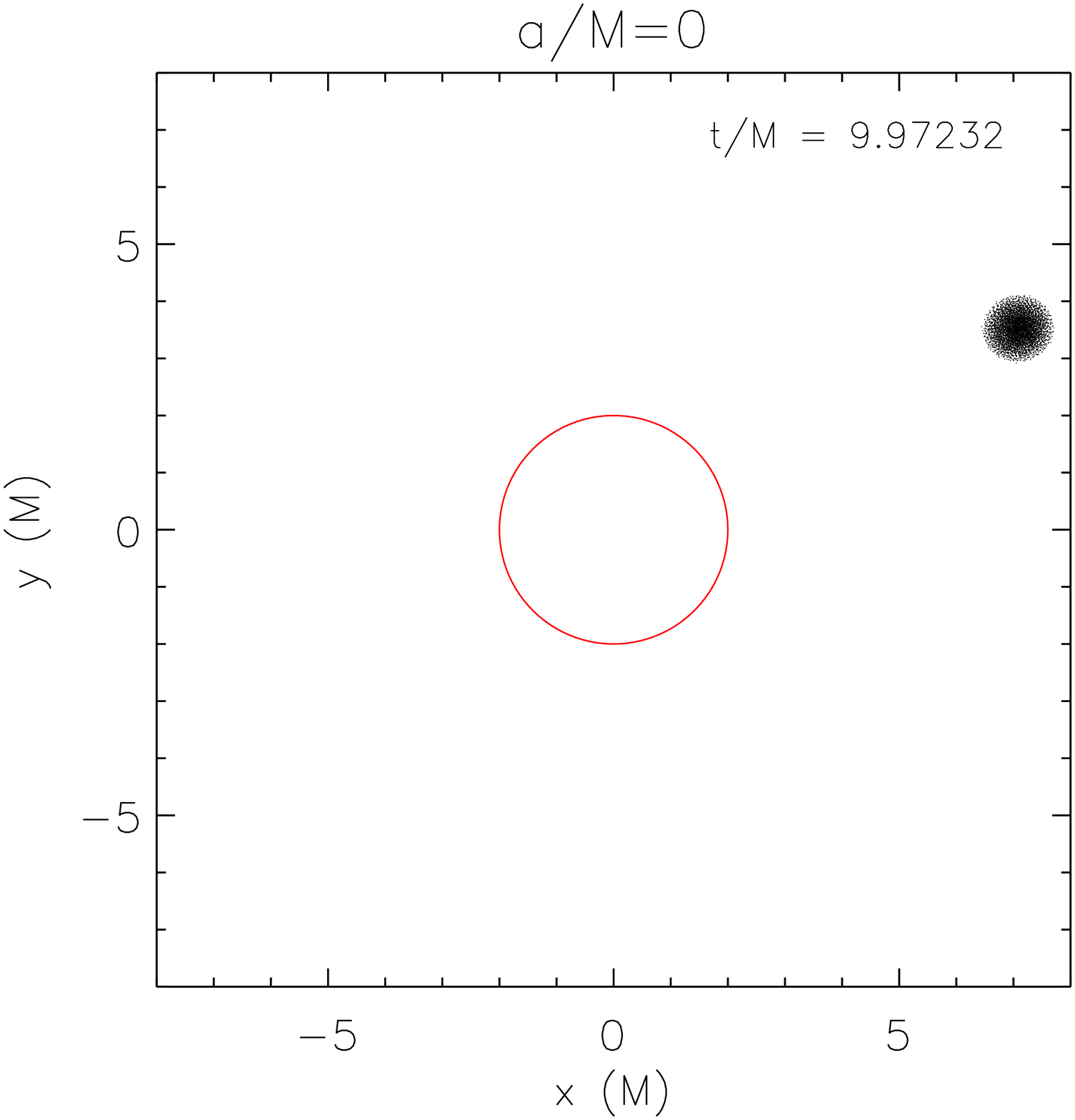}{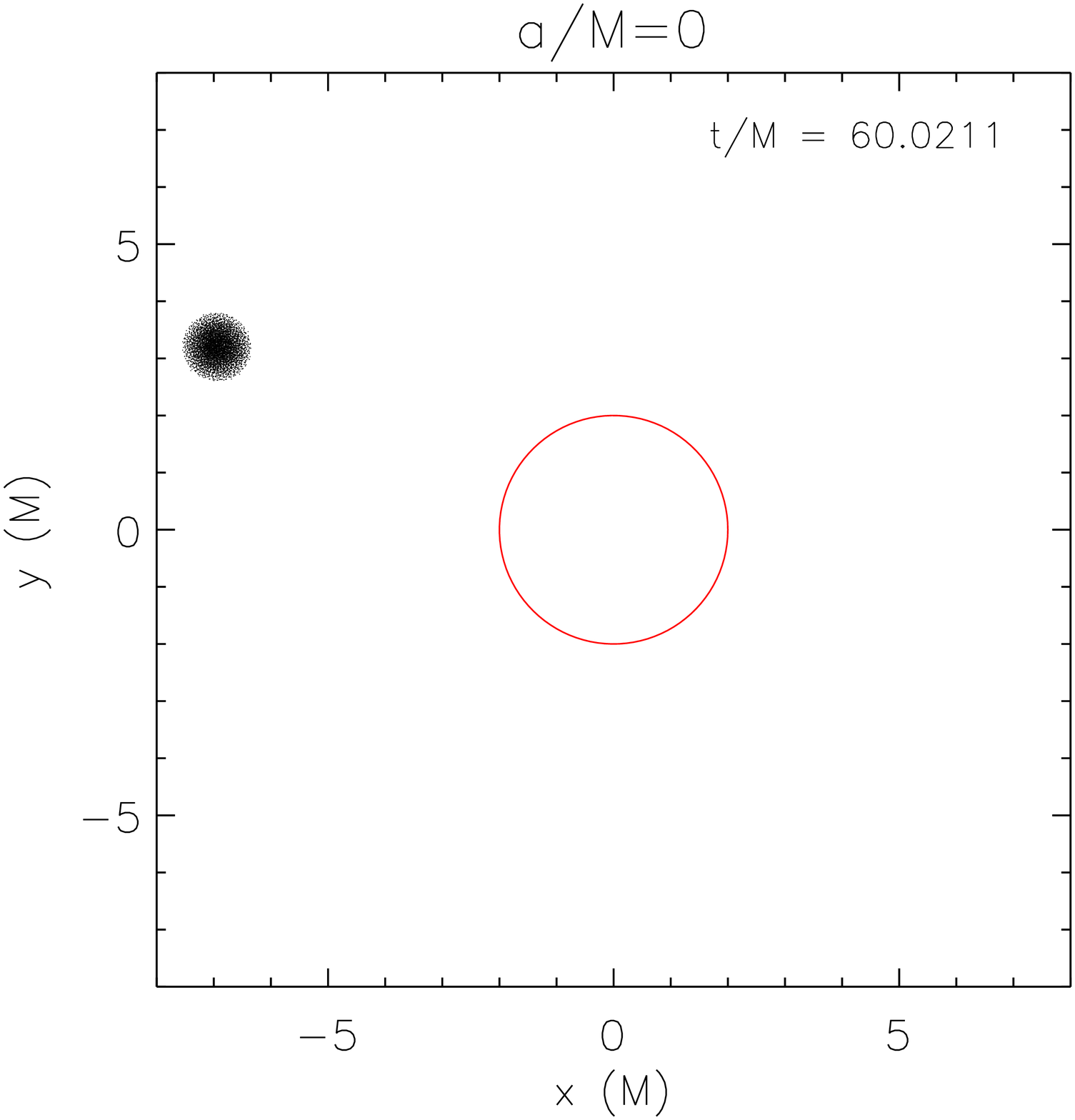}
\plottwo{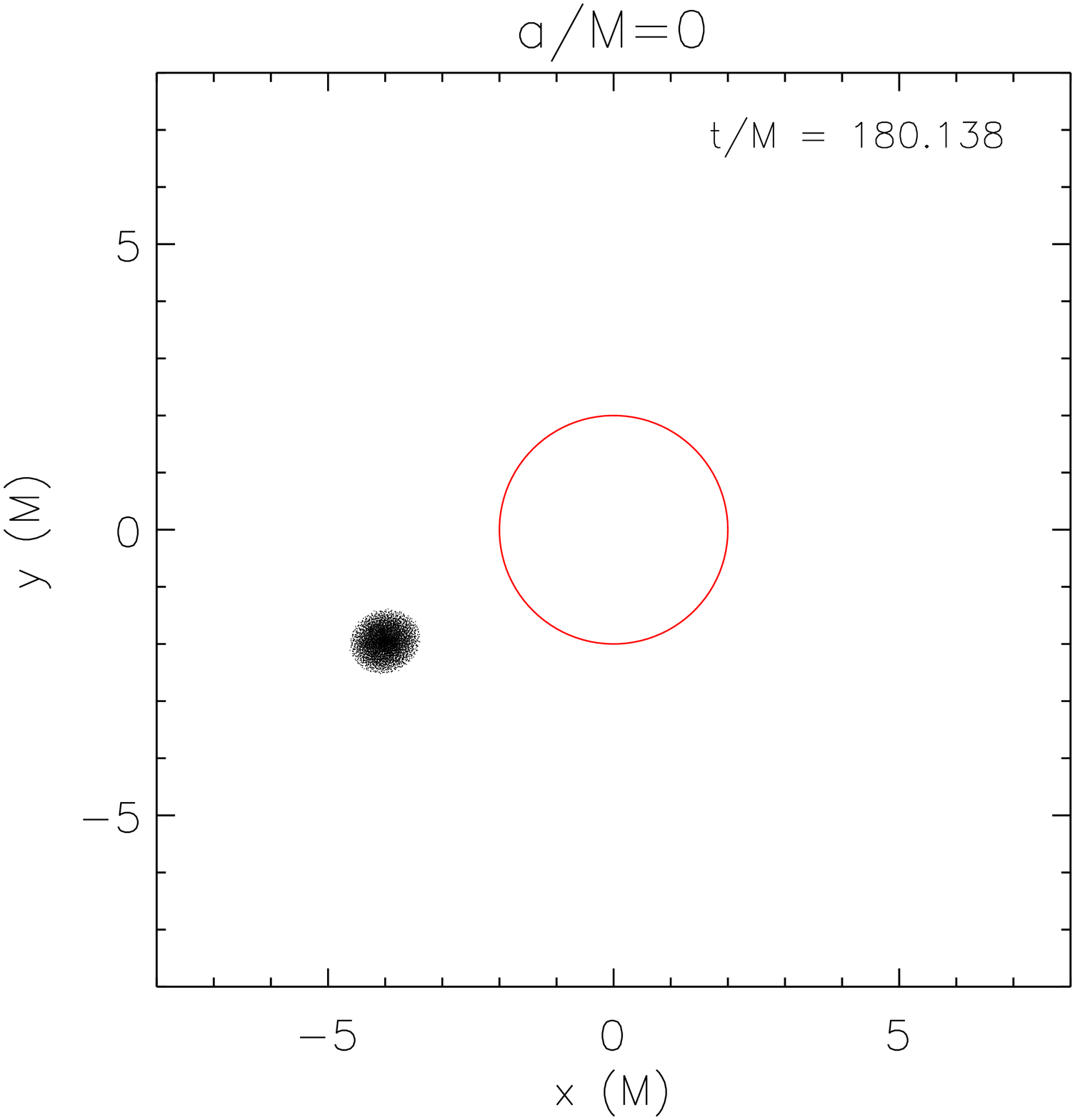}{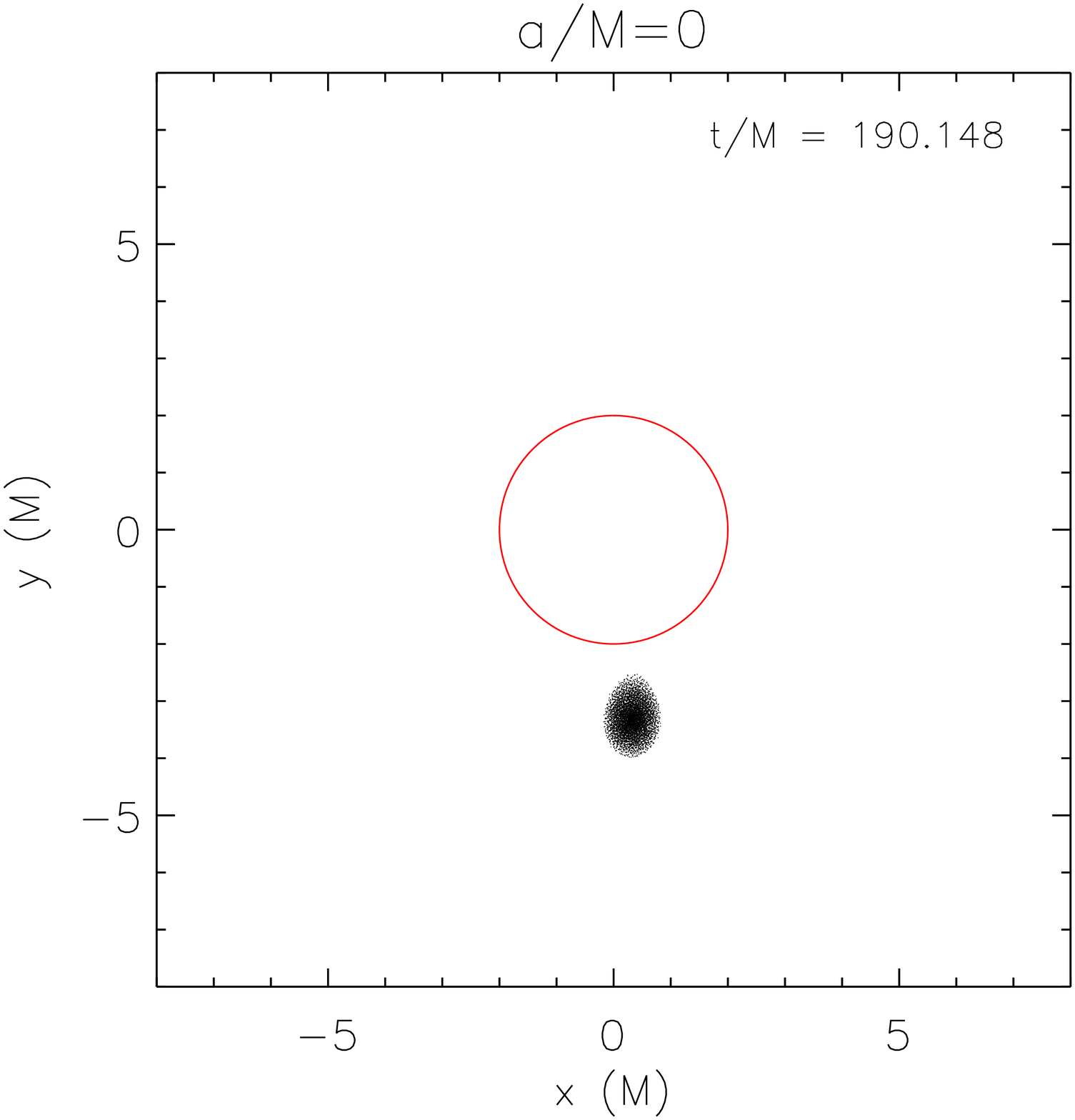}
\plottwo{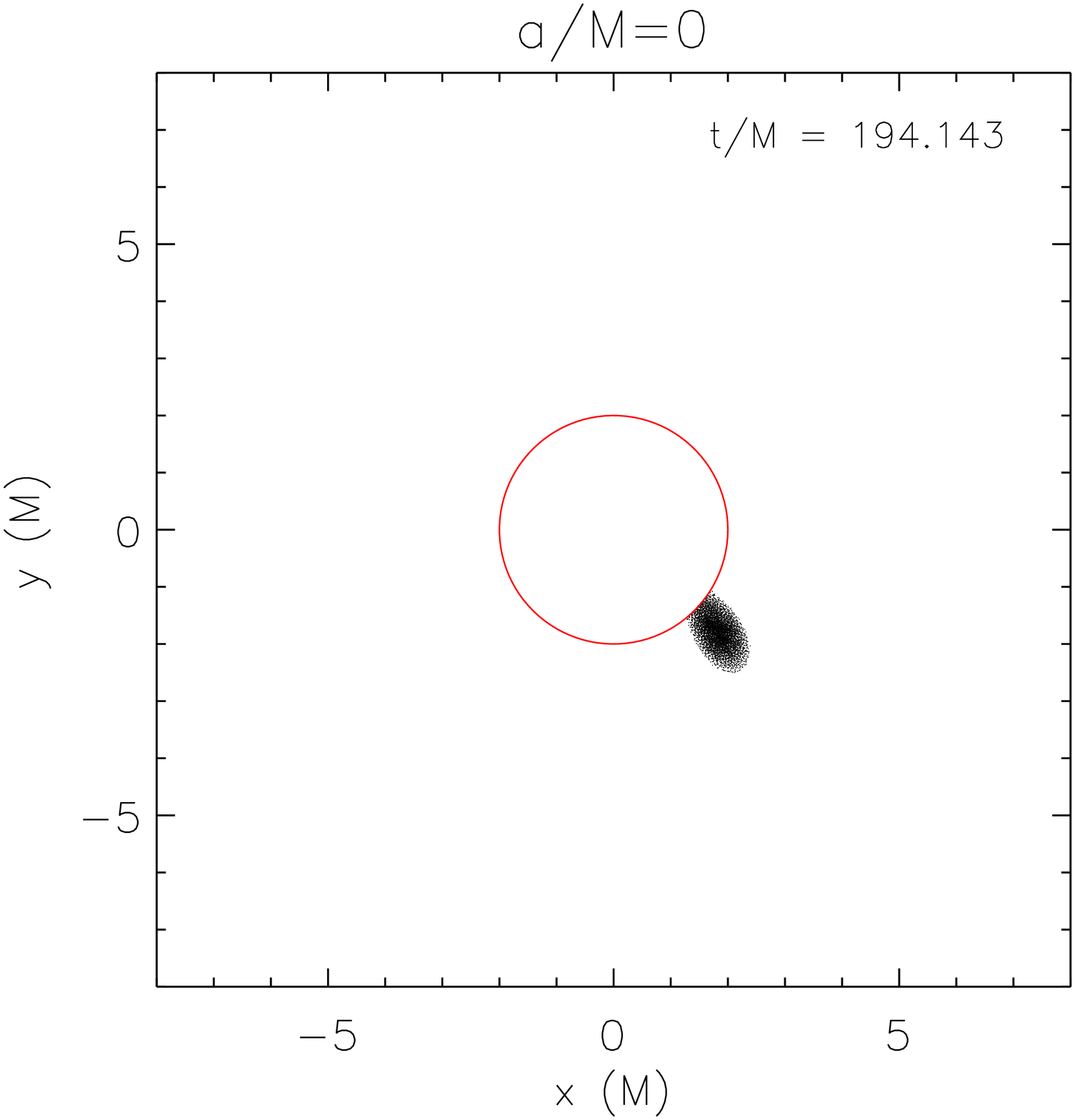}{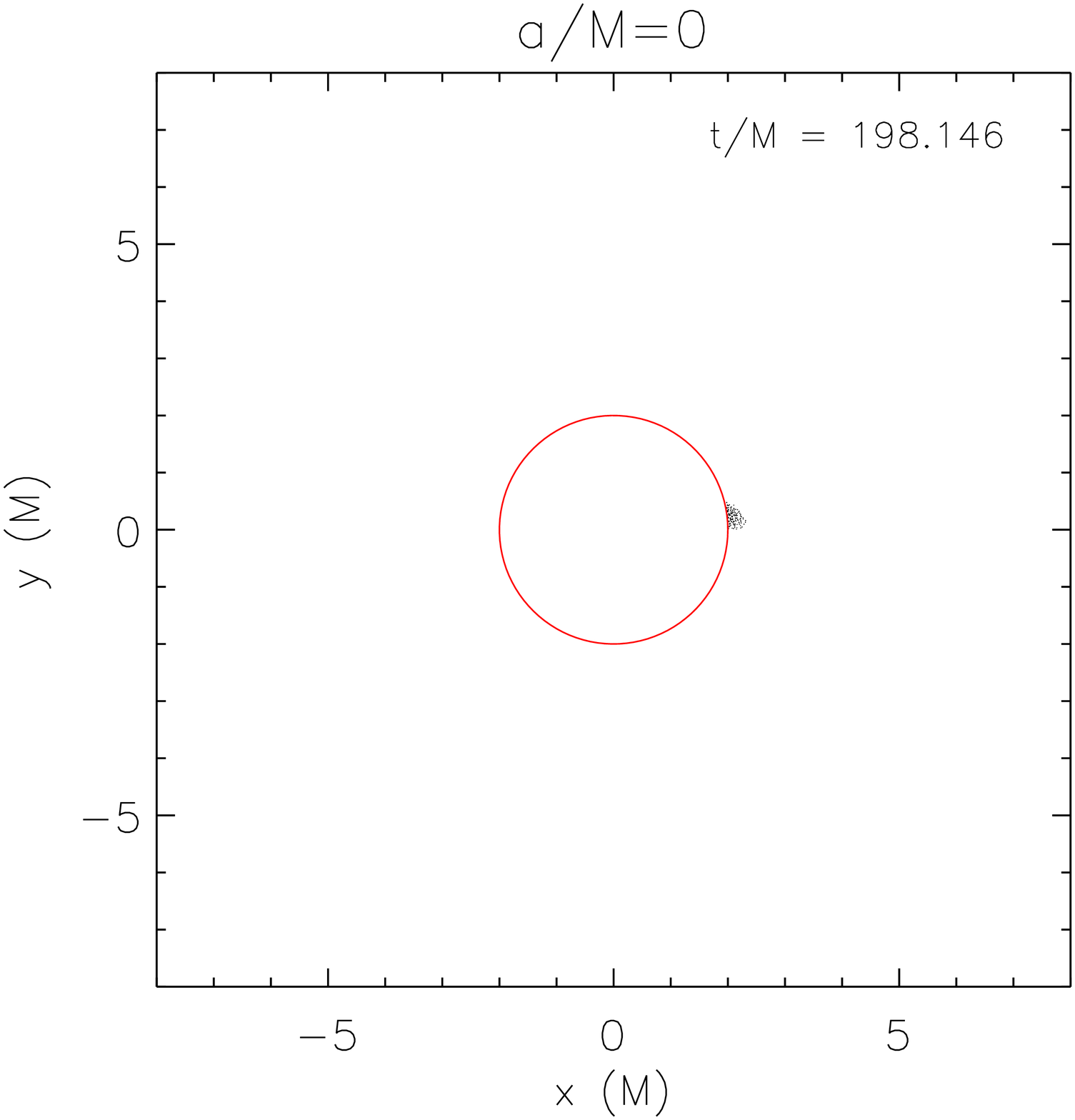}
\caption{Sequence of six snapshots from a simulation of a non-spinning BH--NS merger  (equatorial
projection). The $1.4\,M_\odot$ NS gets completely disrupted by a
non-rotating BH of mass $M=15\,M_\odot$. The NS was initially
placed outside the tidal limit ($r\simeq8\,M$ for this NS of
radius $R\simeq15\,$km with a $\Gamma=2$ polytropic EOS). The NS
fluid is disappearing completely into the BH horizon (at $r=2M$,
indicated by the red circle) in a time $t/M=180$ after the
beginning of the simulation.}
\label{a0_snaps}
\end{center}
\end{figure}

Fig.~\ref{a0_snaps} shows the outcome of the merger for a non-spinning 
BH: the NS, after being completely disrupted and
following an inspiraling orbit, disappears entirely into the BH's
horizon in a time $t/M\simeq180$ after the beginning of
the simulation (where $t$ is  the coordinate time for an observer at inifinity).

\begin{figure}
\begin{center}
\plottwo{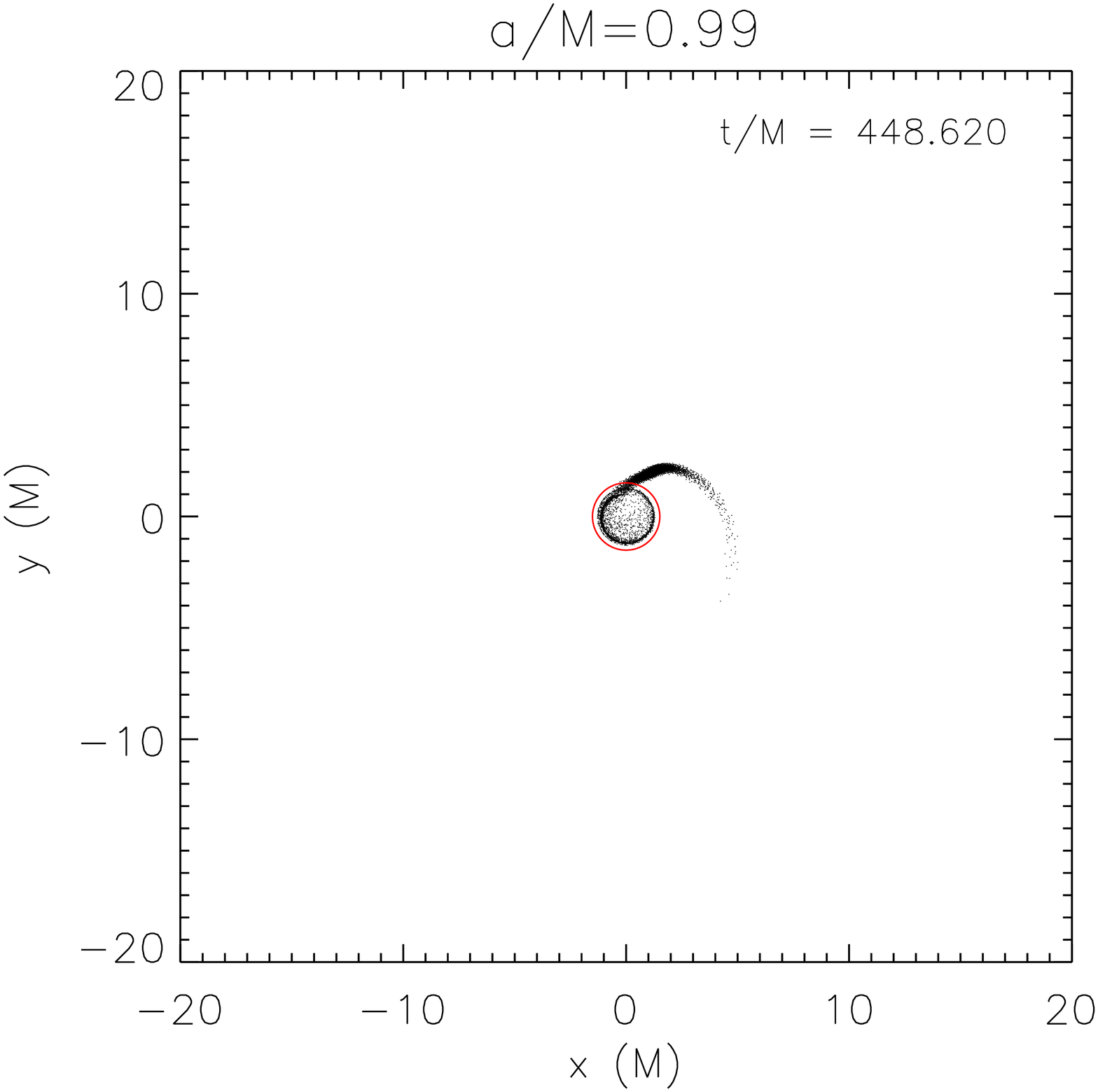}{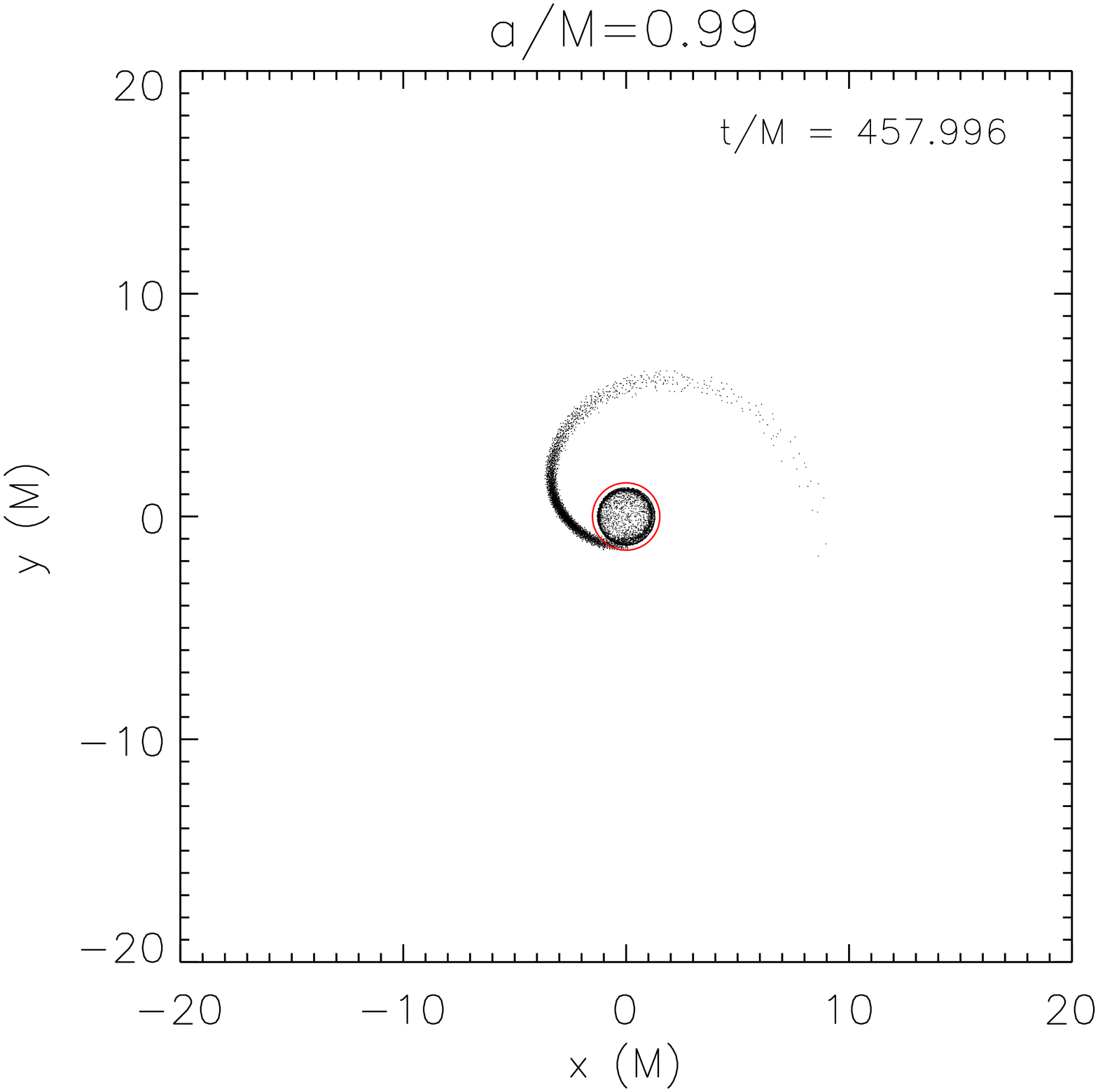}
\plottwo{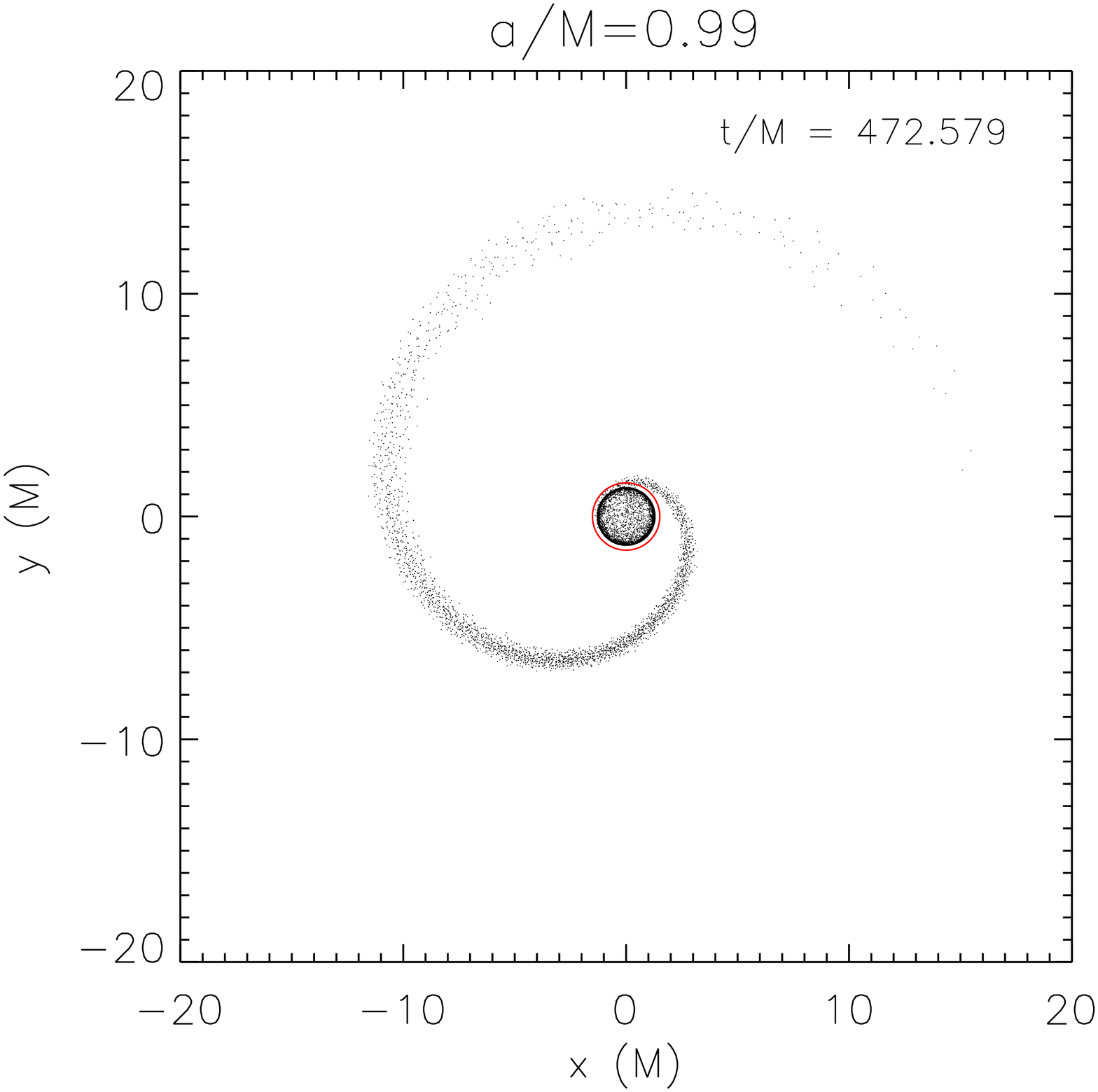}{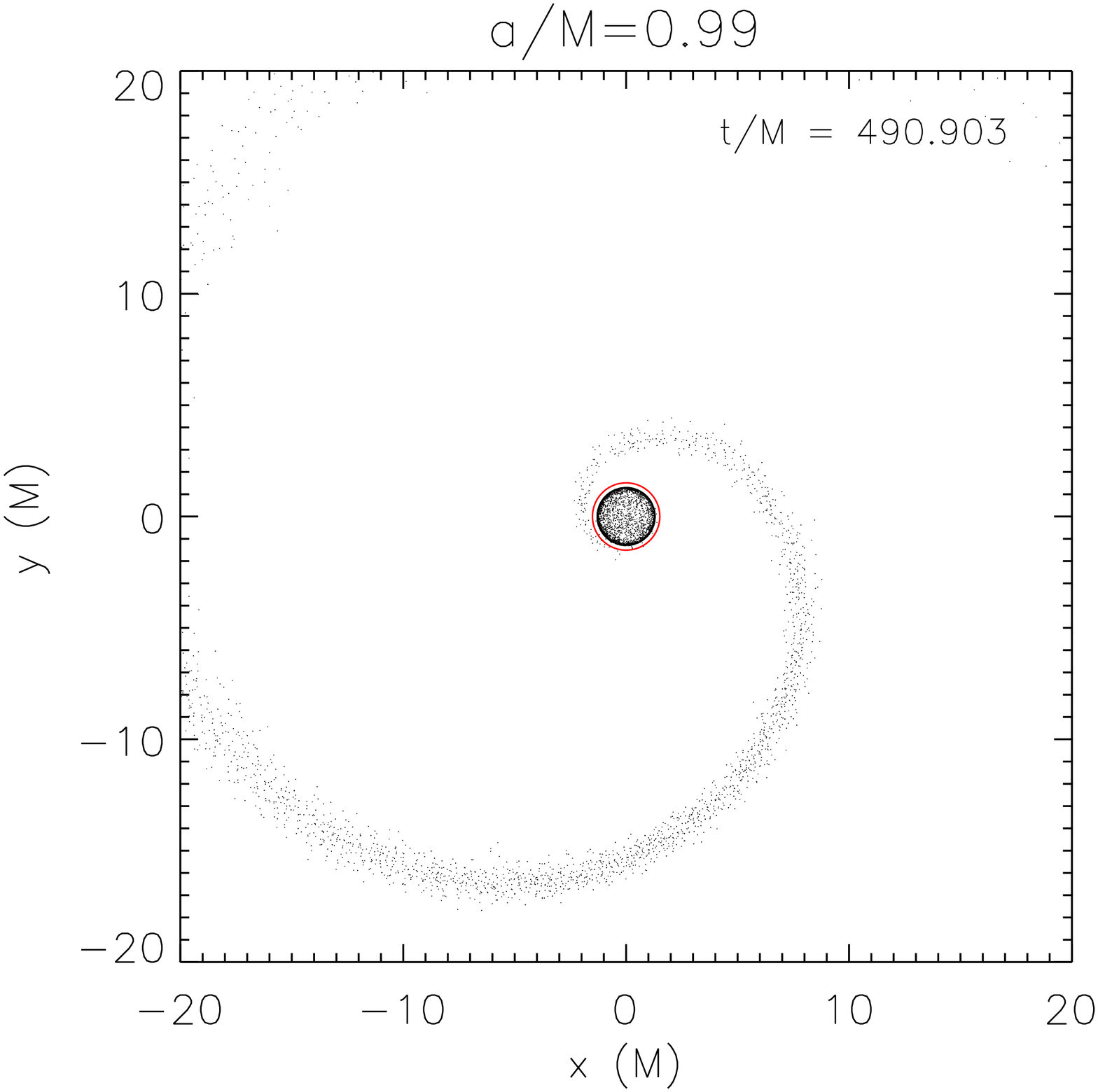}
\caption{Same as in Fig.~\ref{a0_snaps} but for a rotating BH with
$a/M=0.99$. Some of the NS material is still flowing inside the BH's
horizon, but there is now also a tail of expanding material forming. By the end
of the simulation ($t/M \sim 550 $) the infall of material stops completely
and about one third of the NS mass resides outside the horizon, with $\sim 30\%$ 
corresponding to unbound ejected material and the rest to the bound, disk-forming mass.}
\label{a099_snaps}
\end{center}
\end{figure}

\begin{figure}
\begin{center}
\plottwo{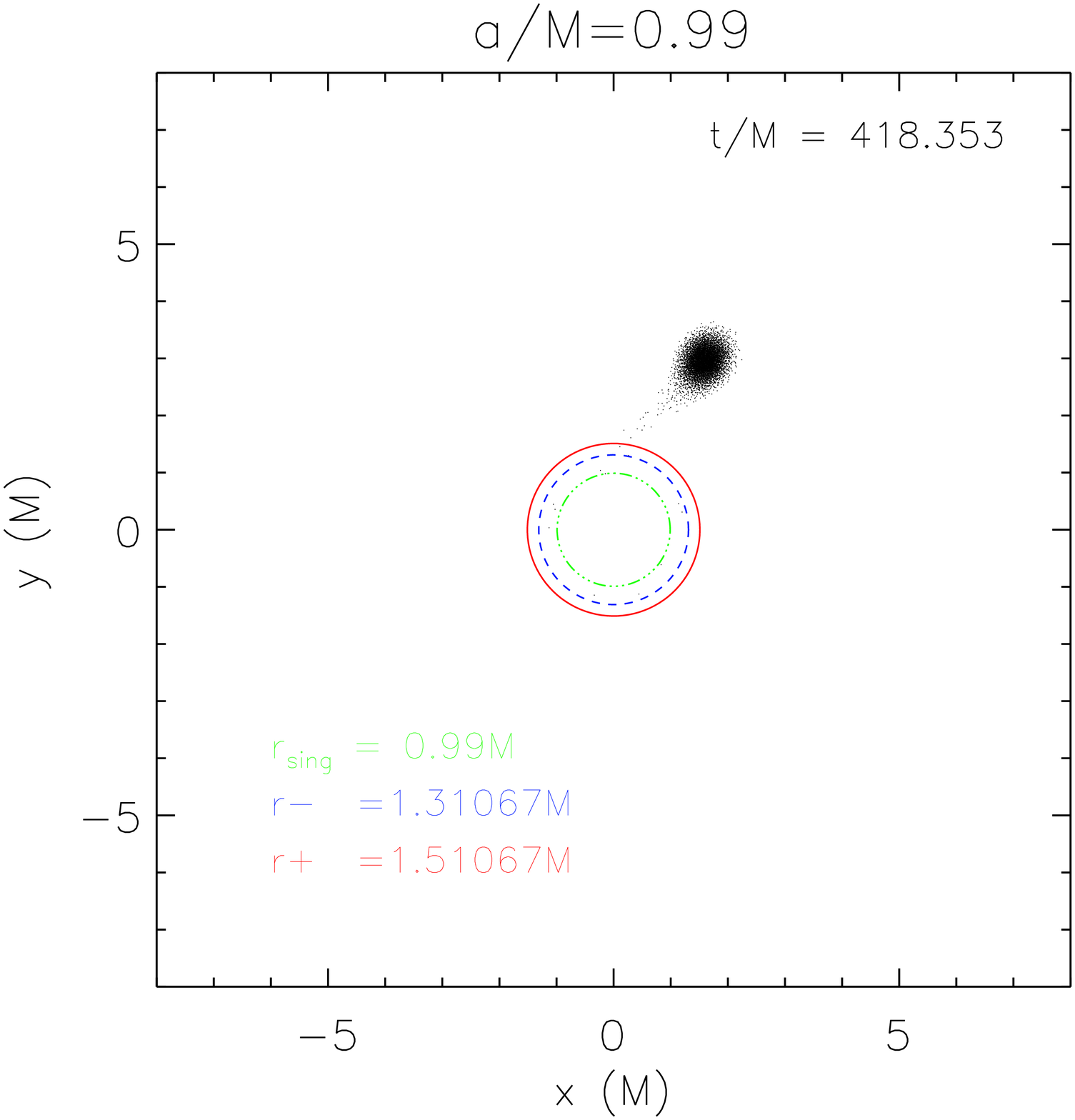}{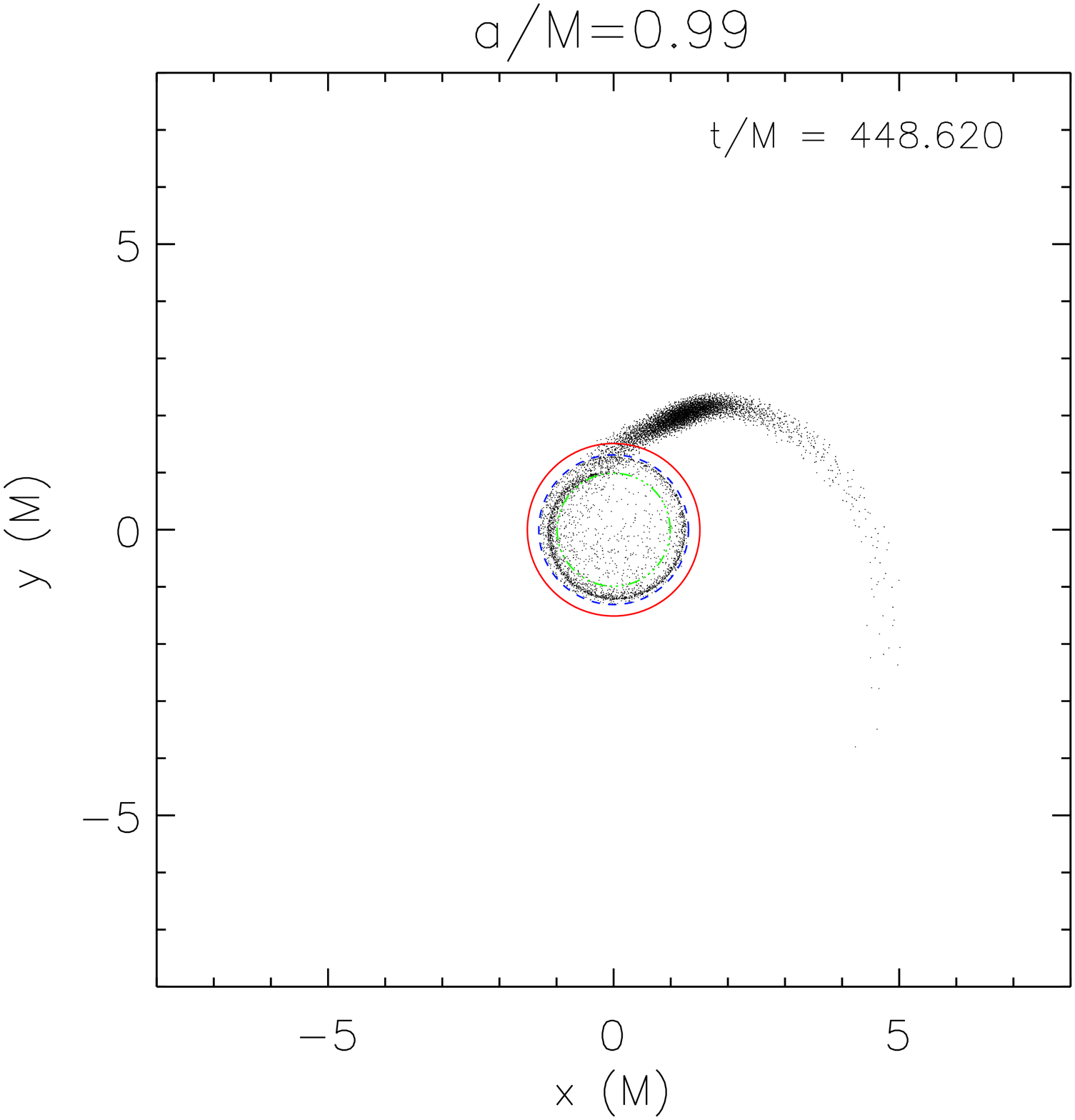}
\plottwo{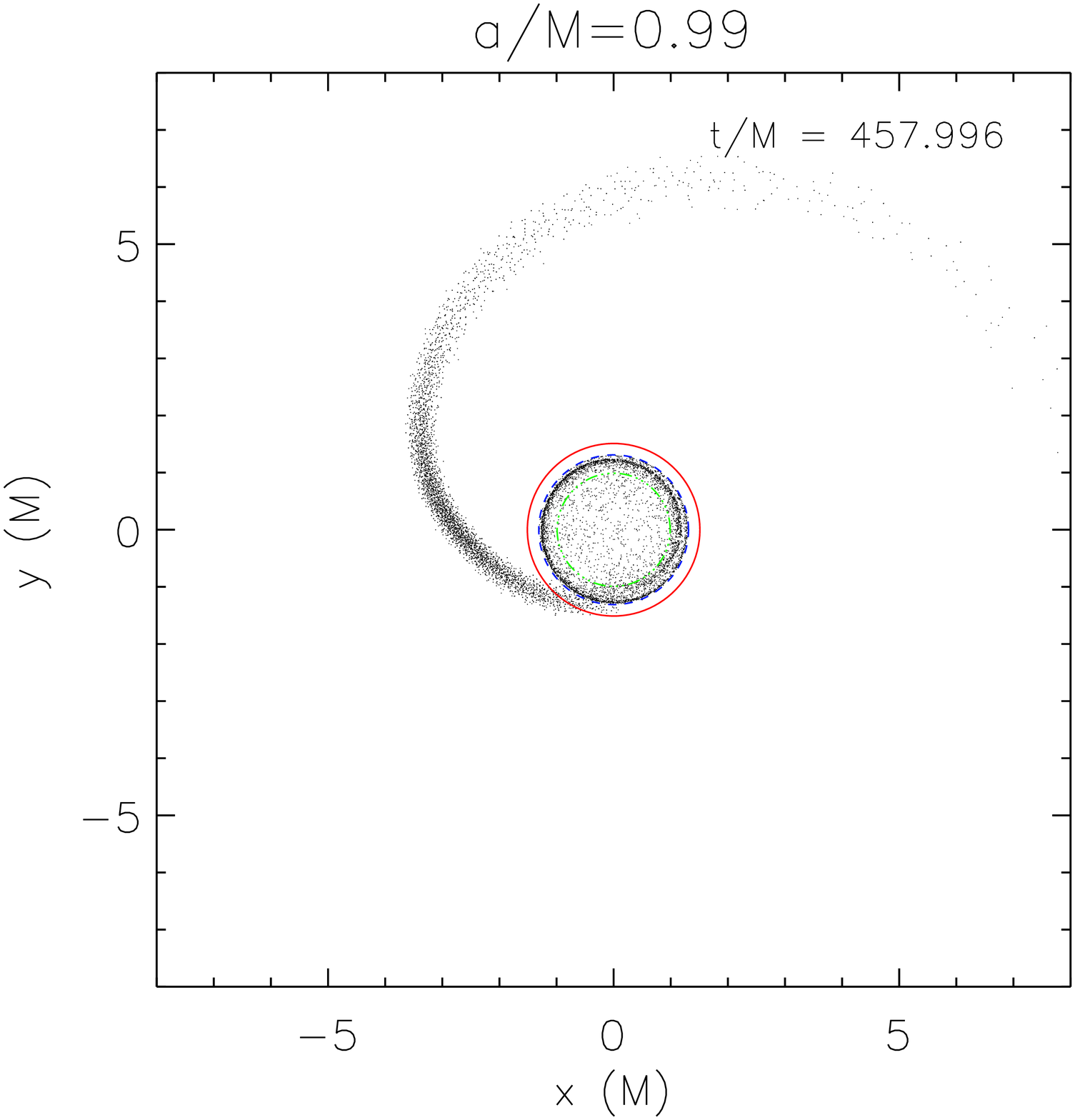}{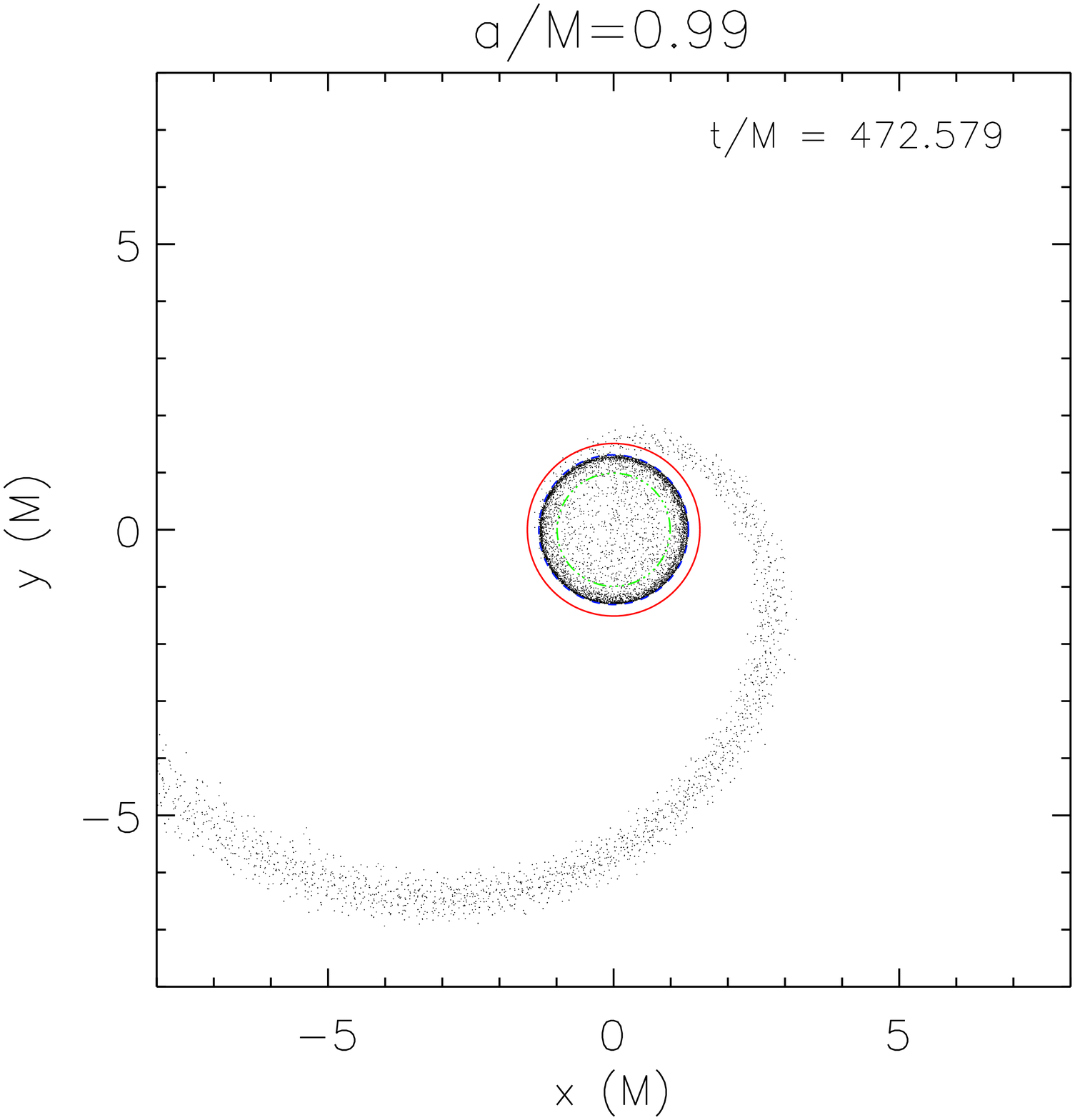}
\caption{A close up view of the central part of four of the snapshots in Fig.~\ref{a099_snaps}. The three
colored circles represent, in K-S coordinates, the BH's future
horizon $r_+$ (blue), the past horizon $r_-$(red) and the ring
singularity at $r=a$ (green).  This illustrates what we discuss
in \S2.1.1: the metric's
singularity  at the inner (past) horizon leads to the
illusory formation of a ring in the equatorial plane with radius
$r=r_-$. }
\label{a099_snaps_zm}
\end{center}
\end{figure}

\begin{figure}[htbp]
\centering \epsfxsize=14.0cm \epsfbox{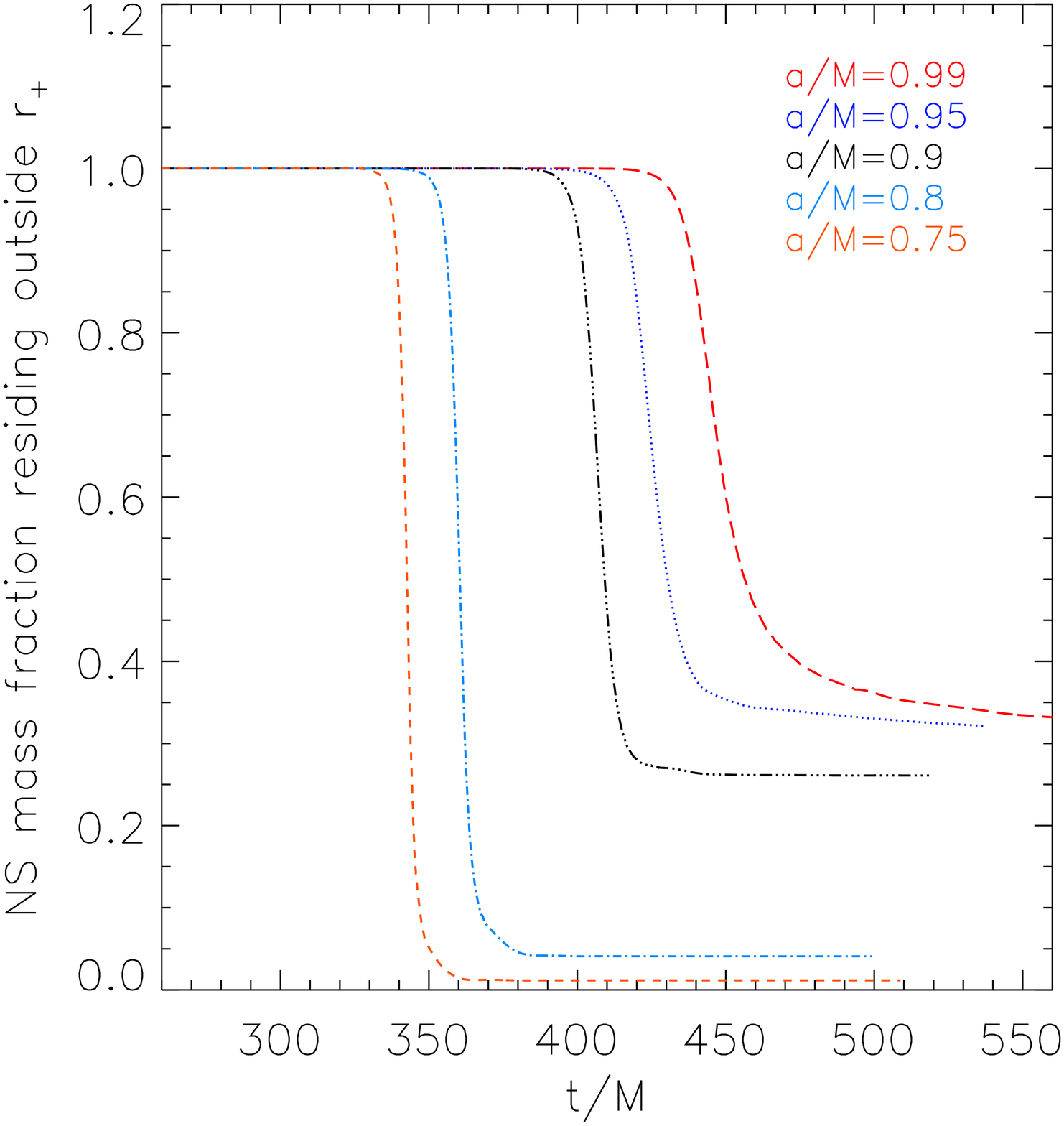}
\caption{Fraction of the NS's initial  mass that resides outside
the BH's future horizon as a function of time, for five different
values of the BH's angular momentum $a/M$ ($0.75 \le a/M \le
0.99$). The survival of NS material outside the BH horizon depends strongly on the BH's angular momentum. As $a/M$ decreases the fraction of surviving mass decreases as well and the merger becomes completely catastrophic for the NS for values $a/M\lesssim0.7$. Also note that
the accretion of material through the BH horizon starts earlier as the BH's angular momentum decreases; this is
 to be expected as the horizon is moving further away from the BH with decreasing $a/M$.} \label{mass_vs_t_all}
\end{figure}

    The result of our simulation for the case of the extremal
Kerr BH ($a/M=0.99$) is shown in Figs.~\ref{a099_snaps} and \ref{a099_snaps_zm}. The NS again gets
completely disrupted and falls toward the co-rotating BH following
an inspiraling orbit while an outwards expanding tail of the
disrupted NS's material is forming at the same time. The NS fluid
starts disappearing into the BH's horizon at $t/M \simeq 400$  with an initially high infall rate
which finally diminishes to an almost zero level. At this point
($t/M=490$) about $33\%$ of the initial NS mass resides
outside the BH's horizon (Fig.~\ref{mass_vs_t_all}). We notice that the disruption takes place 
well outside the ISCO, in contrast to the previous case of a Schwarzschild BH.

We run the same exact type of simulations for different values of
the BH's angular momentum and compare the results. 
Fig.~\ref{mass_vs_t_all} shows the fraction of the initial NS mass
that survives the merger for five different values of $a/M$
($0.75\le a/M\le 0.99$). As the angular momentum of the BH
decreases, so does the fraction of material that survives,
residing outside the BH's horizon. For the case of $a/M=0.1$ (not
included in the latter figure for practical reasons), the
situation is almost the exact same as for the Schwarzschild
(non-spinning) BH: the NS disappears completely into the horizon
(Table \ref{table1} summarizes our results).

 For the simulations with $a/M$ spanning the range $0.1 < a/M < 0.75$, namely for
$a/M=0.2$, $a/M=0.5$ and $a/M=0.6$, we observe that the infall
starts earlier as $a/M$ decreases (consistent with the behavior of
the high $a/M$ mergers). No surviving material exists
for those mergers with low spin for the BH ($a/M<0.7$).

The morphological features of the mergers for the non-maximally
spinning BH are similar to the extremal case, with the difference
that, as $a/M$ decreases the outwards expanding tail tends to
spread less. For the very low $a/M$ cases no such tail is forming.

\begin{figure}
\begin{center}
\plottwo{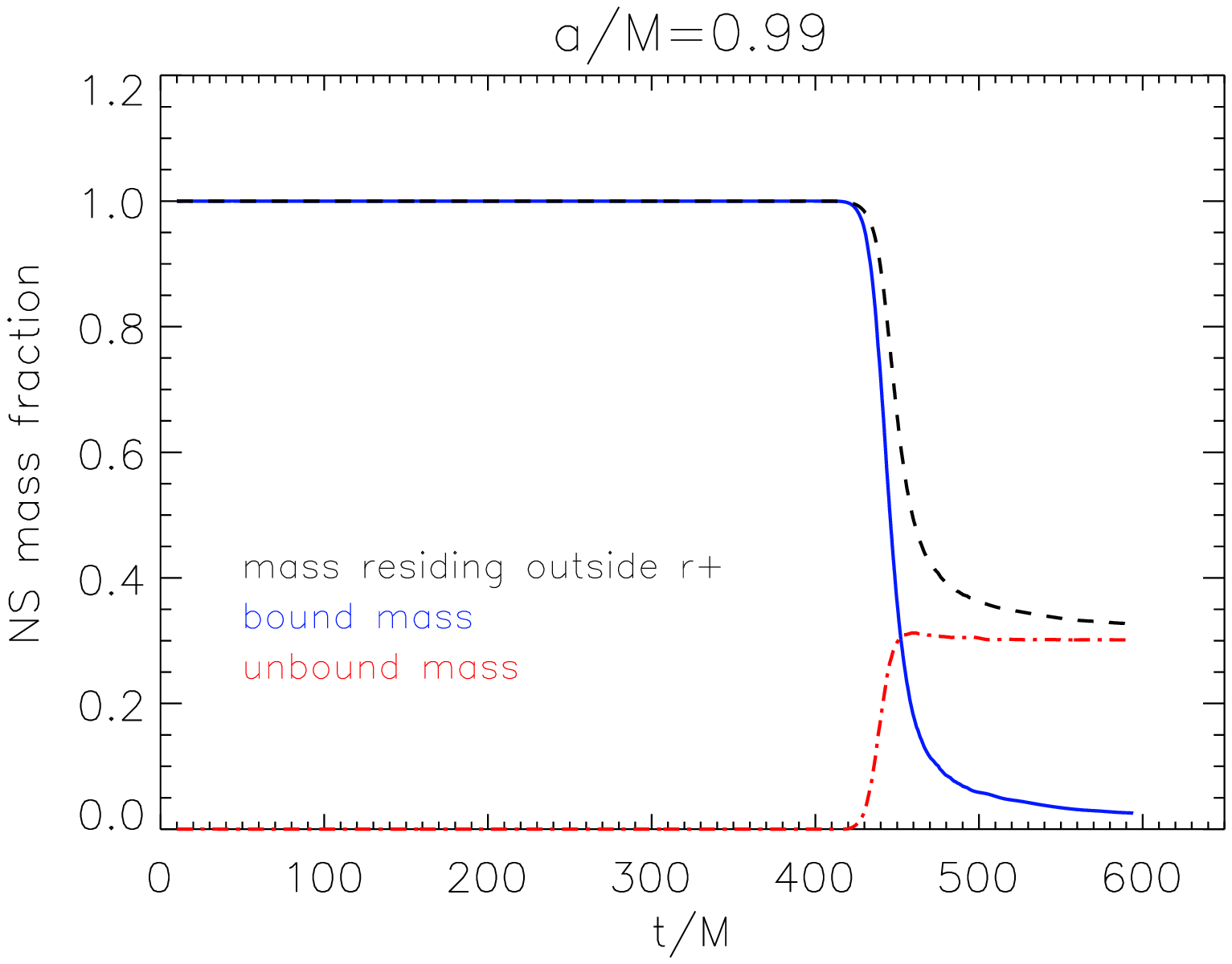}{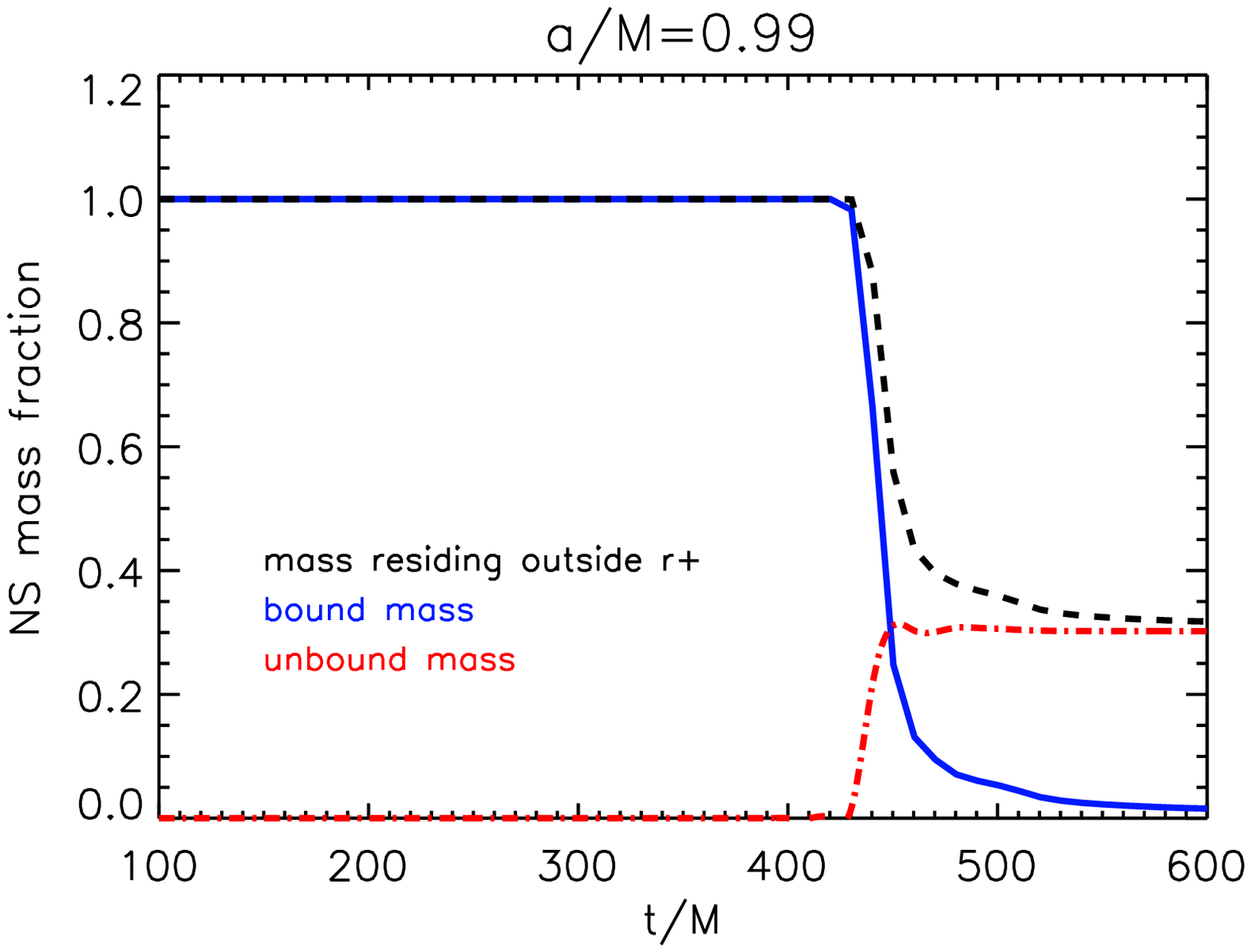}
\plottwo{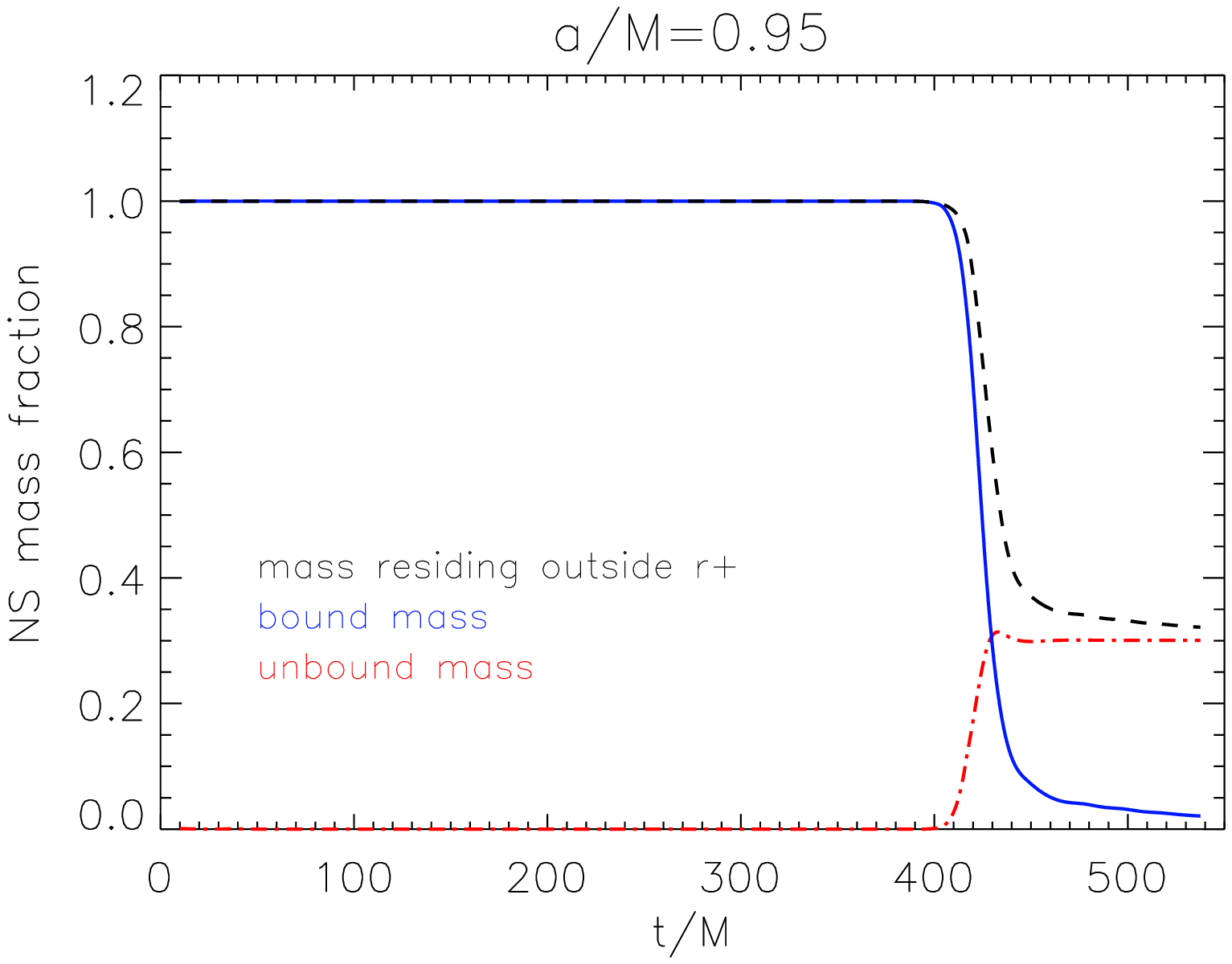}{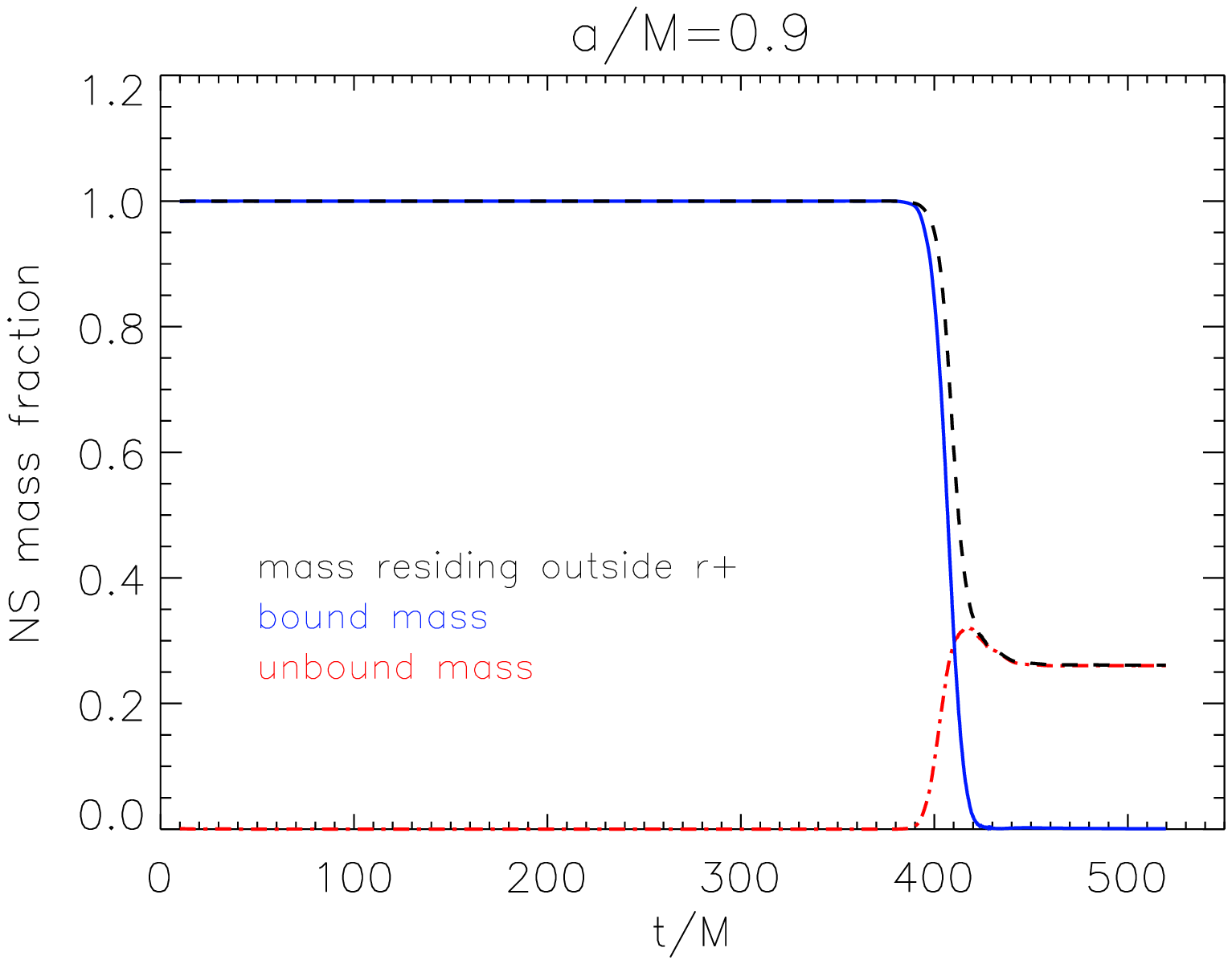}
\plottwo{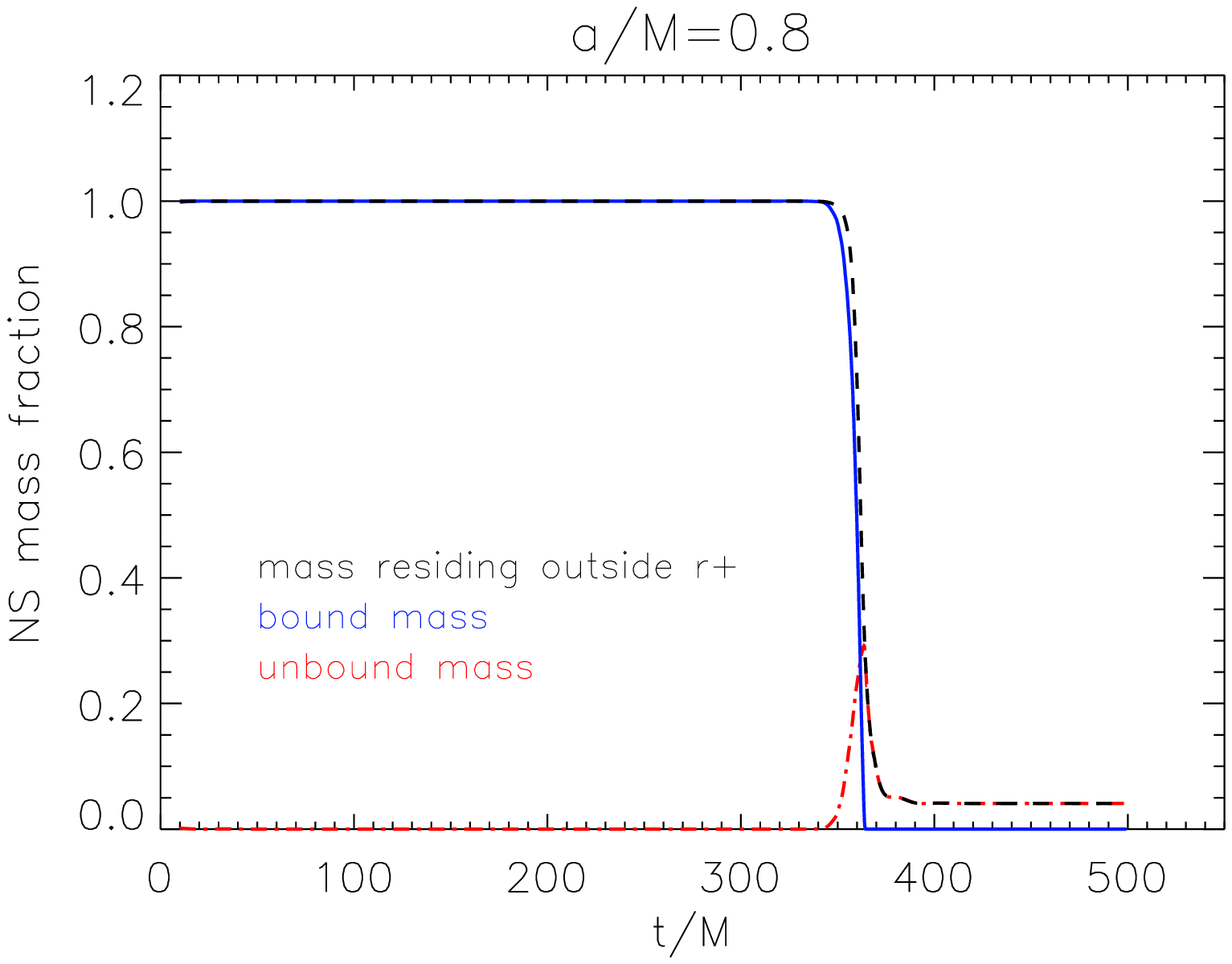}{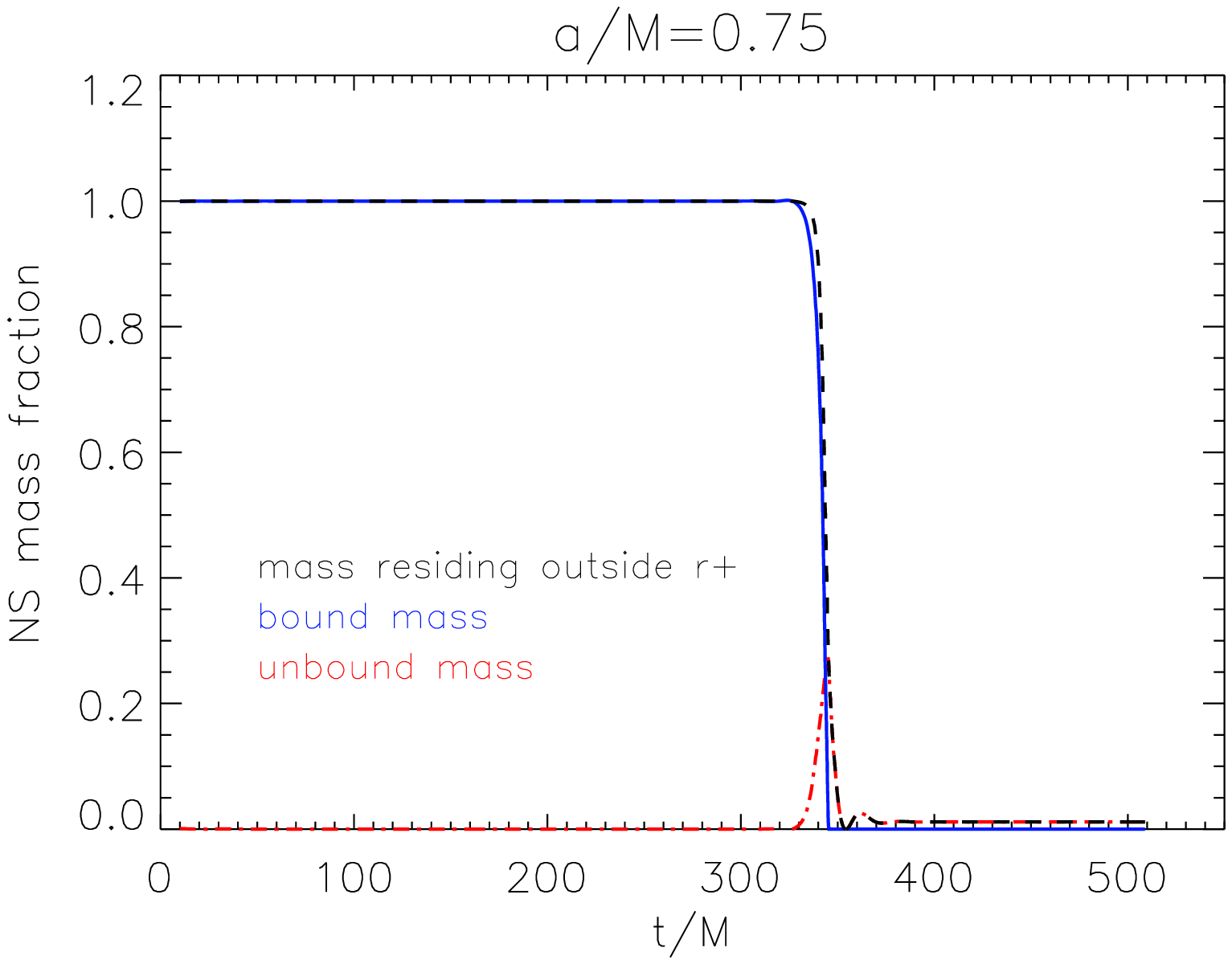}
\caption{ Fraction of the NS's initial mass residing outside the BH's horizon as a function of time for five different simulations, corresponding to five different values of $a/M$ (runs E1-E5). The black, blue and red lines correspond to total, bound and unbound NS material respectively. The top two plots correspond to a maximally spinning BH, using two different resolutions: $10^4$ and $10^5$ SPH particles for the left and right plots, respectively. For values of $a/M<0.95$ the percentage of surviving bound material drops to unresolvable
levels.}
\label{b_unb}
\end{center}
\end{figure}

By the end of the SPH simulations, the  mass fraction that resides
outside the BH's future horizon seems to have reached a stable
value.  Since for $a/M \geq 0.9$ there is a substantial fraction of the
NS mass surviving the merger, one would like to investigate what
the final fate of this material might be, i.e., what percentage of it---if an---will escape or stay bound, forming a stable disk around
the BH. To answer those questions we first calculate the rest
energy (energy as measured by an observer at infinity) of each SPH
particle throughout the whole simulation for runs E1-E5. By
knowing a particle's energy we can determine whether it is bound
or unbound ($e^2 >1$ for unbound and $e^2<1$ for
bound\footnote{for an analytical discussion and a proof of that
see \citet{Wilkins} and \citet[][chap.3, \S19]{Chandra}}, where $e=-(g_{tt}u^t+g_{t\phi}u^{\phi}+g_{tr}u^r)$
 is the energy per unit rest mass for an observer at infinity). 
 Fig.~\ref{b_unb}  shows the variations of bound and unbound mass for
each of our simulations.

\begin{figure}[htbp]
\begin{center}
\epsfxsize=14.0cm \epsfbox{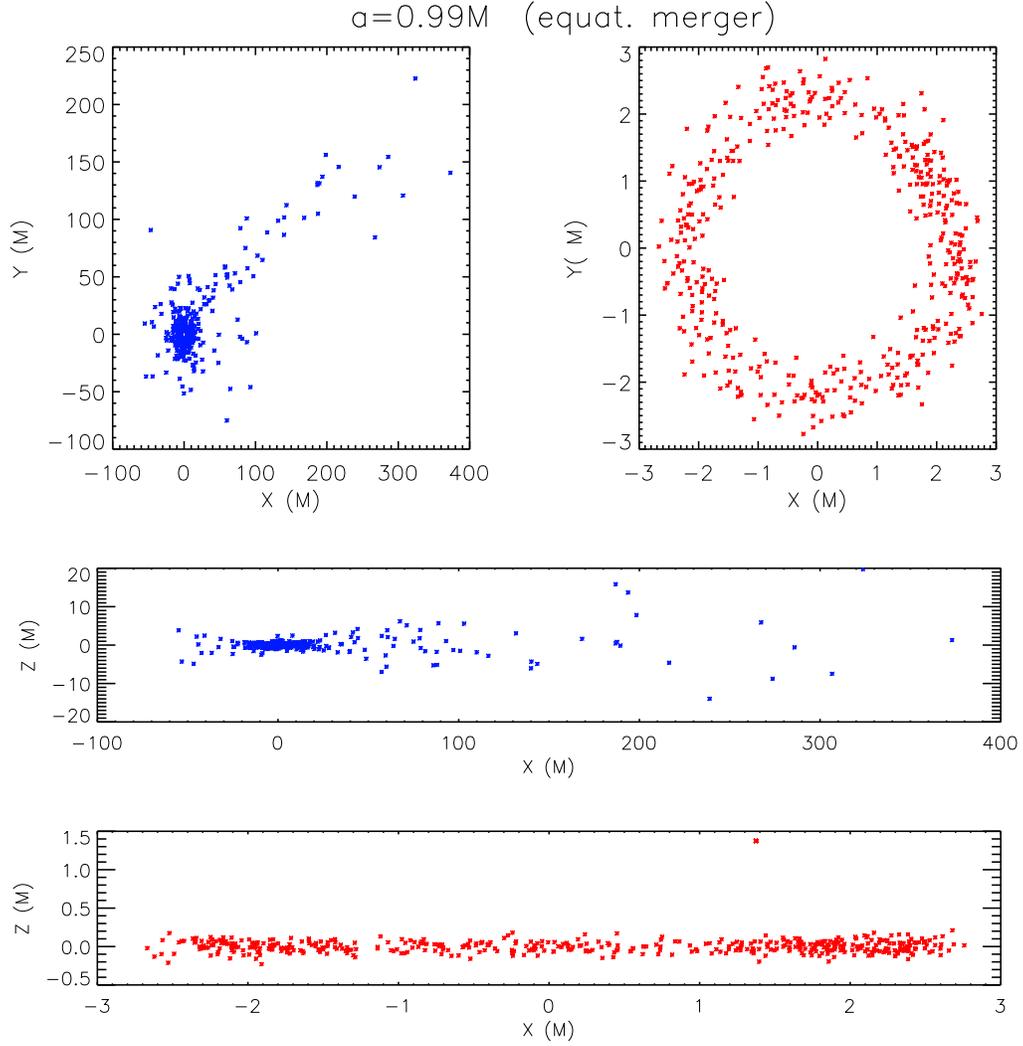} \caption{Spatial
distribution of the apocenters (blue) and the pericenters (red)
for the disk-forming SPH particles of the $a=0.99M$ equatorial
merger (Run E1). The upper left (right) and middle (bottom) panels
correspond to the x-y and x-z plane projections for the apocenters
(pericenters) respectively. The total mass of
the disk is $2.5\%$ of the initial NS mass. The mean pericenter
value (inner radius of the disk) is $r_{peri}\sim 2.5M$ and
the mean apocenter (outer radius of the disk) is
$r_{apo}\sim 30M$. } \label{a099_apo_peri_2D}
\end{center}
\end{figure}

For the two runs with the highest $a/M$ values (runs E1 and E2
from Table 1) we are able to resolve a nonzero fraction of
bound material ($2-3\%$ of the NS mass). For runs E3, E4 and E5 the percentage of bound material drops to unresolvable values (recall that our mass resolution is $m_p=M_{NS}/10^4$). As shown in Fig.~\ref{b_unb} for the runs E1 and E2, both the unbound and bound NS material stabilize at a certain value and persist there for a long time. That leads us to believe that the bound fraction of the surviving mass stays around the BH forming a stable disk. In order to check whether our definition of bound and unbound material is correct, we perform the following test-runs: we evolve the results of the SPH simulations using again the geodesic integrator. For every run (different $a/M$) we set as initial conditions for the geodesic integrator
the last output file from the equivalent SPH simulation. We make
sure to run the geodesic integrator for a sufficiently long time, as it is much
faster than the SPH code. The results of those tests confirm our
expectations. Namely, the material that we recognized as unbound escapes
completely following parabolic trajectories, whereas the particles
that were energetically determined to be bound (for runs E1 and E2
only) remain bound around the BH (and outside the BH's horizon)
following equatorial precessing orbits, with maximum apastron $\sim30M$, minimum periastron of about $\sim 2-3M$ and a very small dispersion in the direction of the BH spin axis $\sim 0.1-0.2M$ (see Fig.~\ref{a099_apo_peri_2D}).

Two important results are seen in Table \ref{table1}: no
significant fraction of bound material survives for mergers with $a/M\lesssim 0.95$ and 
therefore no formation of a stable disk is observed; for the cases with $a /M\lesssim 0.7$ the merger is completely catastrophic for the NS, with no material surviving.

Finally, in the upper two plots of Fig.~\ref{b_unb} we present a comparison between a lower resolution run ($10^4$ particles) and a run with higher resolution ($10^5$ particles) for the case 
of a maximally spinning BH. The left panel corresponds to the low-resolution run and the 
right panel to the high-resolution run. The agreement between the two runs is very good and
this justifies our use of the lower resolution for all runs presented in this paper. 
The higher-resolution run took 248 CPU hours on two dual-core AMD Opteron 2.8 GHz processors, while 
the lower-resolution run took only 26 CPU hours.
For both cases, the simulation resulted in the ejection of about $30 \%$ of the NS mass (unbound material) along with the survival of $\sim 2-3\%$ of the initial NS mass as bound material.

\subsection{Non-equatorial BH--NS Mergers}

Many studies have suggested that a significant NS birth kick is required for the formation of coalescing BH--NS binaries \citep[e.g.,][]{kalogera, lipunov}. Any misalignments between the axis of the BH and the NS progenitor are expected to be canceled during the evolution of the binary prior to the supernova (SN) explosion that is associated with the formation of the NS, due to mass-tranfer phases. Any spin-orbit misalignement  therefore is expected to be introduced by the SN that forms the NS. \citet{kalogera} argues that tilt angles greater than $30^o$ are expected for $30\%-80\%$ of coalescing BH--NS binaries, whereas tilt angles of $50^o-100^o$ are expected for at most $70\%$ of such systems. In order to account for these findings, we set up some simulations for misaligned mergers, covering a wide range of tilt angles from $30^o-180^o$. 

\subsubsection{Setting Up Initial Conditions}

The selection of initial conditions for the equatorial mergers is
straight forward: select an initial radius $r_0$ -outside the
Roche limit- and calculate the angular velocity $\Omega_{\phi}$ needed for the relaxation procedure
using Eqs.~(\ref{Omega}) and (\ref{KSr}). Obviously, Eq.~(\ref{Omega}) does not hold
for non-equatorial orbits, since its derivation assumes that
$\theta=\pi /2$. When moving away from the equatorial plane,
there is a third constant of the motion\footnote{the other two
being the energy $E$ and angular momentum $L_z$, related to the
stationary and axial Killing vectors respectively.} appearing (the
Carter constant $Q$), as a result of the existence of a killing
tensor. (The geodesics equations of the Kerr spacetime are
included in Appendix B). In order to find a stable,
so-called spherical, non-equatorial orbit on which to initially place the
NS, we follow the technique described in \citet{Hugh}. (For an in
depth analysis of the procedure, refer to Hughes 2000  and Wilkins 1971). Here, we point out the basic steps for
finding and setting up numerically the initial conditions for such
an orbit, as presented in \citet{Hugh}.

For a circular orbit $R(\equiv dr/d\tau)=0=R'$, where the prime
indicates differentiation with respect to $\tau$ . Furthermore,
for the orbit to be stable $R''<0$ should also hold. One can
specify a unique orbit by fixing $r_0$ and $L_z$. The conditions
$R=0=R'$ will fix the other two parameters of the orbit, $E$ and $Q$.
The inclination of the orbit (which is a constant of the motion)
can be calculated as : \beq \cos(i)=\frac{L_z}{\sqrt{L_z^2 +Q}}
\label{inclin-i}\eeq (with $i$ being zero on the equatorial plane
where the Carter constant $Q$ is also zero).

The most bound orbit (in terms of energy) for a given radius $r_0$
is the equatorial prograde orbit (in the same sense the retrograde
orbit corresponds to the least-bound orbit). Therefore, one can start by
choosing a radius $r_0$ in the equatorial plane (where $Q=0$) and then,
by solving $R=0=R'$, calculate the energy $E$ and angular momentum
$L_z$ for that orbit. 
Solving analytically the condition   (with $R$
being defined by Eq.~(B1)) one gets for the prograde and retrograde
orbits \citep{Hugh}

\beq E^{pro}=\frac{1-2v^2+pv^3}{\sqrt{1-3v^3+2pv^3}}\eeq
\beq L_z^{pro}=rv \frac{1-2pv^3+p^2v^4}{\sqrt{1-3v^2+2pv^3}}\eeq
\beq E^{ret}=\frac{1-2v^2-pv^3}{\sqrt{1-3v^3-2pv^3}}\eeq
\beq L_z^{ret}=-rv \frac{1+2pv^3+p^2v^4}{\sqrt{1-3v^2-2pv^3}}\eeq

where $v=\sqrt{M/r} $ and $p=a/M$.

With $E$ and $L_z$ numerically known for a given equatorial orbit of
radius $r_0$, one can proceed into finding non-equatorial stable
orbits. The way to do so is by keeping the radius $r_0$ fixed,
decreasing the value of $L_z$ and solving again for the conditions
$R=0=R'$, which now are going to give the energy and Carter
constant. This way one can keep on decreasing the value of $L_z$
until it reaches the value $L_z^{ret}$ or until $R''=0$
(marginally bound). The stability of the new, inclined orbit (with
the inclination given by Eq.~(\ref{inclin-i})) should be checked
with the requirement $R''<0$. The angular velocity of this orbit
can now be numerically determined as $\Omega_{\phi}=d\phi /dt$ by
using the geodesics equations for $d\phi/d\tau$ and $dt/d\tau$
(Eqs.~(B3) and (B4)). The velocity for a stable, spherical,
non-equatorial orbit of radius $r_0$ and inclination $i$ will be
$v_y=r_0 \Omega_{\phi} \cos(i)$

Since we have followed this numerical method to find initial
conditions for the inclined mergers, tuning the $r_0$ and $L_z$
values in order to solve for an exact value of inclination angle
was not trivial to do. In any case we tried to get as close to the
value of the desired inclination angle as it was numerically
possible. For example, the inclined merger labelled to be $30^o$
was in reality $29.6577^o$. In Table \ref{table2} we have listed the actual values of orbital inclinations used in our simulations, but when referring to them either in the text or in plot labels we use the rounded off values. 

\subsubsection{Results}

We have run five simulations of non-equatorial mergers, for five
different inclination angles all with $a/M=0.99$ (Table \ref{table2}). The rest of the characteristics of the
simulations presented in this section, i.e., BH and NS masses,
polytropic index $\Gamma$, number of SPH particles, are as
mentioned in \S 3.1 .

Runs I4 ($i\approx 90^o$) and I5 (retrograde orbit) lead to
the NS being entirely lost into the BH's horizon after spiraling
around it for several orbits. The difference between the outcomes
of those two mergers is that, for run I4 the feeding of the BH
does not take place in the equatorial plane: the NS follows a 3-D
spherical orbit of decreasing radius before it finally vanishes
completely into the BH.

\begin{figure}[htbp]
\begin{center}
\epsfxsize=10.0cm \epsfbox{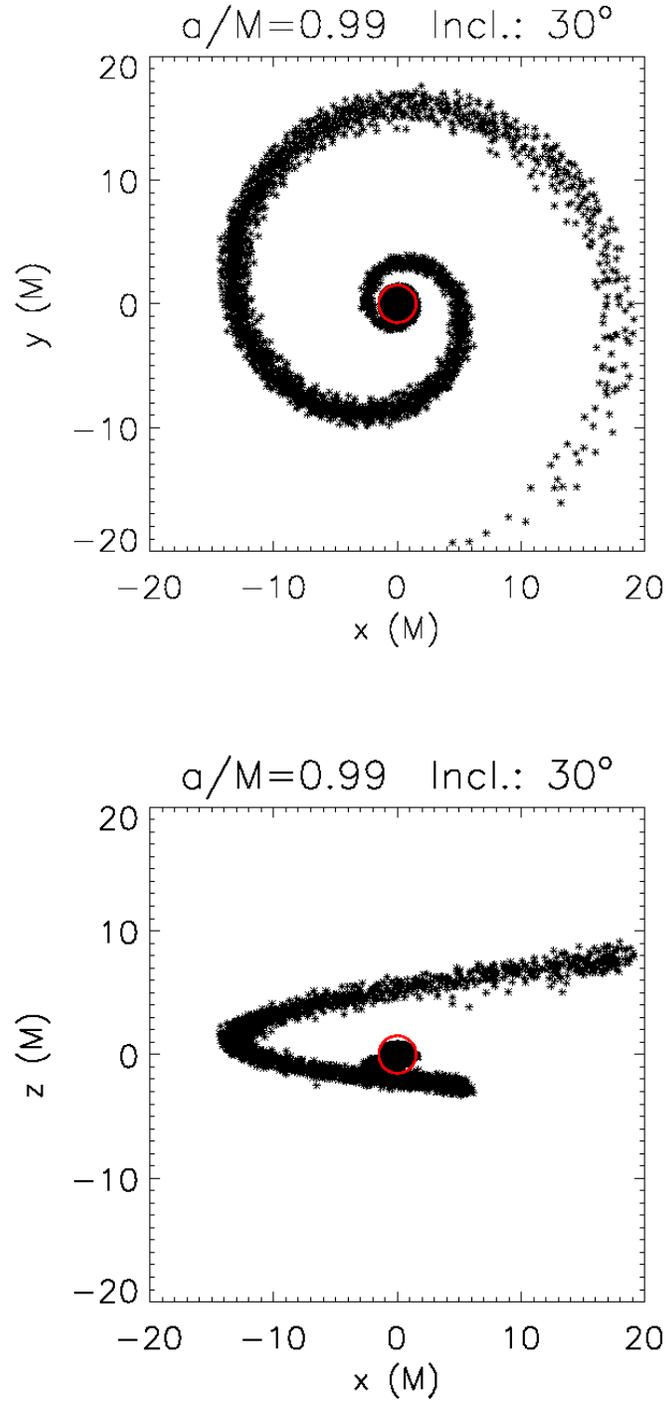}
\caption{Snapshot of Run I1 (up) and its x-z projection (down) towards the end of the simulation.The red circle represents the BH's outer horizon. The material that survives the merger forms an expanding helix. For this particular run $~30\%$ of the initial NS mass gets dynamically ejected and $6\%$ remains bound, forming a stable torus outside the horizon.}
\label{theta60_xy_xz}
\end{center}
\end{figure}

\begin{figure}[htbp]
\begin{center}
\epsfxsize=10.0cm \epsfbox{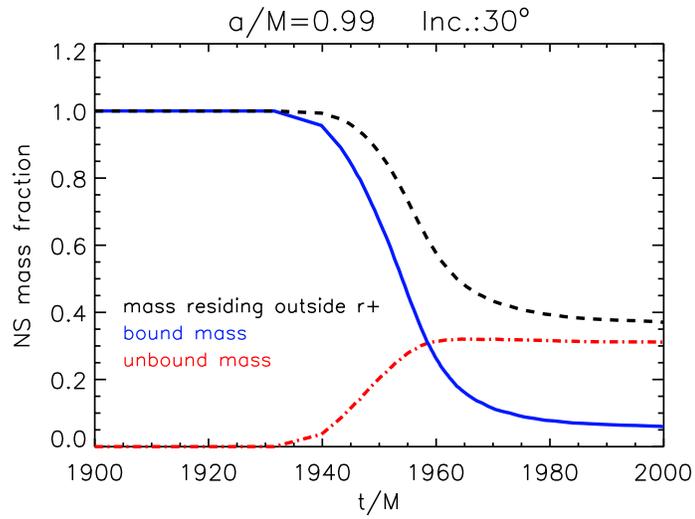}
\caption{Fraction of the NS mass that resides outside the BH's
future horizon for Run I1. At time $t/M=1950$ after the beginning
of the simulation, the surviving material stabilizes at $\sim
37\%$. $31\%$ corresponds to unbound escaping mass, while $6\%$
stays bound and forms a torus around the BH's horizon.}
\label{theta60_mass_vs_t}
\end{center}
\end{figure}

\begin{figure}[htbp]
\begin{center}
\epsfxsize=14.0cm \epsfbox{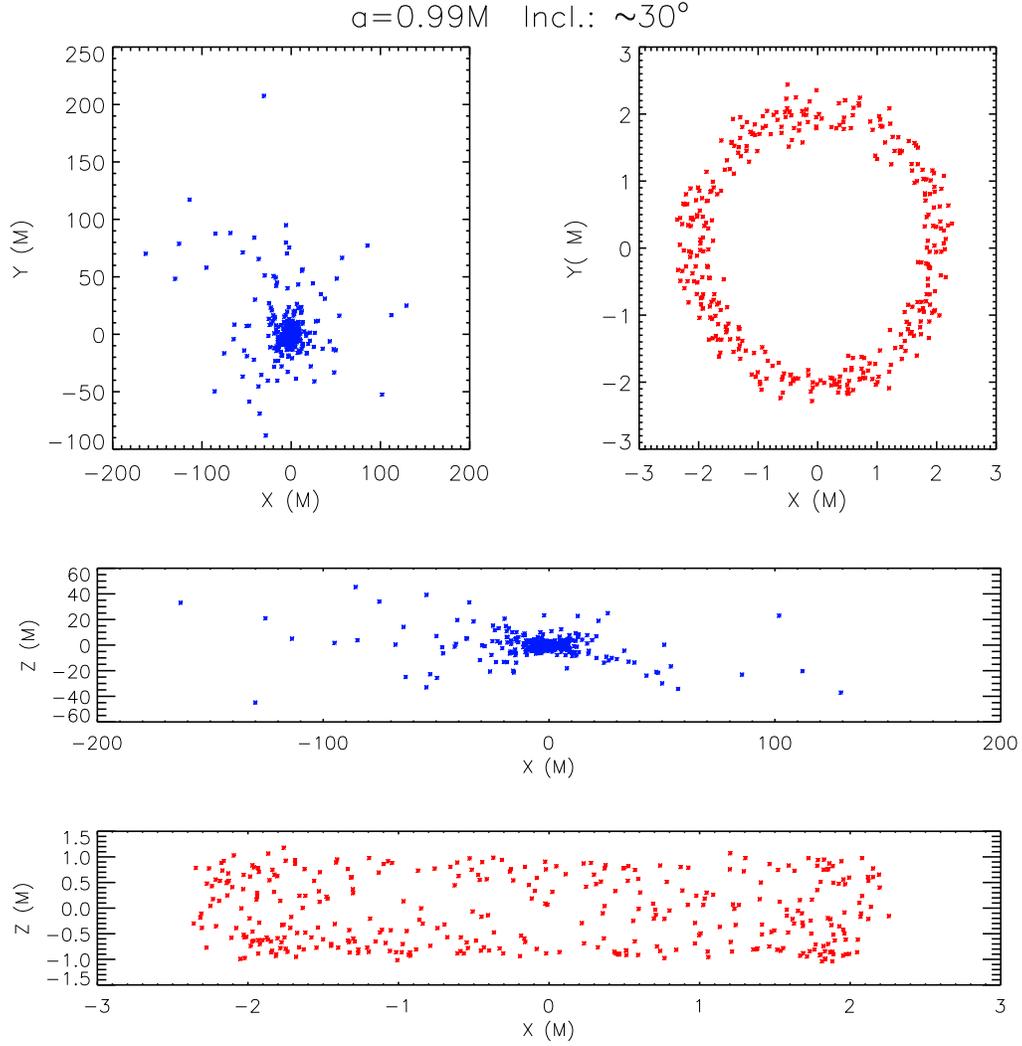}
\caption{Spatial distribution of the apocenters (blue) and the
pericenters (red) for the bound SPH particles of Run I1. Note that in inclined mergers 
the bound mass forms a thick torus rather than a disk (as for equatorial mergers;
compare Fig.~\ref{a099_apo_peri_2D}).
The upper left (right) and middle (bottom) panels correspond to
the x-y and x-z plane projections for the apocenters (pericenters)
respectively. The total mass of
the torus is $6\%$ of the initial NS mass. The mean pericenter
value (inner radius of the torus) is $r_{peri}\sim 2\,M$
and the mean apocenter (outer radius of the torus) is
$r_{apo}\sim 25M$. } \label{theta60_a099_apo_peri_2D}
\end{center}
\end{figure}

The outcome of Run I1 is presented in Figs.~\ref{theta60_xy_xz}
and \ref{theta60_mass_vs_t}. As in the equatorial mergers of high
$a/M$ presented in \S 3.1, the merger results in the formation
of an expanding tail of the NS's material, with the innermost part
of the helix 'feeding' the BH. For this case the helix does not
lay on the equatorial plane , although the feeding point does. By
the end of the simulation almost $40\%$  of the NS mass remains
outside the BH's horizon with most of it unbound and only about $
6\%$ bound. We followed the procedure described already in
\S 3.1 and used the geodesic integrator to further evolve the
results of the SPH simulation as a test of the validity of our
definition of bound and unbound mass and to check the spatial
distribution of the bound material around the BH. The whole
unbound part of the surviving material escapes outwards, whereas
the bound mass forms a stable torus outside the BH's horizon. 
Fig.~\ref{theta60_a099_apo_peri_2D} shows the spatial distribution of
pericenters (in red - upper right and lower panels) and apocenters (in blue - upper left and middle panels) on the x-y and x-z planes. Every particle is depicted at the moment of its own
apocenter (or pericenter) (i.e., those plots do not correspond to a snapshot
of specific time of the equivalent merger).

\begin{figure}[htbp]
\begin{center}
\epsfxsize=10.0cm \epsfbox{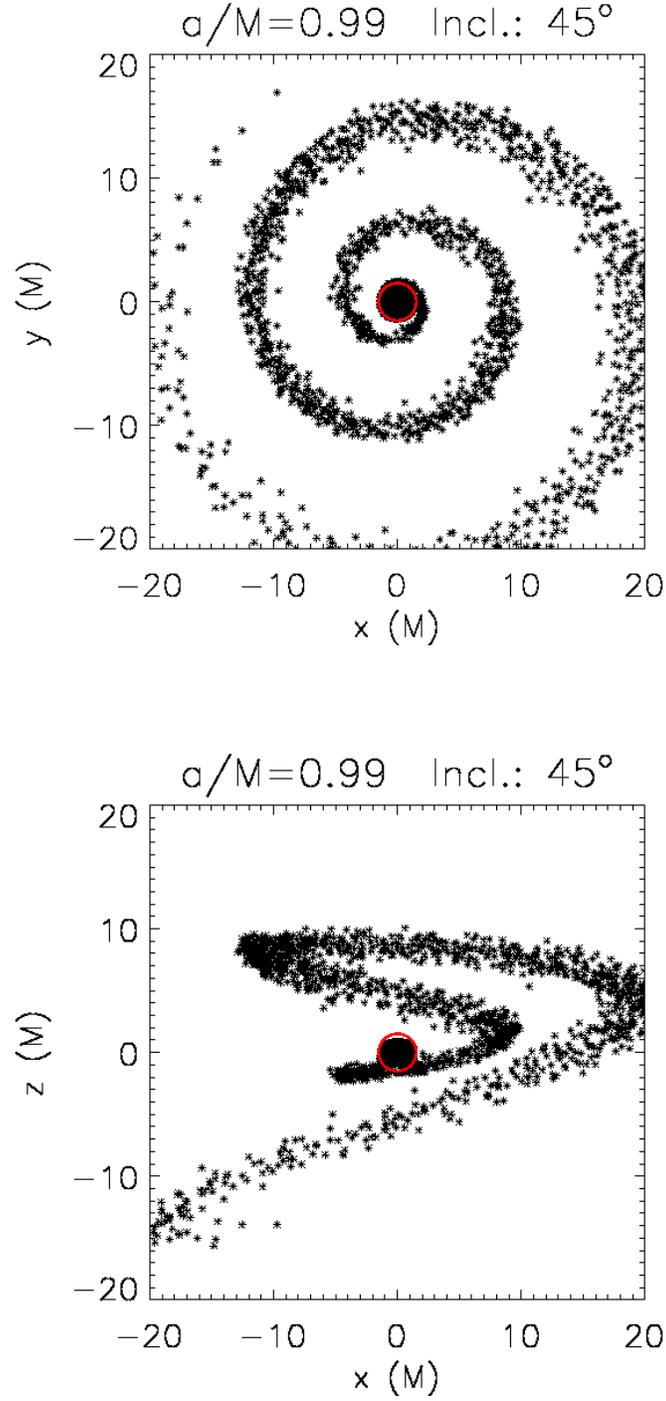}
\caption{Snapshot from Run I2 (upper panel) and its x-z projection (lower panel)
towards the end of the simulation.The red circle represents the
BH's outer horizon. By the end of the simulation an outwards expanding helix has formed, ejecting $25\%$ of the initial NS mass away from the BH. The fraction of surviving bound material was too small to be resolved in this calculation.} \label{theta45_xy_xz}
\end{center}
\end{figure}

\begin{figure}
\begin{center}
\epsscale{0.9}
\plottwo{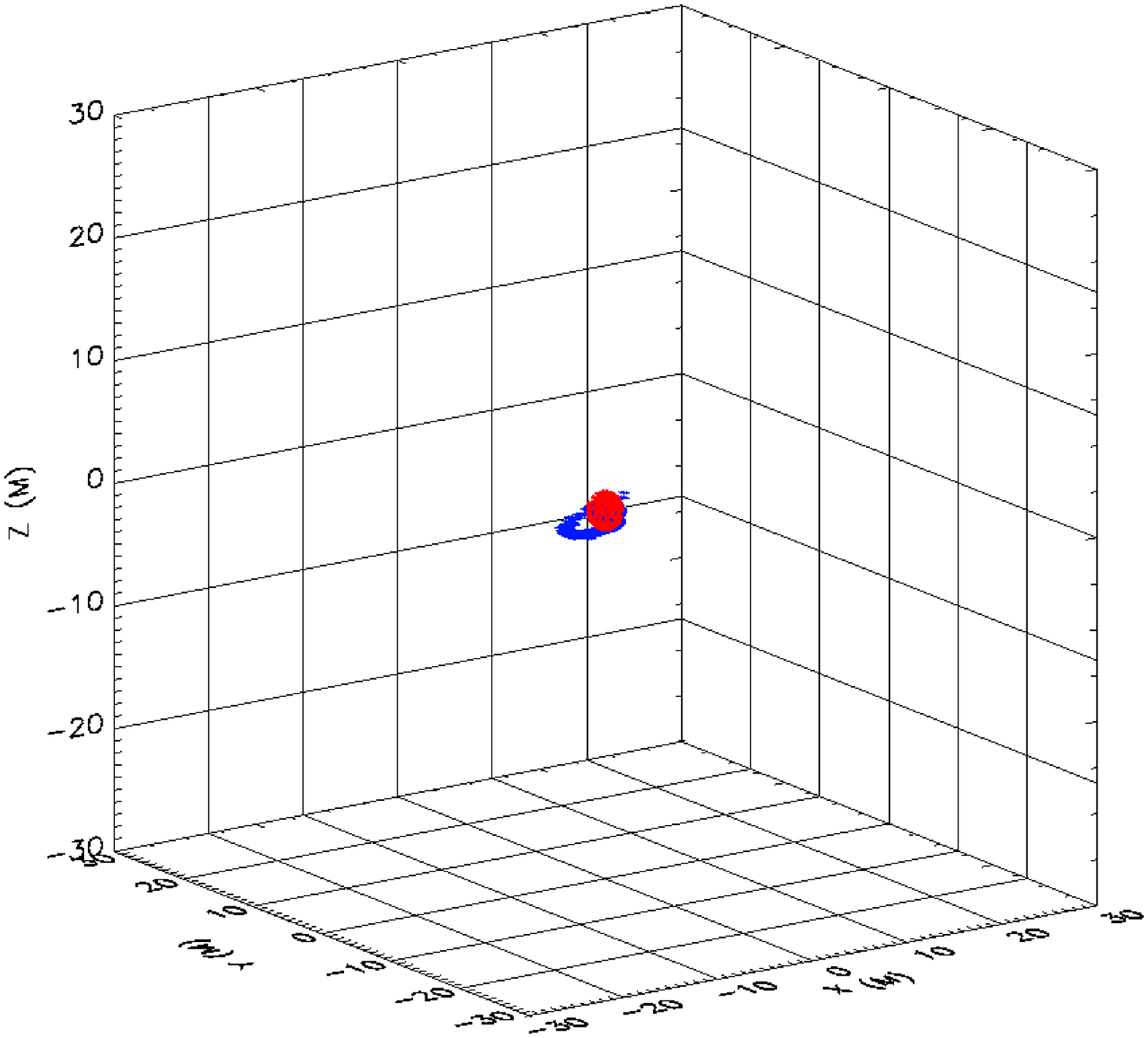}{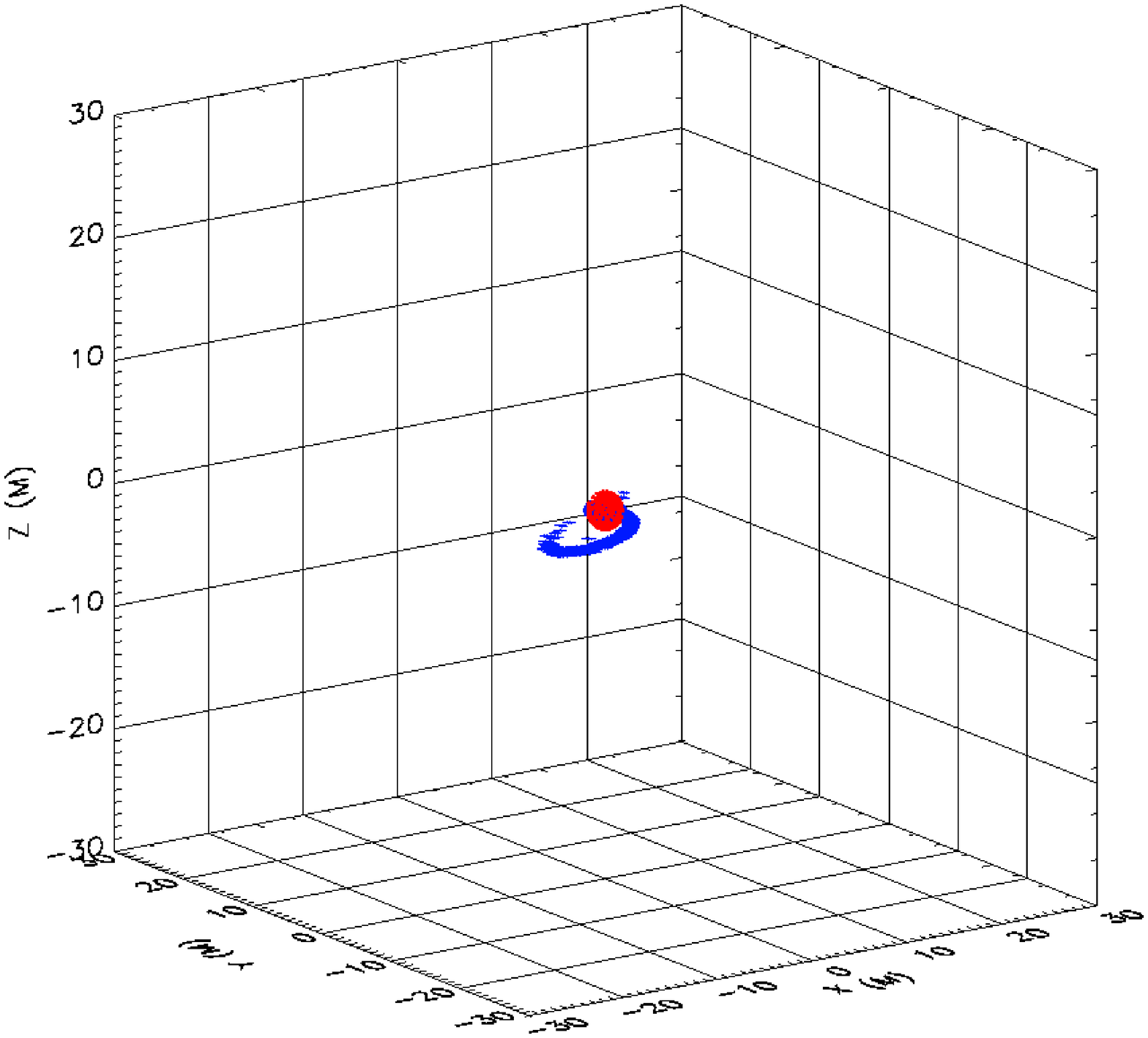}
\plottwo{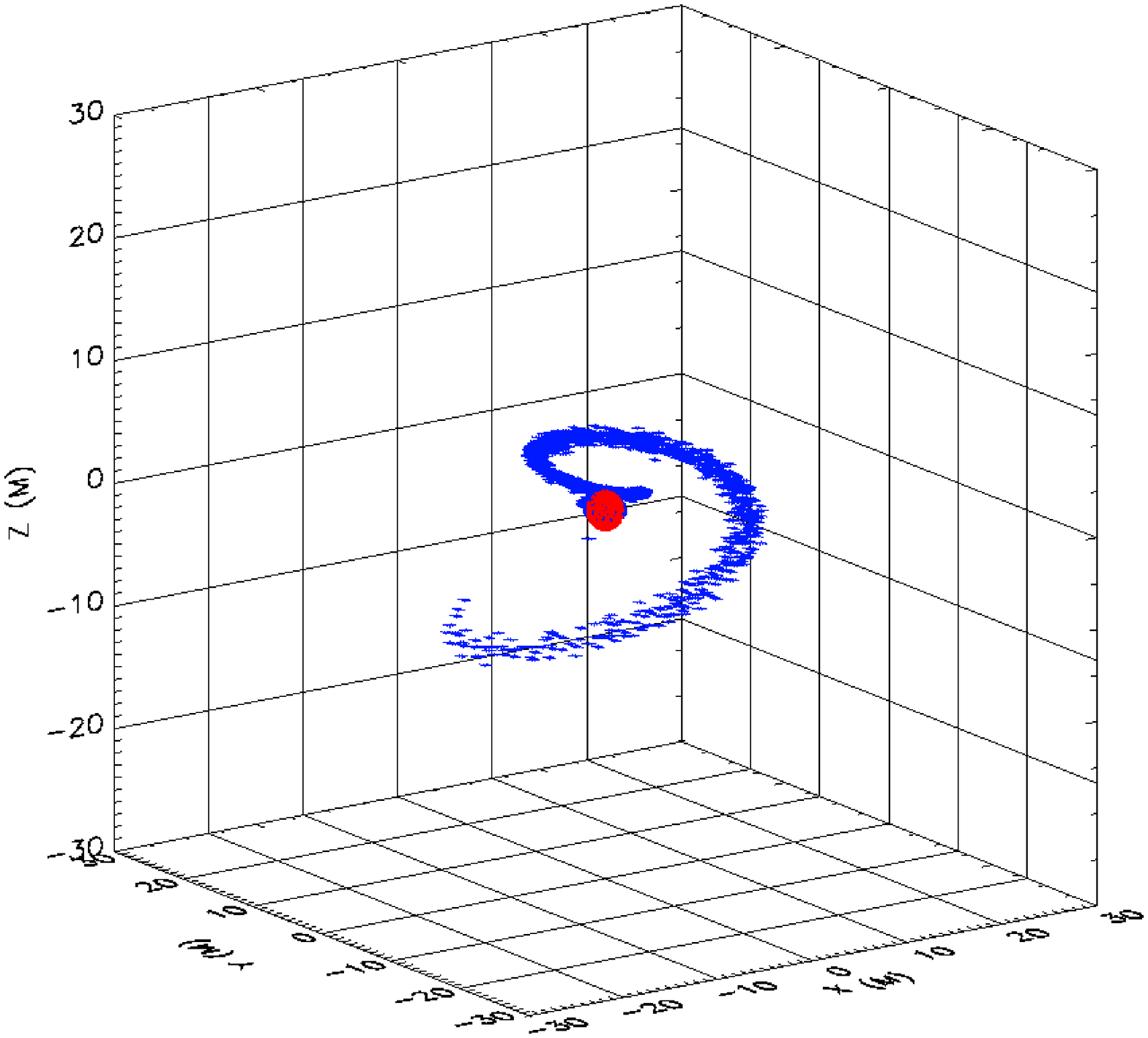}{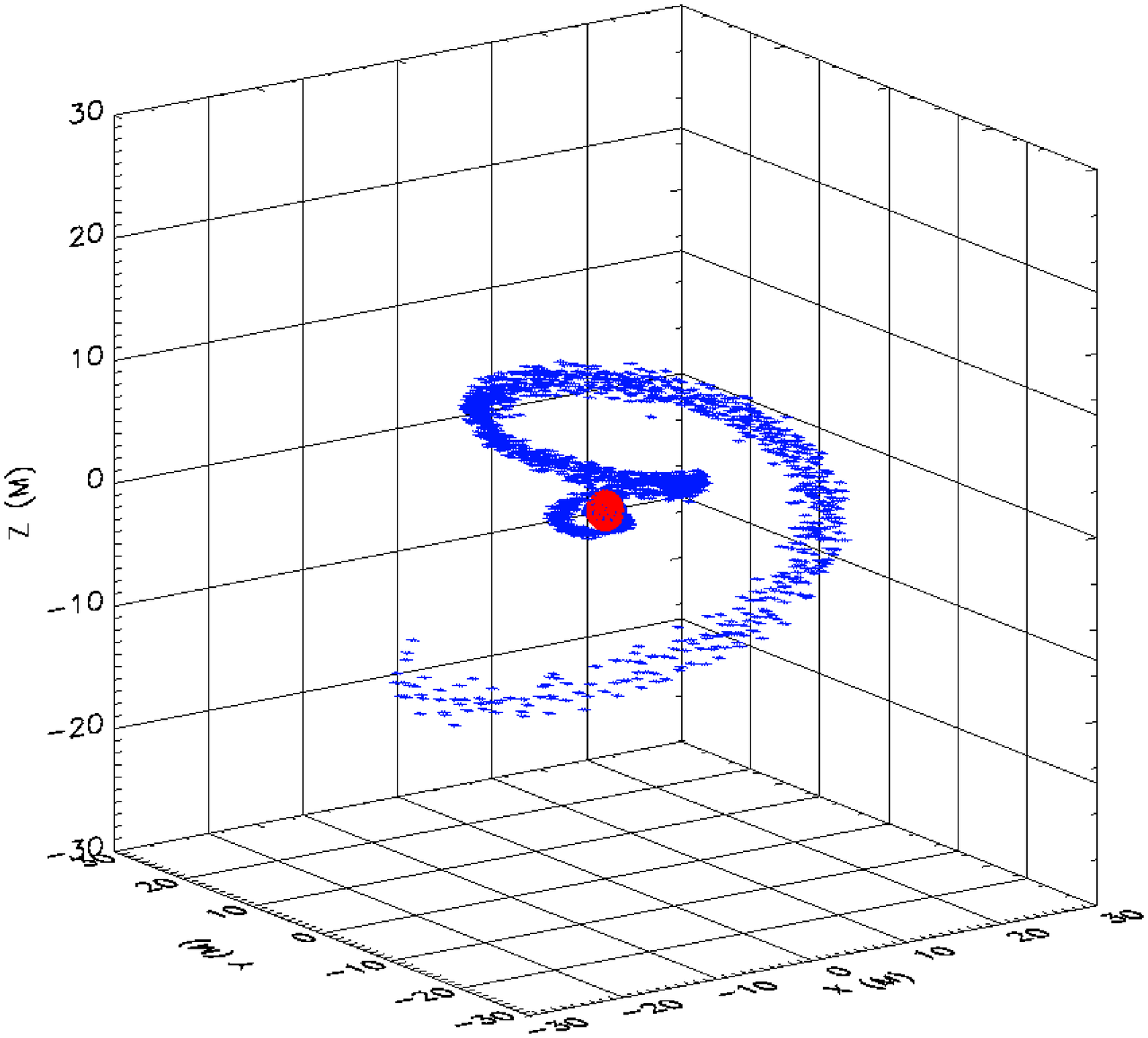} \caption{Four 3-D
snapshots from Run I2 towards the end of the simulation. The NS has
already been disrupted and starts shedding  material into the BH's
horizon. An expanding helix of unbound material forms at the same
time, resulting in the ejection of $25\%$ of the NS mass.}
\label{theta45_a099_3D}
\end{center}
\end{figure}

\begin{figure}[htbp]
\begin{center}
\epsfxsize=10.0cm \epsfbox{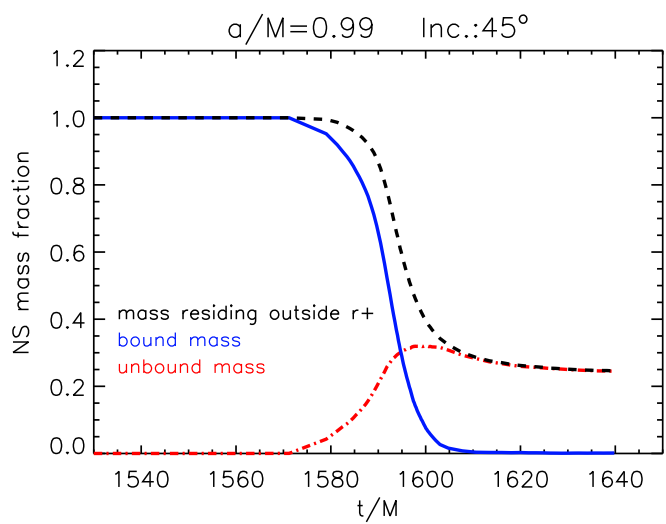}
\caption{Fraction of the NS mass that resides outside the BH's
future horizon for Run I2. At time $t/M=1600$ after the beginning
of the simulation, the surviving material stabilizes at $ 25\%$
almost all of which is unbound and therefore escaping outwards.}
\label{theta45_mass_vs_t}
\end{center}
\end{figure}

Run I2 corresponds to a merger with initial inclination of
$\sim45^o$. As the infall of the NS's material into the BH starts,
at about $t/M=1500$ after the beginning  of the simulation, an
outwards expanding tail is forming (Figs.~\ref{theta45_xy_xz} and
\ref{theta45_a099_3D}) and by the end of the simulation $25\%$ of
the initial NS mass (Fig.~\ref{theta45_mass_vs_t}) is left unbound
and escaping. No bound, disk-forming material was resolved in this
case.

\begin{figure}[htbp]
\begin{center}
\epsfxsize=10.0cm \epsfbox{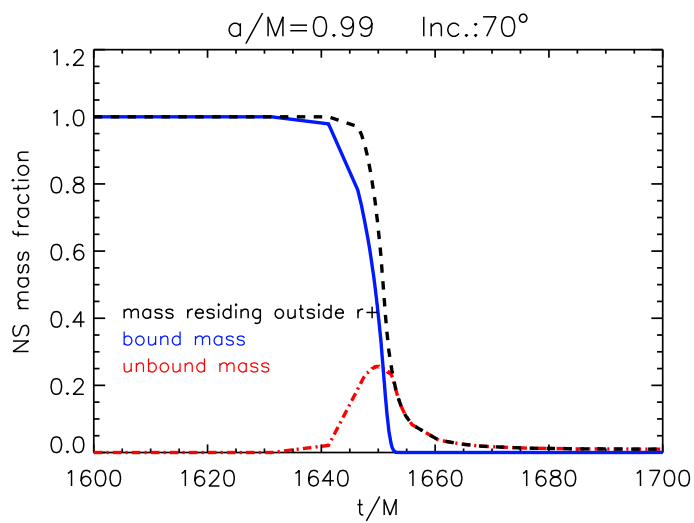}
\caption{Fraction of the NS's mass that resides outside the BH's
future horizon for Run I3. At tine $t/M=1640$ after the beginning
of the simulation, the whole NS mass has crossed the BH's outer
horizon.} \label{theta20_mass_vs_t}
\end{center}
\end{figure}

For an inclination of $\sim 70^o$ (Run I3) the outcome of the
merger is no different than that of Runs I4 and I5: the whole NS
disappears into the BH's horizon. The morphological features of
this merger, however, are somewhere in between those of the low
inclination mergers, where there is material surviving outside, and the higher
inclination ones, which ended up being  completely catastrophic
for the NS. After the NS orbits around the BH following a
spherical orbit of decreasing radius, it starts accreting into the
BH  with the feeding point being well above the equatorial plane.
As the infall continues, the part of the NS that is still outside
the BH's horizon moves towards the equatorial plane and at the
last stages of the merger, when the feeding point has sufficiently
approached the equatorial plane, a small, expanding, spiraling
tail forms which  is energetically unable to escape or expand
significantly and ends up being accreted into the BH as well. At
the end of the simulation there is no surviving mass (Fig.~\ref{theta20_mass_vs_t}).

The difference between equatorial and inclined mergers in terms of
the surviving bound mass seems to be the vertical (z-axis)
distribution of the NS debris. In the equatorial case the bound
material remains in an equatorial disk ($ -0.15M \le z \le 0.15M$)
whereas in the inclined case the disk is replaced by a torus with $
-1M \le z \le 1M$, and this torus is also more
massive than the disk in equatorial mergers.

We note that the geodesic integrator runs are taken just as an
approximation and a qualitative test of the further evolution of
the two groups of surviving material (bound and unbound), for the
cases where hydrodynamics is of no importance and the SPH
particles can be treated as free particles; e.g., for unbound
particles at great distances from the BH.

\section{Gravitational Wave Signals}

We present here the GWs extracted for some of the simulations described in the previous section. We have used the standard quadrupole formula, in which the two
polarizations of the waveform are given by
\beq  rh_{\times}=2\ddot{I}_{xy} \label{rh} \eeq
\beq rh_{+}=\ddot{I}_{xx}-\ddot{I}_{yy}  \label{rp}\eeq
where $r$ is the distance to the observer, $h_{\times}$ and $h_+$ the cross and plus polarization modes, and $\ddot{I}_{ij}$ the second time derivative of the second mass moment.

\begin{figure}
\begin{center}
\epsscale{0.9}
\plottwo{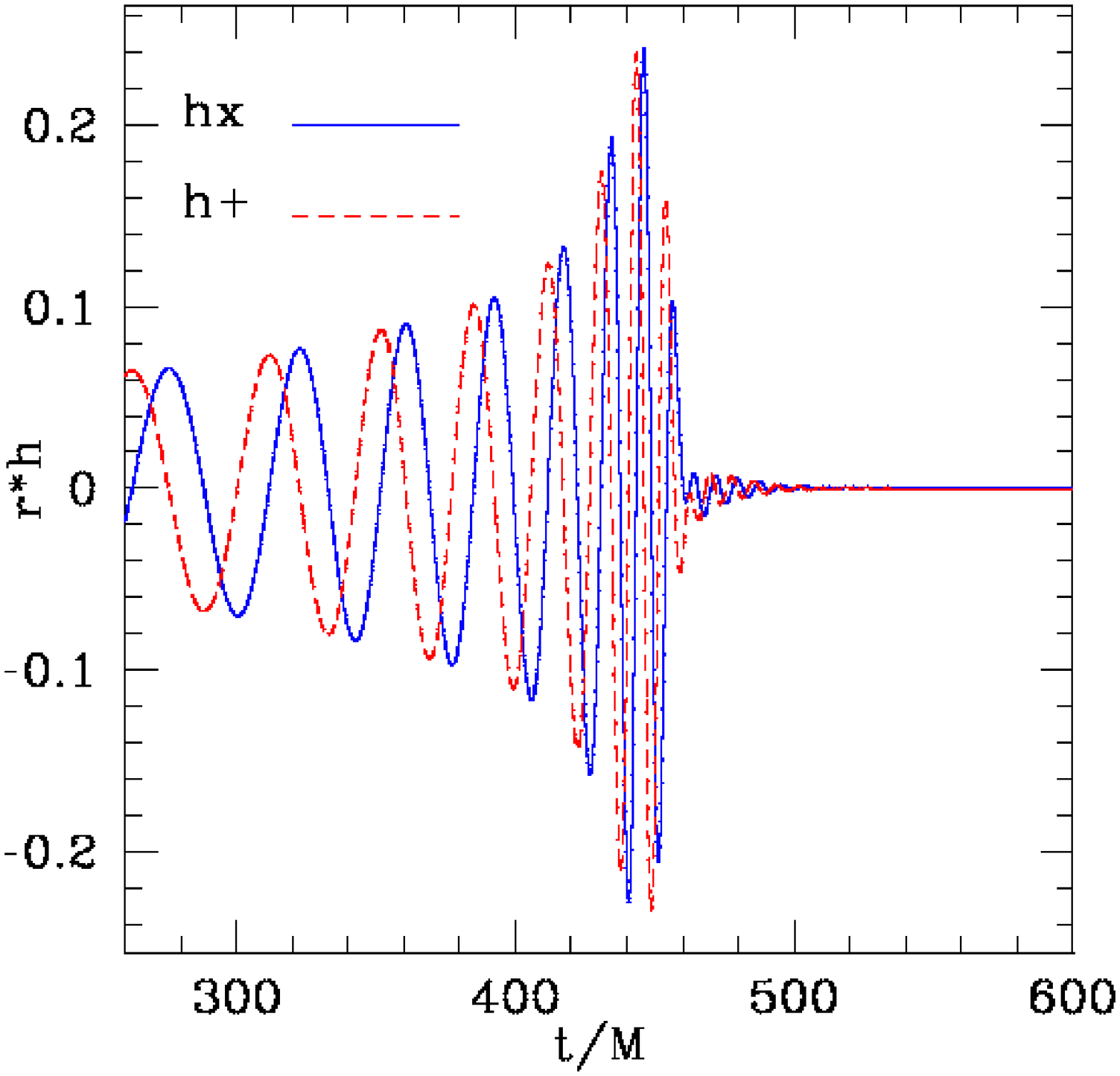}{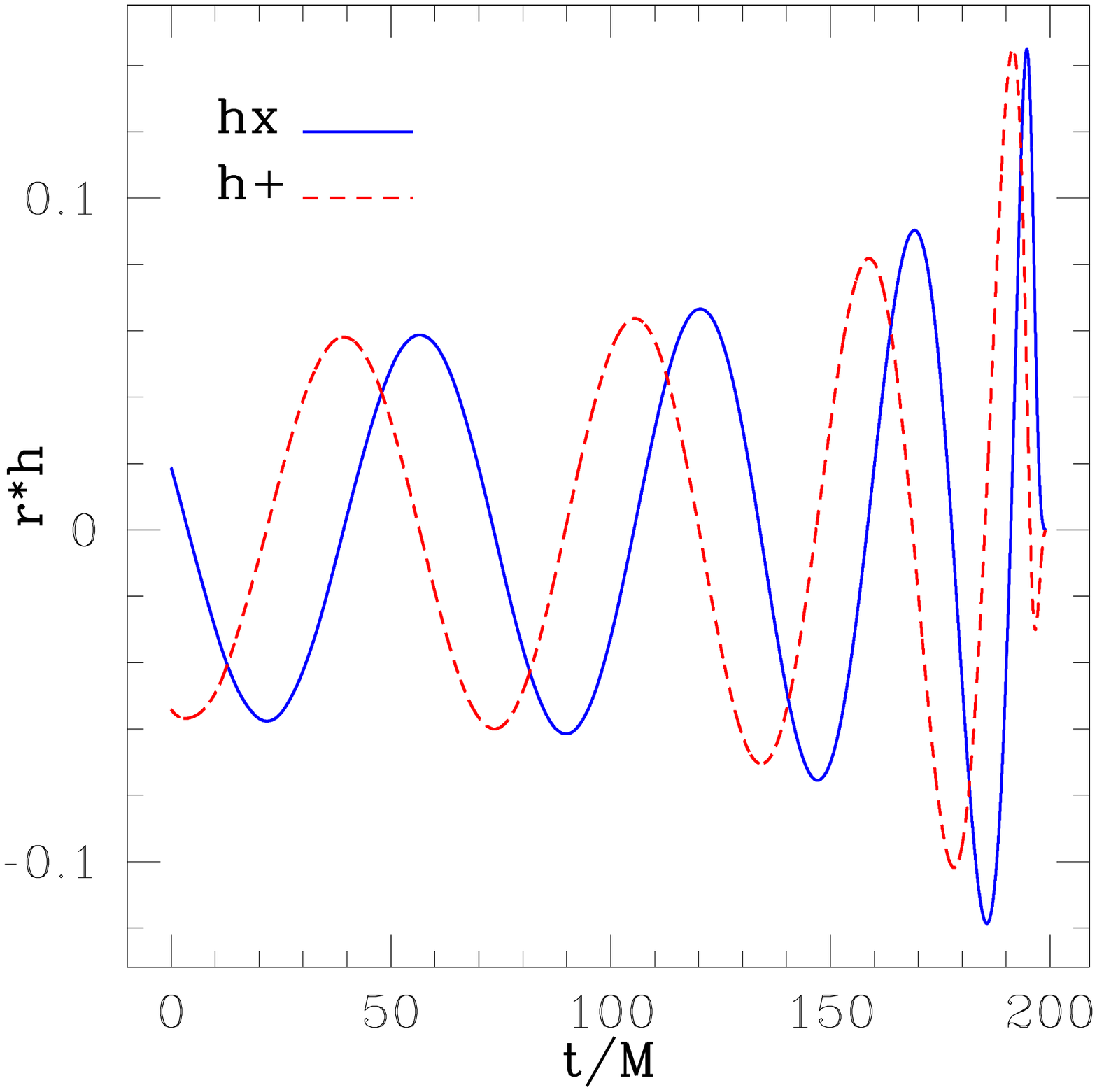}
\plottwo{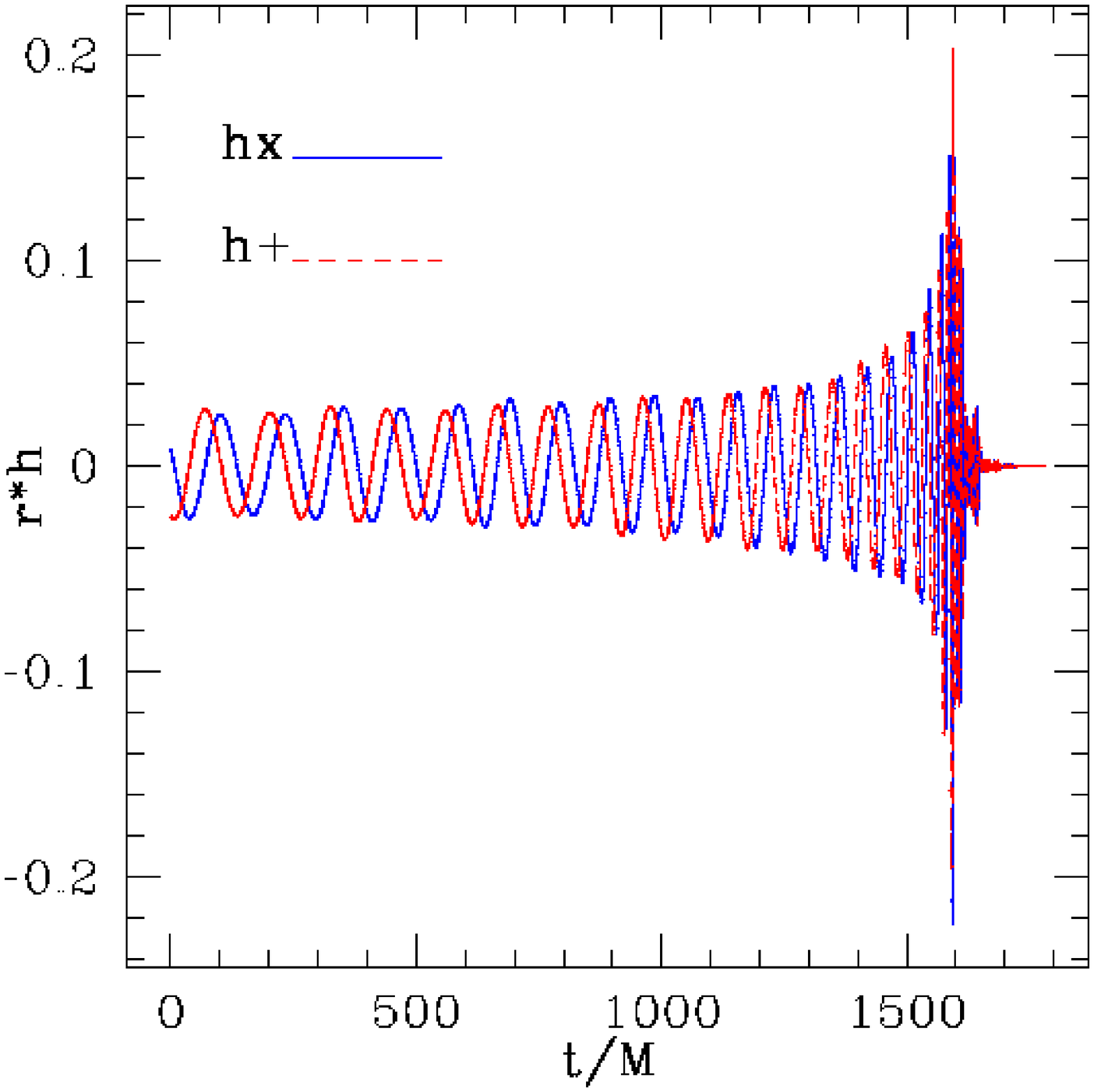}{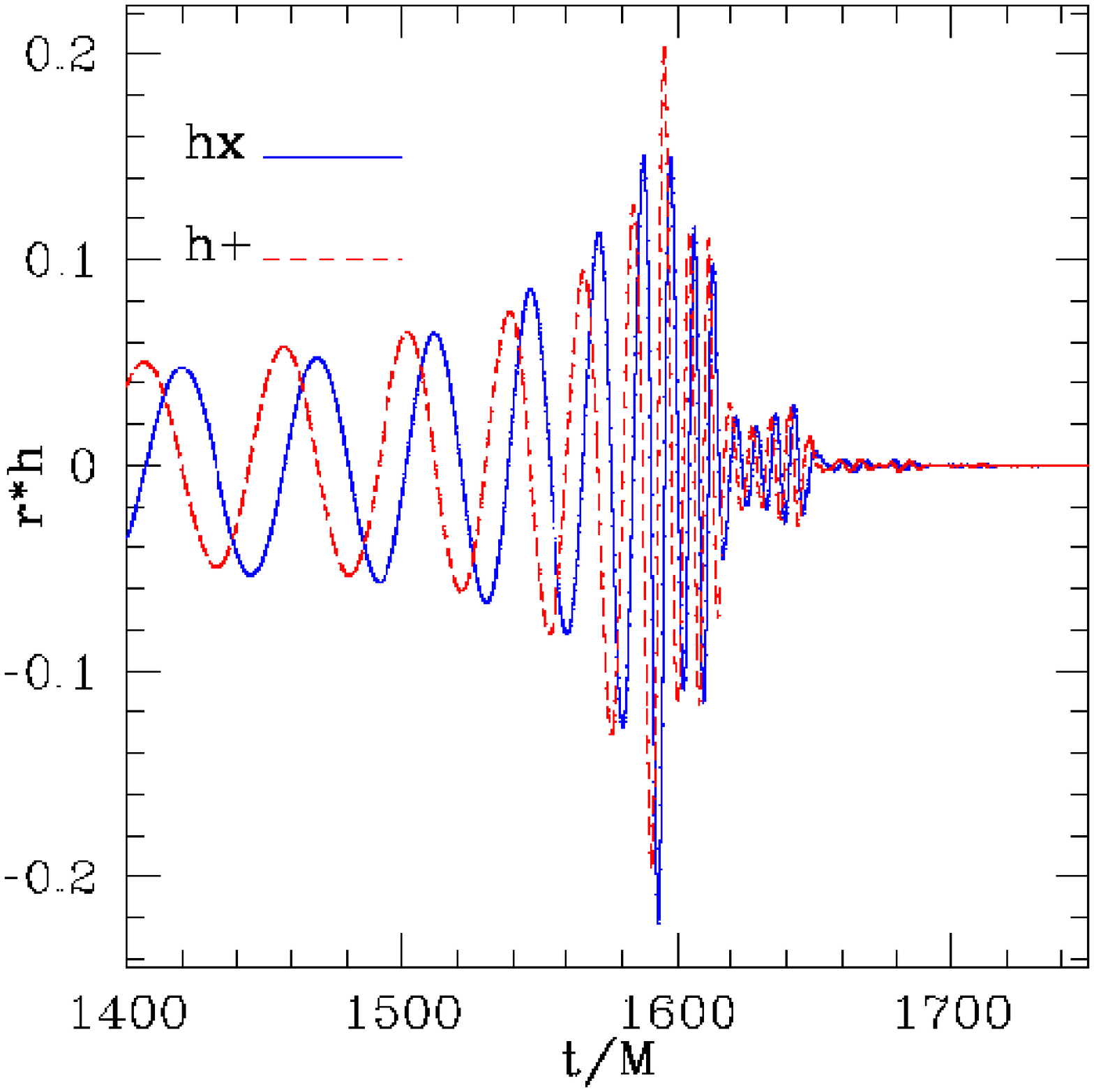} 
\plottwo{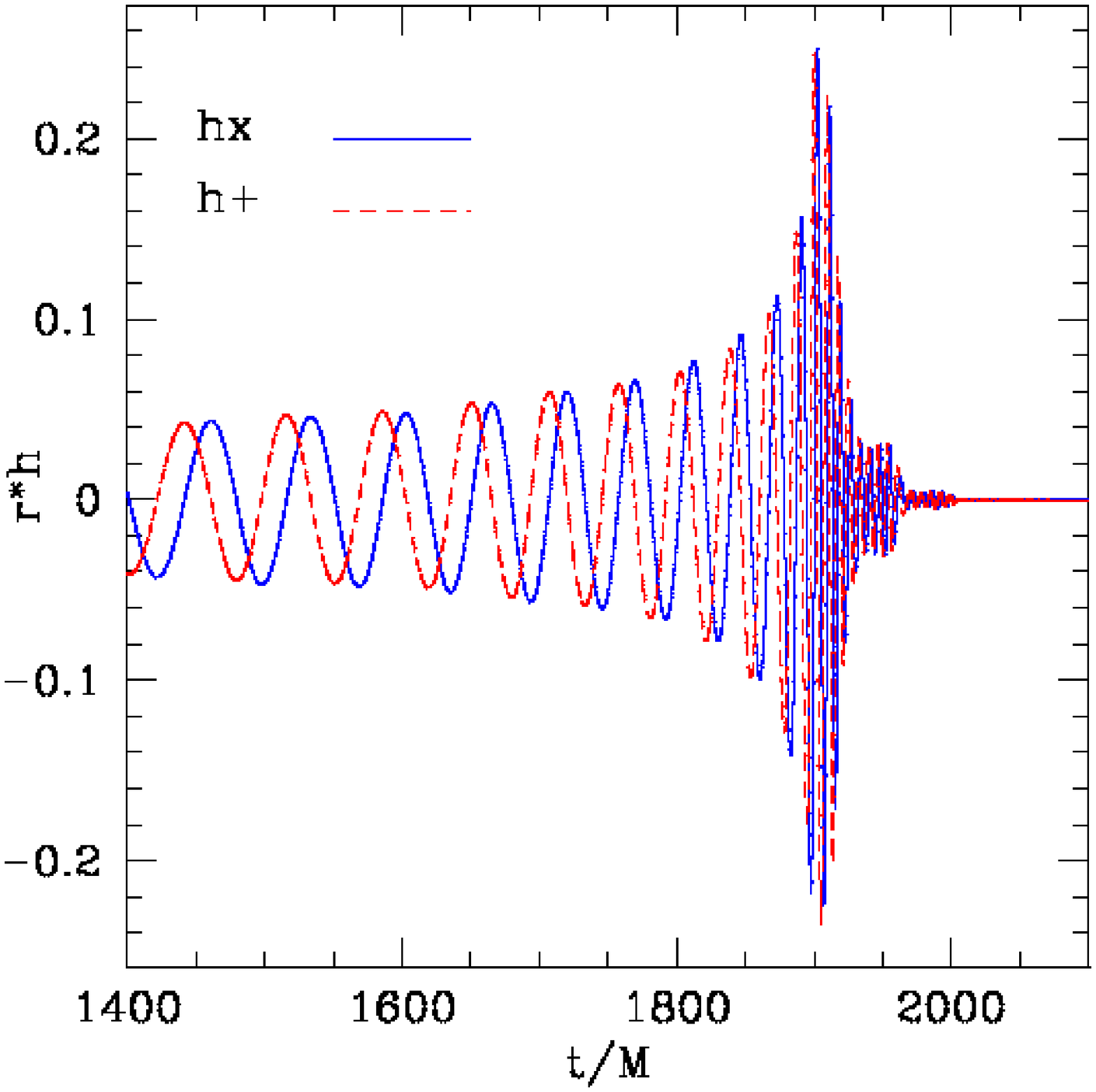}{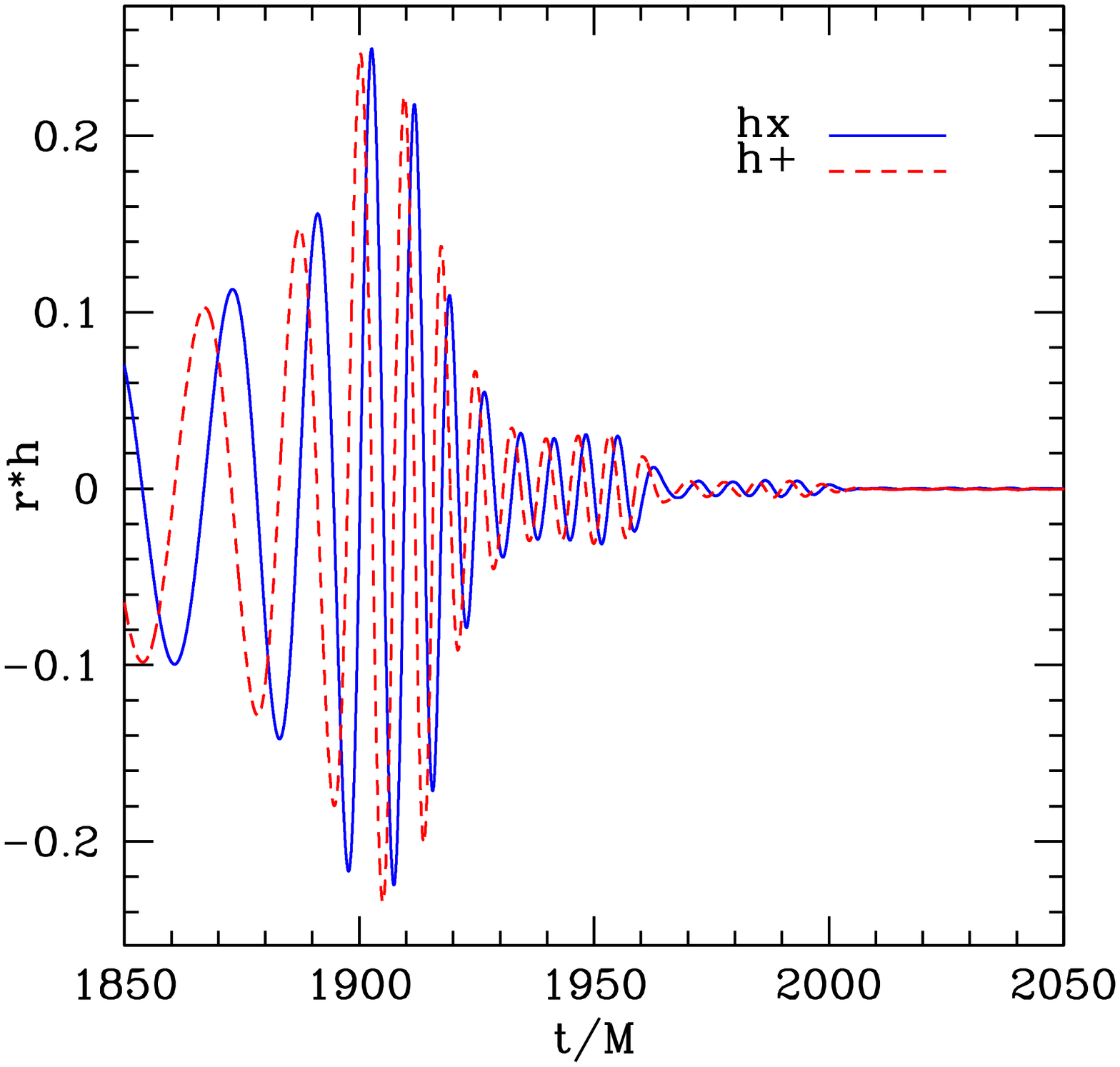}
\caption{GW signals extracted from some of the simulations. The two upper plots correspond to runs E1 (left) and E10 (right). The middle panel plots are for run I2 with the right side showing a blowup of the left plot focussing on the later stages of the evolution. The lower panel shows the GW signal for run I1. Again the right side shows  a blowup of the plot on the left.}
\label{gw}
\end{center}
\end{figure}

Fig.~\ref{gw} shows the GWs extracted from runs E1, E10, I2 and I1. 
The upper left plot corresponds to Run E1 (equatorial merger of a maximally spinning BH). As the inspiral of the NS towards the BH continues, the amplitude and frequency of the waveform is increasing, to reach its maximum just before the NS gets disrupted and starts shedding mass into the BH. From this point on, the amplitude decreases very rapidly and the signal eventually vanishes. 
 The GWs for an equatorial merger of a non-spinning BH (run E10 in Table 1) is shown in the upper right plot of Fig.~\ref{gw}. Again we see the chirp signal reaching its maximum just before the NS is lost into the BH's horizon. For a non-spinning BH the disruption of the NS is taking place very close to the horizon  (as discussed above) and it just barely starts getting 
 tidally distorted before plunging almost intact into the BH. As a result, we do not see the gradual diminishing of the wave amplitude as observed in the spinning case.
 The middle and lower panel plots of Fig.~\ref{gw} show the GWs for Runs I2 and I1, two simulations for the merger of a maximally spinning BH with a NS set in an inclined orbit of inclination $45^o$ and $60^o$, respectively. Again we observe the characteristic chirp signal as the NS is inspiralling, with the frequency of the signal decreasing as the wave amplitude is increasing and reaches its maximum just before the NS disruption takes place. Additionally, the precession of the orbital plane is seen as a secondary chirp signal of smaller amplitude. At this point, the main core of the newly disrupted NS is found orbiting on a different plane than the rest of the NS material that has just started forming an outwards moving tail.
 
The GW signals shown in Fig.~\ref{gw} are, at best, qualitatively correct,
for two reasons. First, recall that the inspiral part of the calculations presented here
is affected by the artificial strengthening of radiation reaction that we have used in order to drive the initial stages of our simulations (as described in \S 2.3).  As a result, the early parts of the GW signals suffer a time compression that would not be present in reality. Second, the simple quadrupole formula of Eqs.~\ref{rh} and \ref{rp} applies strictly to the case of a quasi-Newtonian source, while in reality the NS fluid is moving highly relativistically in the strong field region around a BH. Our intention for now is merely to provide a qualitative preview of the GW signals produced by these BH--NS mergers, and not to derive quantitatively accurate waveforms. As part of our future work  
we plan to develop a more sophisticated and more accurate treatment of the radiation  
reaction and GW extraction, which will then allow us to study in greater detail
how the merger's parameters produce distinctive features in the GW signals and energy spectrum.

\section{Discussion and Summary}

We have performed simulations of BH--NS mergers with mass ratio $q\simeq 0.1$ and polytropic index $\Gamma=2$ using a 3-D 
relativistic SPH code. We investigated equatorial mergers for various values of the BH spin and we also 
carried out simulations for inclined mergers (considering a tilted orbital plane for the NS with respect to the BH's 
equatorial plane) in the case of an extremal Kerr BH. We find that the outcome of the merger depends strongly 
on the spin of the BH and the inclination of the orbit. More specifically, for equatorial mergers, the survival of NS 
material is possible only for mergers with $a/M>0.7$ and, in that case, the percentage of surviving material 
increases with increasing BH spin, varying from about $33\%$ to $\sim1\%$ for $a/M$ decreasing from 0.99
to 0.7. Complete disruption of the NS happens for all values of $a/M$. Most of the surviving material gets ejected 
from the vicinity of the BH and only a few percent stays bound to the BH,  forming a relatively thin stable disk outside 
the BH's horizon. Only for very high BH spins ($a/M>0.95$) do we see a substantial fraction ($\sim 3\%$) of the 
disrupted NS material remaining bound. 

The outcome of inclined mergers (for fixed $a/M=0.99$)  shows strong dependence on the value of the inclination
angle. For sufficiently low inclinations ($<60^o$) there is always a large fraction of the NS mass surviving the 
merger, up to $\sim40\%$, depending on the inclination angle. Moreover, for these mergers the formation of a thick 
stable  torus of substantial bound mass is also strongly inclination-dependent. Whenever the inclination exceeds 
$40-45^o$ the fraction of bound material drops to levels unresolvable by our present calculations, although there 
is still a significant fraction ($\sim 25\%$) of the material that is being ejected.

Given the sensitive dependence of our results on the BH spin, one might wonder
whether the accretion of NS material onto the BH could lead to significant spin up, 
with a corresponding change in the spacetime metric which is not accounted for in our 
SPH code (where the BH spin parameter is assumed fixed during the entire evolution). 
For each one of the runs presented in \S 3 we have calculated the change in the BH spin 
by adding to the initial angular momentum of the BH the angular momentum of the total 
accreted mass ({\em a posteriori\/}). For Runs E1-E5 the total change in the BH spin after 
the accretion was less than $0.1$. More precisely, for those runs with $0.7< a/M <0.95 $ the final BH was spun up by less than $0.1$, while for Runs  E1 and E2 the final BH spin decreased to $a/M=0.965$ and $a/M=0.946$, respectively. Those changes are small enough that we think
they would not have affected the final results of the simulations. 
For the runs with BH spin $0 \leq a/M \leq 0.6$ the  change in the final BH spin varies between $0.27$ for a non-spinning BH and $\sim 0.1$ for a BH with initial $a/M=0.6$. Even for these 
cases the change in the BH spin will not alter the result of the merger, as all simulations with 
BH spin $<0.7$ resulted in the accretion of the entire NS mass by the BH.

Another possible caveat concerns our Newtonian treatment of the NS fluid self-gravity.
We should stress here the point made already  in \citet{faber06b}, that a more accurate,
 relativistic treatment of the NS fluid's self-gravity could lead to even faster mass 
 transfer to the BH, as it might result in the expansion of the NS on a dynamical scale 
 during the mass loss phase. Although this effect could have significant impact on the outcome
  of some mergers, we have observed in all our simulations that the complete disruption of the NS lasts typically less than one dynamical time from the onset of mass loss and therefore we believe that the adoption of a simple Newtonian treatment of self-gravity in our code is justified. 

Our findings have direct implications for the viability of BH--NS mergers as progenitors of short GRBs.  The possibility of short GRBs being powered by such mergers depends on the  characteristics of the binary, mainly the BH spin and the orbital inclination. Although not 
all BH--NS mergers seem to be promising as progenitors for short GRBs, binaries with fast 
spinning BHs and zero to moderate inclinations appear to  form  a disk or torus around 
the BH, massive enough to power a short GRB when later accreted by the BH. Therefore, 
only a fraction of BH--NS binaries could be associated with the production of short GRBs. The question of whether this fraction is high enough to explain the observed short GRBs has yet to be answered as it depends on uncertain quantities such as the BH spin and mass distributions, as
well as the orbital inclination distribution for BH--NS binaries. Current  population synthesis codes try to answer such questions and give better constraints on the distributions of those quantities. A
recent discussion of this question in connection to short GRBs is given in \citet{bel_ran}. The 
fact that the short GRB progenitors need not belong to a homogeneous group and that 
NS--NS binary mergers could be equally viable progenitors of short GRBs is also an open 
issue currently under investigation.

The complete tidal disruption of the NS in all the merger calculations that we
presented in this paper is in qualitative agreement with previous Newtonian
studies of BH--NS mergers with {\em soft\/} EOS \citep{rosswog,lee99b}.
In our simulations that involved a Schwarzschild (non-spinning)
BH, no NS material was observed being ejected or forming a disk
around the BH. The accretion of the whole NS onto the BH is very
prompt for this case. This is to be expected,
as for the particular mass ratio ($q\simeq0.1$) considered in our
simulations the tidal limit is found inside the ISCO for $a/M=0$
and thus we would not expect to find a stable disk forming around
the BH. The ejection of material and the formation of an accretion disk
were observed only for mergers with highly spinning BHs.

\begin{figure}[htbp]
\begin{center}
\epsfxsize=15.0cm \epsfbox{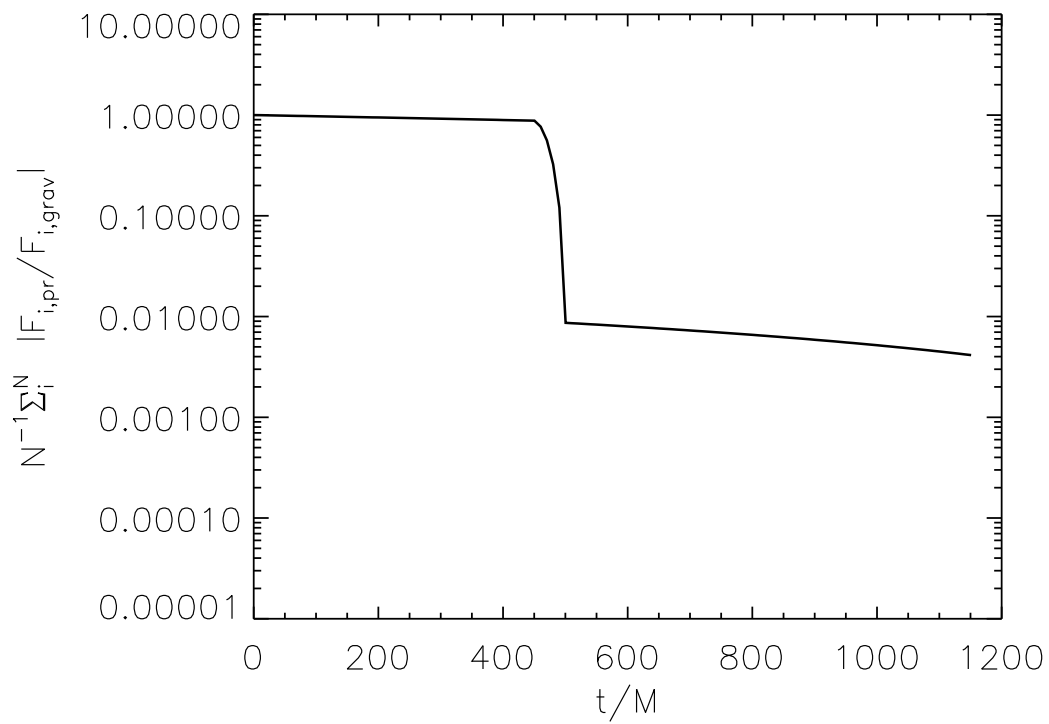}
\caption{The averaged ratio of pressure gradient forces to gravitational forces as a function of time for our highest resolution run. Right after the disruption of the NS ($t/M \sim 400$) the pressure forces become negligible in comparison to the gravitational forces. As a result the fluid's EOS plays no role anymore in the dynamics of the merger and the decompressed material follows essentially ballistic trajectories.  } \label{forces}
\end{center}
\end{figure}

 In Paper II we will extend our simulations to include different mass  ratios,
although still in the limit where
the use of a fixed background is justified (i.e., for BH masses much  
larger than the NS mass), and we will explore the effects of changing the NS EOS and relaxing  our assumption of an initially synchronized NS. 
In this context we would like to add a brief comment about the importance of the NS EOS for the dynamics of the merger right after the disruption of the NS. Fig.~\ref{forces} shows the averaged ratio of the pressure gradient forces over the gravity forces as a function of time (for the run with higher resolution presented in \S 3.1). We see clearly that right after the disruption of the NS (at $t/M \sim 400$) the pressure forces become insignificant and the decompressed NS material essentially follows ballistic trajectories, indicating that the details of the EOS no longer play any role. Thus one need not worry that the decompressed material should be described by a different EOS. Only the high-density EOS describing the interior of the NS prior to disruption  may have a significant effect on the outcome of the merger. Ultimately, the adoption of a full numerical  relativistic scheme, where no simplifying assumptions on the background spacetime metric are made, would  
be ideal, as it would allow us to explore mergers for binaries with higher mass ratios. Although the  many Newtonian studies of these mergers have provided a qualitative overview of their possible outcomes, there  
is a consensus that relativistic effects will play a very significant role in the hydrodynamics 
of the NS disruption \citep[as already shown by][for the non-spinning BH case]{shibatab}, 
and therefore in understanding the GW emission and the possible connection of such mergers to short GRBs.

\paragraph{Acknowledgments}
This work was supported by NSF Grants PHY-024528 and PHY-0601995 at Northwestern University. 
We thank Scott Hughes and Josh Faber for useful discussions and the anonymous 
referee for constructive comments. 
PL acknowledges the support of the Center for Gravitational Wave Physics at Penn State, funded by the NSF 
under Cooperative Agreement PHY-0114375, and NSF grants PHY-0244788 and PHY-0555436 
\clearpage

\appendix

\section{K-S form of the Kerr metric}

Kerr presented his solution for the first time  \citep{Kerr63} in the
following format

\[ ds^2=(\tilde{r}^2+a^2\cos^2\theta)(d\theta ^2+\sin^2\theta
d\phi^2)+2(du+a\sin^2\theta d\phi)\times(d\tilde{r}+a\sin^2\theta
d\phi)\] \beq
-(1-\frac{2M\tilde{r}}{\tilde{r}^2+a^2\cos^2\theta})\times(du+a\sin^2\theta
d\phi)^2 \label{kerr1} \eeq.

By using: $$u=t+\tilde{r}$$
          \beq \tilde{r}\cos\theta =z \eeq
          $$(\tilde{r}-ia)e^{i\phi}\sin\theta =x+iy$$

Eq.~(\ref{kerr1}) can be transformed to an asymptotically flat
coordinate system \citep{Kerr63}.

First let's apply $u=t+\tilde{r}$ on Eq.~(\ref{kerr1}). This will
lead to the more familiar form

\[ ds^2=-(1-\frac{2M\tilde{r}}{\Sigma})dt^2+(1+\frac{2M\tilde{r}}{\Sigma})d\tilde{r}^2+ \Sigma
d\theta^2\]\beq
(\tilde{r}^2+a^2+\frac{2M\tilde{r}}{\Sigma}a^2\sin^2\theta)\sin^2\theta
d\phi^2 + \frac{4M\tilde{r}}{\Sigma}dt \, d\tilde{r}\eeq
\[+\frac{4M\tilde{r}}{\Sigma}a\sin^2\theta d\phi \, dt +(1-\frac{2M\tilde{r}}{\Sigma})2a\sin^2\theta
d\phi \, d\tilde{r}\] where $$\Sigma=\tilde{r}^2+a^2\cos^2\theta$$
and $\tilde{r}$ is defined by : \beq
\tilde{r}^4-(\rho^2-a^2)\tilde{r}^2-a^2z^2=0 \eeq with
$\rho^2=x^2+y^2=z^2$.

Now applying the rest of the transformations in (A2), one gets

\beq
ds^2=dx^2+dy^2+dz^2-dt^2+\frac{2M\tilde{r}^3}{\tilde{r}^4+a^2z^2}(k)^2
\label{KS_cart}\eeq

where \beq k=[\frac{\tilde{r}(xdx+ydy)+a(x dy-u
dx)}{\tilde{r}^2+a^2}+\frac{z}{\tilde{r}}dz+dt] .\eeq

Let us make a useful parenthesis here: to understand better why the
transformation rules (A2) were chosen, it is constructive to start
with Eq.~(\ref{kerr1}) and bring it to the following form

\[ ds^2=[-du^2+2du \, d\tilde{r}+\Sigma d\theta^2 +2a\sin^2\theta d\tilde{r} \, d\phi
+(\tilde{r}^a+a^2)\sin^2\theta d\phi^2]\] \beq
+\frac{2M\tilde{r}}{\Sigma}(du+a\sin^2\theta d\phi^2)
\label{kerr1b}.\eeq

Eq.~(\ref{kerr1b}) can be interpreted as following: the terms not
containing the mass  m, give the flat space metric in some
coordinate system, while the last term can be expressed in terms
of the null (tangent with respect to $\eta_{\alpha \beta}$)
vector, $l_{\alpha}$
\[-l_{\alpha}dx^{\alpha}=du+a\sin^2\theta d\phi\]

and therefore the line element $ds^2$ can be written in the form

\beq ds^2=
(ds^2)_{flat}+\frac{2M\tilde{r}}{\Sigma}(l_{\alpha}dx^{\alpha})^2\eeq

and the metric is 
\beq g_{\alpha \beta}=\eta_{\alpha \beta} +\frac{2M\tilde{r}}{\Sigma}l_{\alpha}l_{\beta}. \eeq

This is actually the original way which Kerr followed to discover
his solution.\footnote{Any metric of the form $g_{\alpha
\beta}=\eta_{\alpha \beta} +Hl_{\alpha}l_{\beta}$, with H a scalar
and $l_{\alpha}$ a null vector field, is called a Kerr-Schild
metric.}

The idea now is to find those transformations that will take the
part of Eq.~(\ref{kerr1b}) that is contained in the brackets to
the standard representation of the Minkowski  space. This will
lead us to the already mentioned transformation rules (A2).

Notice that the metric (\ref{KS_cart}) is analytical everywhere
except at $x^2 + y^2 =a^2$ (or else at $\rho=a$ and $z=0$).

\section{Geodesics of the Kerr spacetime in B-L coordinates}

The Kerr geodesics equations \citep{Misner}
\beq \Sigma^2 \left (\frac{dr}{d\tau} \right
)^2=[E(r^2+a^2)-aL_z]^2-\Delta[r^2+(L_z-aE)^2+Q]\equiv R \eeq

\beq \Sigma^2 \left (\frac{d\theta}{d\tau} \right
)^2=Q-\cot^2\theta L_z^2-a^2\cos^2\theta (e-e^2)\equiv \Theta \eeq

\beq \Sigma \left (\frac{d\phi}{d\tau}\right )=\csc^2\theta L_z+aE
\left (\frac{r^2+a^2}{\Delta}-1 \right )
-\frac{a^2L_z}{\Delta}\eeq

\beq \Sigma \left (\frac{dt}{d\tau} \right ) =E \left
[\frac{(r^2+a^2)^2}{\Delta}-a^2\sin^2\theta \right ]+aL_z \left
(1-\frac{r^2+a^2}{\Delta} \right )\eeq

\section{$\Omega_\phi$ for equatorial orbits in K-S coordinates}

The metric in K-S coordinates

\[ ds^2=-(1-\frac{2M\tilde{r}}{\Sigma})dv^2 +\Sigma
d\theta^2+2dv\,d\tilde{r}-\frac{4aM\tilde{r}\sin^2\theta}{\Sigma}dv\,d\tilde{\phi}'
-2a\sin^2\theta d\tilde{r}\,d\tilde{\phi}'\]
\beq+(\tilde{r}^2+a^2+\frac{2M\tilde{r}a^2\sin^2\theta}{\Sigma})\sin^2\theta\,
d\tilde{\phi}'^2 \label{KS}\eeq

where
\beq \tilde{r}^4-(x^2+y^2+z^2-a^2)\tilde{r}^2-a^2z^2=0 \eeq
\beq \Sigma=\tilde{r}^2+a^2\cos^2\theta .\eeq
Restricting the metric on the $\theta=\pi/2$ plane

\beq ds^2=-(1-\frac{2M}{\tilde{r}})dv^2 +2dv\,d\tilde{r}-\frac{4aM}{\tilde{r}}dv\,d\tilde{\phi}'
-2a d\tilde{r}\,d\tilde{\phi}'+(\tilde{r}^2+a^2+\frac{2Ma^2}{\tilde{r}})d\tilde{\phi}'^2. \label{KS2} \eeq

Two Killing vectors are associated with the $v-$ and $\tilde{\phi}'-$ invariance or Eq.~(\ref{KS2})

\beq \vec{\xi}=(1,0,0,0) \label{xi} \eeq

\beq \vec{\eta}=(0,0,0,1) \label{eta} \eeq

and the defined conserved quantities

\beq e\equiv -\vec{\xi} \cdot \vec{u} \label{e} \eeq

\beq l\equiv \vec{\eta} \cdot \vec{u}  \label{l}\eeq

with $e$ and $l$ being the conserved energy per unit rest mass and conserved angular momentum per unit rest mass respectively. $\vec{u}$ is the four-velocity.

From Eqs.~(\ref{e}) and (\ref{l}) and using $u^{\tilde{r}}=0$ 

\beq e=-g_{vv}u^{v}-g_{v\tilde{\phi}'}u^{\tilde{\phi}'} \label{ee}\eeq 

\beq l=g_{\tilde{\phi}'v}u^{v} +g_{\tilde{\phi}' \tilde{\phi}'}u^{\tilde{\phi}'}. \label{ll} \eeq

From Eqs.~(\ref{ee}) and (\ref{ll})

\beq u^v \equiv \frac{dv}{d\tau}=\frac{1}{\tilde{r}^2-2M\tilde{r}+a^2}[(\tilde{r}^2+a^2+\frac{2Ma^2}{\tilde{r}})e-\frac{2Ma}{\tilde{r}}l] \label{dvdtau}\eeq
\beq u^{\tilde{\phi}'} \equiv \frac{d\tilde{\phi}'}{d\tau}=\frac{1}{\tilde{r}^2-2m\tilde{r}+a^2}[(1-\frac{2M}{\tilde{r}})l+\frac{2Ma}{\tilde{r}}e] .\label{dphidtau} \eeq

$\Omega_{\tilde{\phi}'}$ is  defined as
\beq  \Omega_{\tilde{\phi}'}\equiv \frac{d\tilde{\phi}'/d\tau}{dv/d\tau}=\frac{(1-\frac{2M}{\tilde{r}})l/e+\frac{2Ma}{\tilde{r}}}{(\tilde{r}^2+a^2+\frac{2Ma^2}{\tilde{r}})-\frac{2Ma}{\tilde{r}}l/e}. \label{omega}  \eeq

We need to substitute for l/e in Eq.~(\ref{omega}). Heading for that we can make use of the normalization condition

\beq \vec{u} \cdot \vec{u}=-1,\eeq
 to find

\beq  \frac{e^2-1}{2}=\frac{1}{2}\frac{d\tilde{r}}{d\tau}+[-\frac{M}{\tilde{r}} + \frac{l^2-a^2(e^2-1)}{2\tilde{r}^2}-\frac{M(l-ae)^2}{\tilde{r}^3}].  \label{norma}\eeq

Based on Eq.~(\ref{norma}), one can define the effective potential

\beq V_{eff}=-\frac{M}{\tilde{r}} + \frac{l^2-a^2(e^2-1)}{2\tilde{r}^2}-\frac{M(l-ae)^2}{\tilde{r}^3}. \eeq

For a circular orbit of radius $\tilde{r}_0$, the effective potential has a minimum at $\tilde{r}_0$
\beq \frac{\vartheta V_{eff}}{\vartheta \tilde{r}}(\tilde{r}_0)=0 \label{cond1}\eeq
and also from  Eq.~(\ref{norma})
\beq V_{eff}(\tilde{r}_0)=\frac{e^2-1}{2}. \label{cond2} \eeq

From Eqs.~(\ref{cond1}) and (\ref{cond2}) one can solve for $l/e$:
start from Eq.~(\ref{cond1})

\[ \frac{\vartheta V_{eff}}{\vartheta \tilde{r}}=0 \Rightarrow \]

\beq  \tilde{r}= \frac{-a^2(e^2-1)+l^2+\sqrt{(a^2-a^2e^2+l^2)^2-12(l-ae)^2M^2}}{2M}. \label{r0} \eeq

Now, from Eqs.~(\ref{cond2}) and (\ref{r0}) solve for $l/e$

\beq \frac{l}{e}=\frac{-\tilde{r}^{3/2}(a^2+\tilde{r}(\tilde{r}-2M))+aM^{1/2}(a^2+\tilde{r}(3\tilde{r}-4M))}{M^{1/2}(a^2M-\tilde{r}(\tilde{r}-2M)^2}. \label{l/e} \eeq

Substituting Eq.~(\ref{l/e}) in Eq.~(\ref{omega}) will give
\beq  \Omega_{\tilde{\phi}'}=\pm \frac{M^{1/2}}{\tilde{r}^{3/2}\pm aM^{1/2}} \label{OMEGA} \eeq

with the upper and lower signs corresponding to co-rotating and counter-rotating orbits.

\clearpage

\begin{deluxetable}{c c c c c c c c c c c}
\tablecaption{Runs and Results for Simulations of Equatorial Mergers\label{table1}}
\tablewidth{0pt}
\tablehead{
\colhead{\textbf{Run}} & \colhead{\textbf{E1}} & \colhead{\textbf{E2}} & \colhead{\textbf{E3}} & \colhead{\textbf{E4}} &
\colhead{\textbf{E5}} & \colhead{\textbf{E6}} & \colhead{\textbf{E7}} &
\colhead{\textbf{E8}} & \colhead{\textbf{E9}} &
\colhead{\textbf{E10}}
}
\startdata
 \textbf{a/M} & 0.99 & 0.95 & 0.9 & 0.8 & 0.75 & 0.6 & 0.5 & 0.2 & 0.1 & 0 \\
 \textbf{Total NS mass outside $r_+$} & 33 \% & 32\% & 26\% & 4\% &
1\% & 0\% & 0\% & 0\% & 0\% & 0\% \\  
\textbf{Bound NS mass outside $r_+$} & 2.5\% & 2\% & 0\% & 0\% & 0\% & 0\% & 0\% & 0\% & 0\% & 0\% \\
\enddata
\end{deluxetable}

\begin{deluxetable}{crrrrr}
\tablecaption{Initial Conditions and Results for Inclined Mergers\label{table2}}
\tablewidth{0pt}
\tablehead{
\colhead{\textbf{Run}} & \colhead{\textbf{I1}} & \colhead{\textbf{I2}} & \colhead{\textbf{I3}} & \colhead{\textbf{I4}} &
\colhead{\textbf{I5}} 
}
\startdata
\textbf{a/M} & 0.99 & 0.99 & 0.99 & 0.99 & 0.99 \\
$\mathbf{r_0 (M)}$  & 12 & 11.5 & 12 & 13 & 10 \\
\textbf{inclination angle ($^o$)} & 29.6577 & 45.031 & 70.3989 & 89.9494 & 180 \\
\textbf{Total NS mass outside $r_+$} & 37\% & 25\% & 0\% & 0\% & 0\% \\
\textbf{Bound NS mass outside $r_+$} & 6\% & 0\% & 0\% & 0\% & 0\%\\
\enddata
\end{deluxetable}

 \clearpage

\clearpage

\begin{figure}
\begin{center}
\epsscale{1}
\plottwo{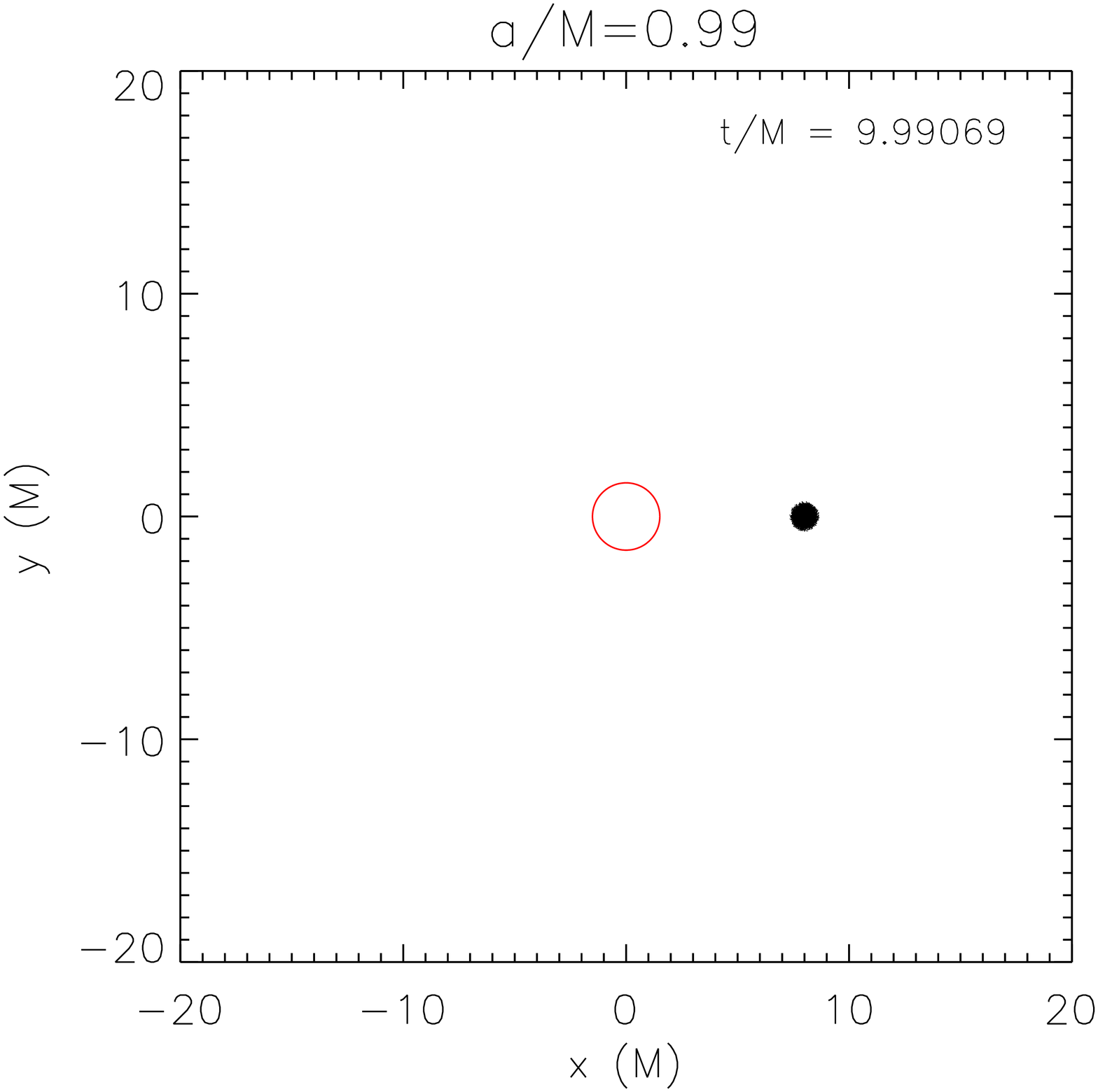}{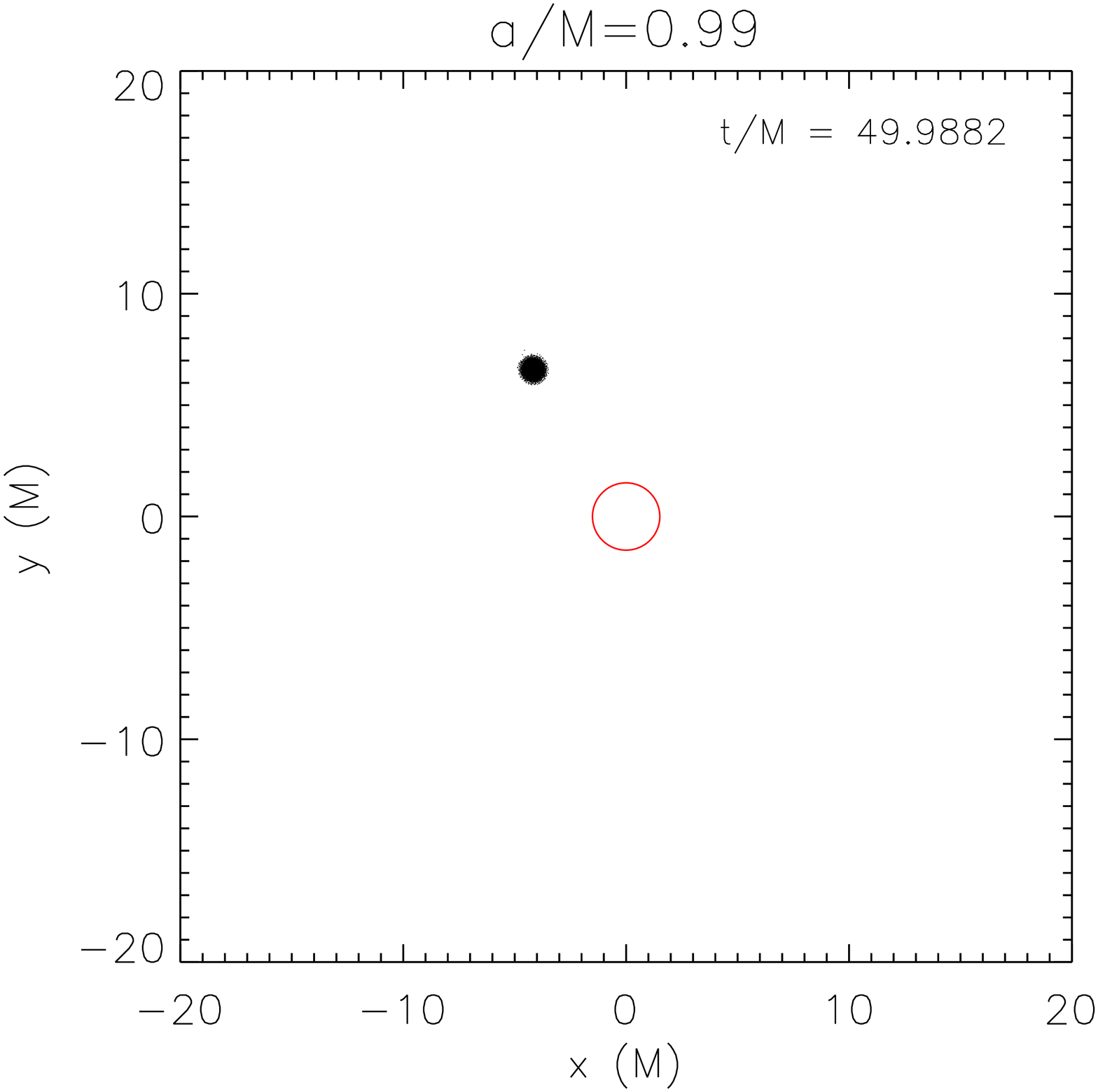}
\plottwo{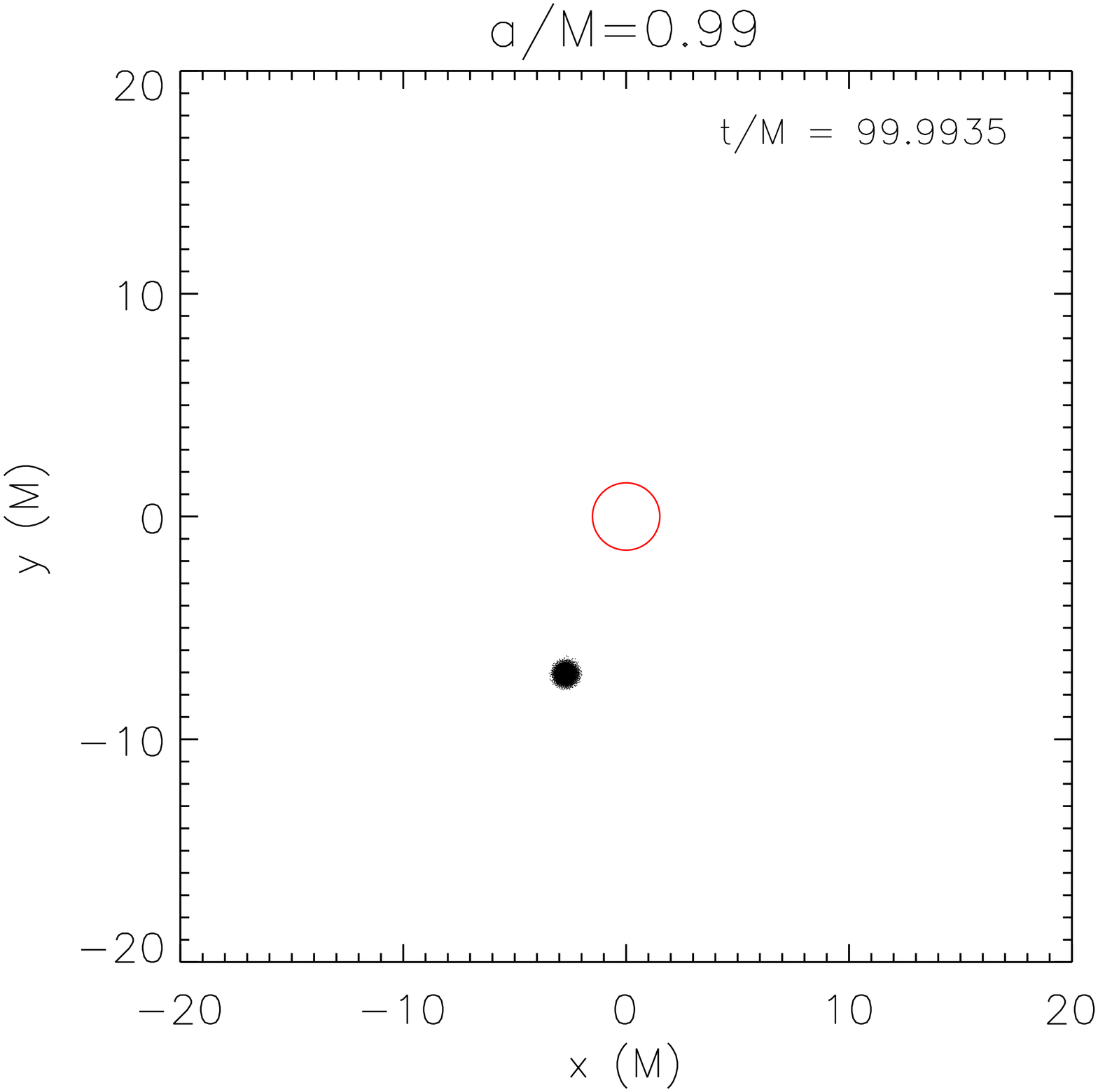}{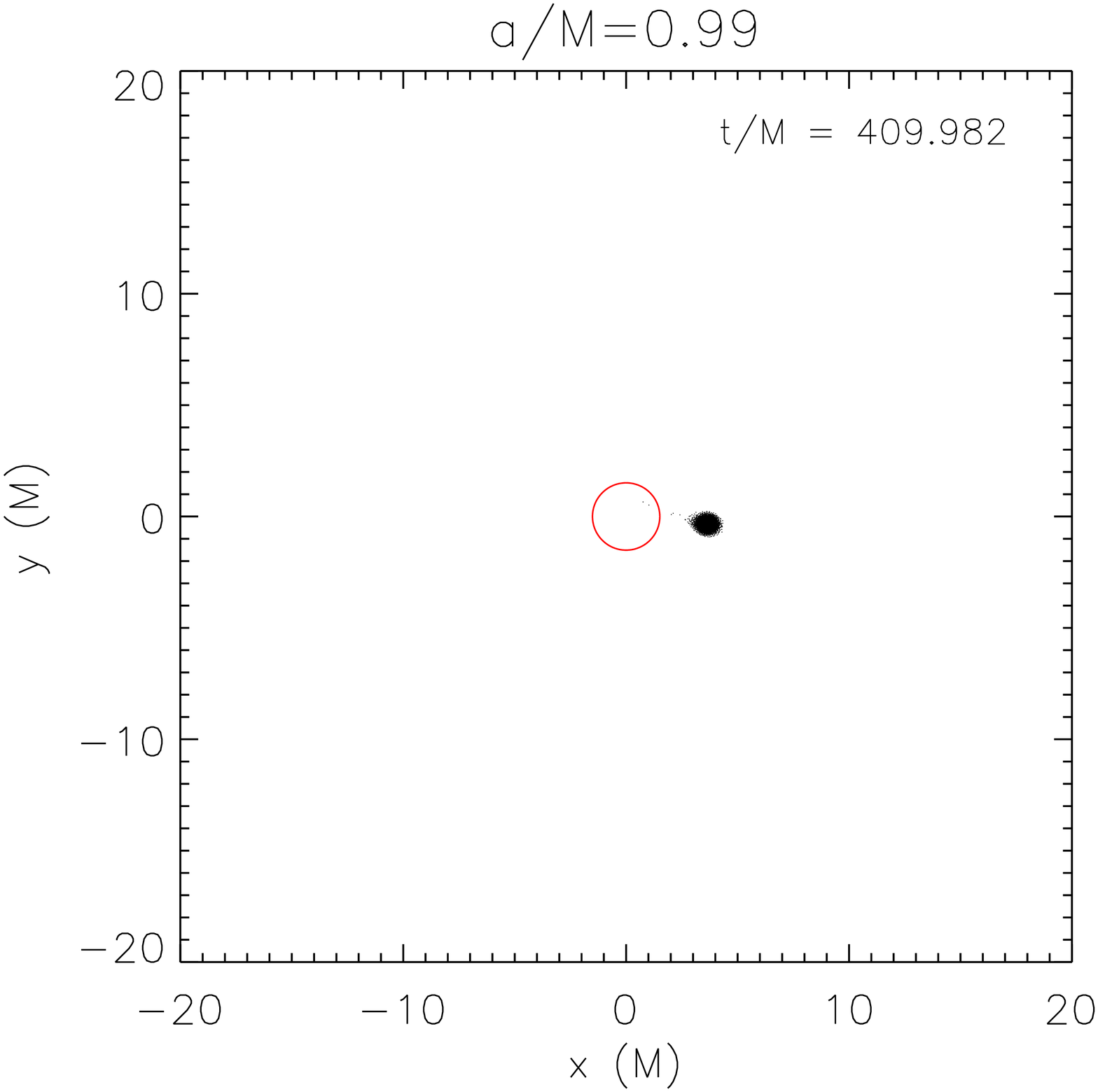}
\plottwo{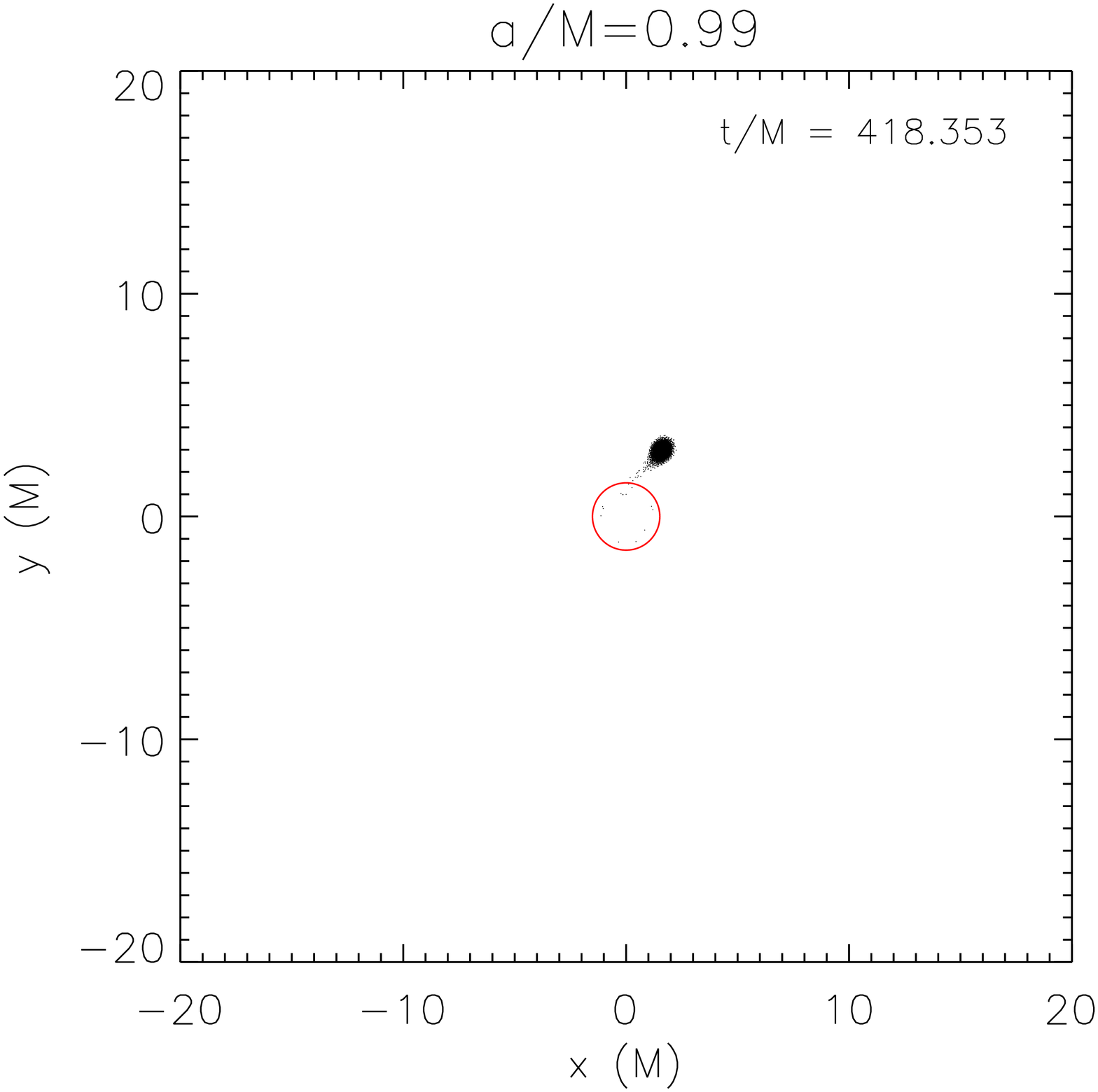}{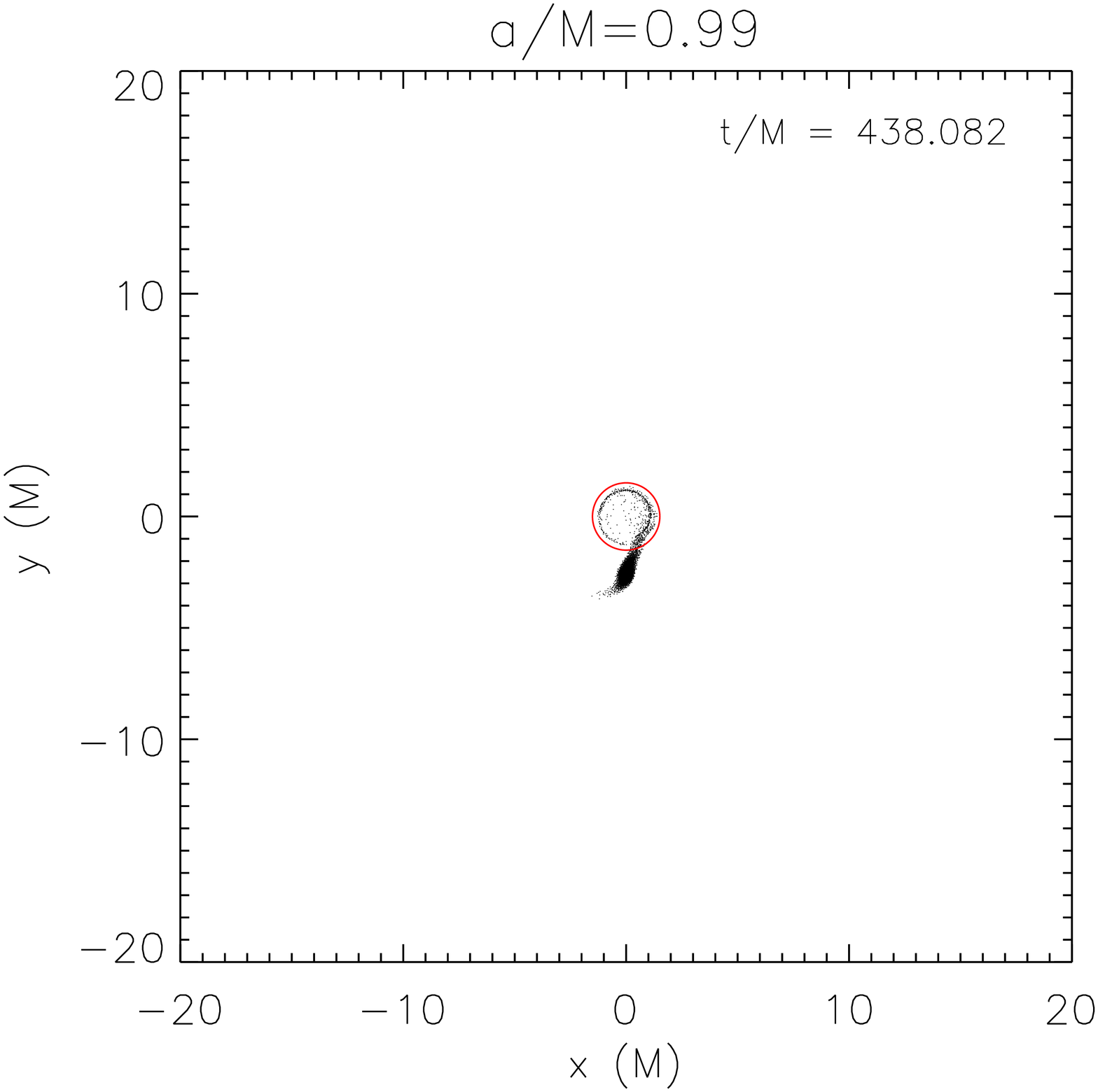}
\end{center}
\end{figure}

\end{document}